\documentclass[letterpaper,12pt]{article}

\usepackage{amssymb}

\usepackage{tikz}
\usetikzlibrary{arrows.meta}
\usepackage{amsmath,amsfonts,amssymb,amsthm,epsfig,xspace,graphics,graphicx}
\usepackage[symbol]{footmisc}
\usepackage{float}

\usepackage[hmargin=1in,vmargin=1in]{geometry}
\usepackage{amsmath,amsfonts,amssymb,amsthm,epsfig,xspace,graphics,graphicx}
\usepackage{tikz}
\usepackage[round,authoryear]{natbib}
\usepackage[normalem]{ulem}
\usepackage{longtable}
\usepackage{array}
\usepackage{xfrac}
\usepackage{enumerate}
\usepackage{mathtools}

\newcommand{\mat}{\textrm{\tiny R}}%
\newcommand{\pards}[2]{\mbox{$\dfrac{\partial #1}{\partial {#2 }}$}}
\newcommand{\trans}{{\mskip-2mu\scriptscriptstyle\top}} 
\newcommand{\Tr}{\hbox{\rm tr}\mskip2mu}

\newcommand{\Div}{\hbox{\textrm{Div}}\mskip2mu}    
\newcommand{\id}{{\bf 1}}

\newcommand{\calB}{\mathcal{B}}%
\newcommand{\calC}{\mathcal{C}}%
\newcommand{\calP}{\mathcal{P}}%
\newcommand{\calS}{\mathcal{S}}%

\newcommand{\bfb}{{\bf b}}\newcommand{\bfB}{{\bf B}}%
\newcommand{\bfC}{{\bf C}}%
\newcommand{\bfd}{{\bf d}}%
\newcommand{\bfe}{{\bf e}}%
\newcommand{\bfF}{{\bf F}}%
\newcommand{\bfj}{{\bf j}}%
\newcommand{\bfn}{{\bf n}}%
\newcommand{\bfR}{{\bf R}}%
\newcommand{\bft}{{\bf t}}\newcommand{\bfT}{{\bf T}}%
\newcommand{\bfu}{{\bf u}}%
\newcommand{\bfv}{{\bf v}}\newcommand{\bfV}{{\bf V}}%
\newcommand{\bfx}{{\bf x}}\newcommand{\bfX}{{\bf X}}%

\newcommand{\Def}{\overset{\text{def}}{=}}
\newcommand{\grad}{\text{grad}\,}
\newcommand{\bfchi}{\boldsymbol{\chi}}%

\begin{document}



\title{Modeling coupled electrochemical and mechanical behavior of soft ionic materials and ionotronic devices}


\author{Nikola Bosnjak\thanks{Corresponding author: nikola.bosnjak@cornell.edu}
, Max Tepermeister and Meredith Silberstein\\
\\
Cornell University\\ Sibley School of Mechanical and Aerospace Engineering\\
Ithaca, NY 14850,  USA}

\maketitle


\begin{abstract}
Recently there has been an increase in demand for soft and biocompatible electronic devices capable of withstanding large stretch. Ionically conductive polymers present a promising class of soft materials for these emerging applications due to their ability to realize charge transport across the polymer network, while preserving the desired mechanical and chemical features.  As opposed to electron transfer in traditional electrical conductors, the charge transport across these polymers is achieved through ion migration. When such materials are used in combination with electrical systems, they are known as ionotronic devices. The ability to simulate device performance based on its material composition and geometry would accelerate and improve ionotronic device design. The main challenge in developing reliable simulation capabilities for ionically conductive polymers is the complex and coupled electro-chemo-mechanical behavior.  In this work we address this challenge by introducing a multiphysics framework incorporating the coupled effects of ion transport, electric fields and large deformation.  The utility of the developed multiphysics model is showcased by simulating representative ion transport problems and the operation of soft ionotronic devices.

\end{abstract}




\section{Introduction}

Soft and deformable ionic conductors, such as hydrogels and polyelectrolytes, present a promising class of materials for integration into soft circuits and ionotronic devices. Traditional electric circuits are mainly composed of hard and rigid electrically conductive materials, and are therefore ill suited for applications requiring large deformation. Flexible conductors fall into one of four categories: (i) hard conductors that flex by bending/unbending \citep{kim2007complementary,lee2012very,song2014superstable}, (ii) liquid metal encapsulated within a soft elastomer matrix \citep{markvicka2018autonomously,yan2019solution}, (iii) electrically conducting polymers \citep{rivers2002synthesis,liu2018electrically}, and  (iv) ionically conductive materials \citep{chen2014highly,shi2018highly}.  Soft ionically conducting materials typically consist of polymer networks with mobile ionic groups (e.g., dissociated salts); charge transport across the network occurs by motion of these ions. These soft ionically conducting materials are promising because of environmental responsivity, intrinsically large strain to failure, tunability, and biocompatiblity.


The applications of soft ionically conductive polymers are vast, from healthcare and soft robotics to sensors \citep{kim2007electroactive,rus2015design,lacour2016materials,yang2018hydrogel}. Soft ionically conductive polymers are utilized in design of soft circuits, where they are employed as soft cables \citep{yang2015ionic,odent2017highly}, diodes \citep{cayre2007polyelectrolyte,gabrielsson2012ion,wang2019stretchable} and transistors \citep{tybrandt2010ion,shin2014highly}.  They also find application in soft actuators, allowing for design of  artificial muscles \citep{jung2010electro,han2018,morales2014electro} and biomimetic robots \citep{yeom2009biomimetic,li2017fast}. Implementing these materials as sensors in soft robotics helps maintain the desired mechanical properties as opposed to more traditional rigid sensors \citep{sun2014ionic,robinson2015integrated,liu2020poly,wang2021highly}.  Moreover, energy harvesting devices based on soft ionic conductors, such as the ones published by \citet{aureli2009energy,tiwari2010disc,hou2017flexible,zhou2017biocompatible}, and \citet{kim2020ionoelastomer}, allow for conversion of mechanical deformation into electric potential and current. In energy conversion technologies, such as batteries and fuel cells, these materials are typically referred to as polymer electrolytes or polymer electrolyte membranes \citep[cf. e.g., ][]{wan2019ultrathin,chen2017improving,banerjee2004nafion}. In this context they provide a mechanical barrier in addition to the ionic conductivity and electrical insulation necessary in a liquid electrolyte (solution with dissociated ions). 

\begin{figure}[hb!]
\begin{tabular}{cc}
\begin{tikzpicture}
\node[inner sep=0] (image) at (0,0){\includegraphics[width=0.45\linewidth]{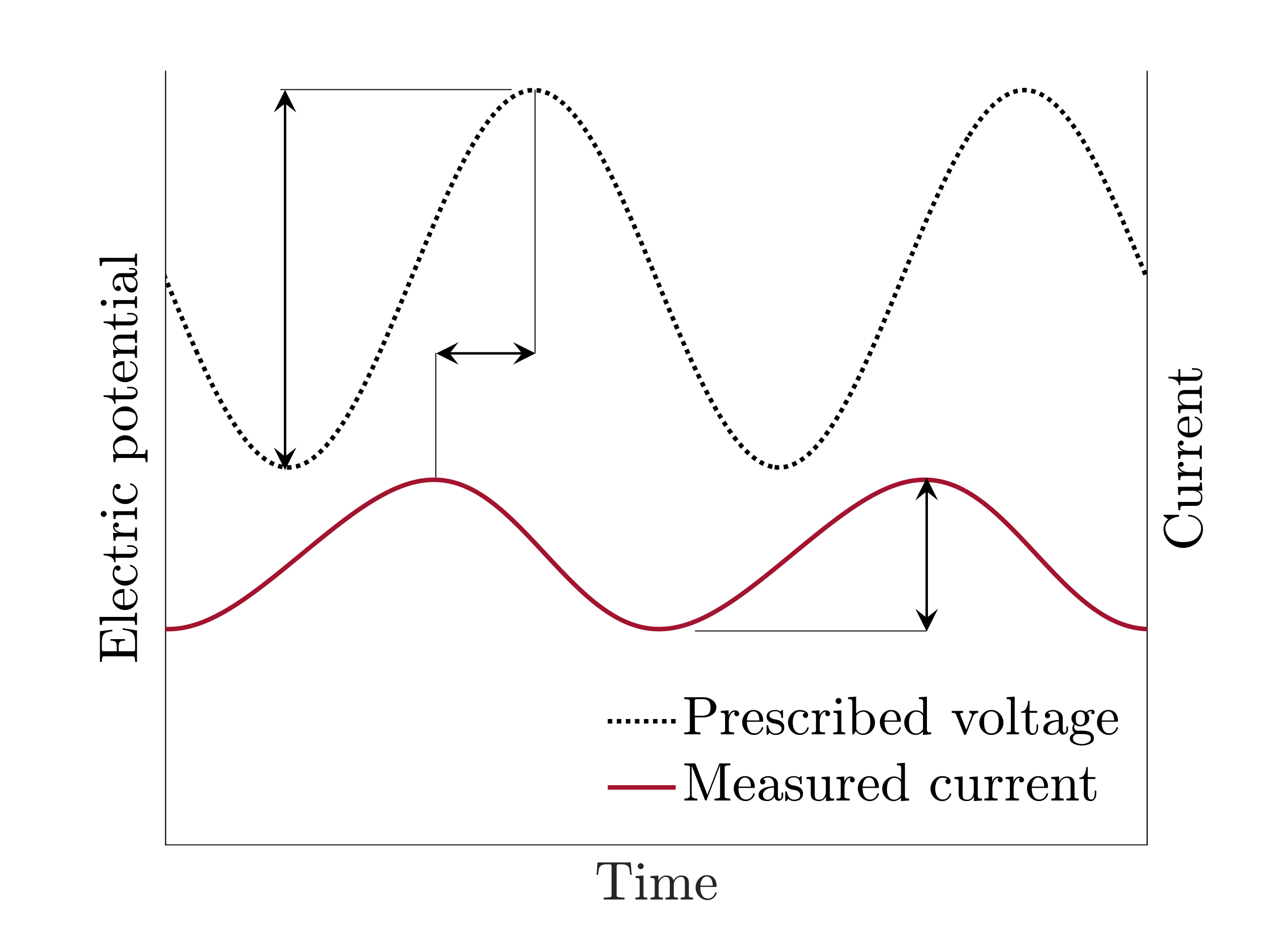}};

\draw (-1.65,1.5) node {\footnotesize $2|$V$|$};

\draw (-.85,1.05) node {\footnotesize$\alpha\delta$};

\draw (1.4,-.5) node {\footnotesize 2$|$I$|$};

\end{tikzpicture}
&
\begin{tikzpicture}
\node[inner sep=0] (image) at (0,0){\includegraphics[width=0.45\linewidth]{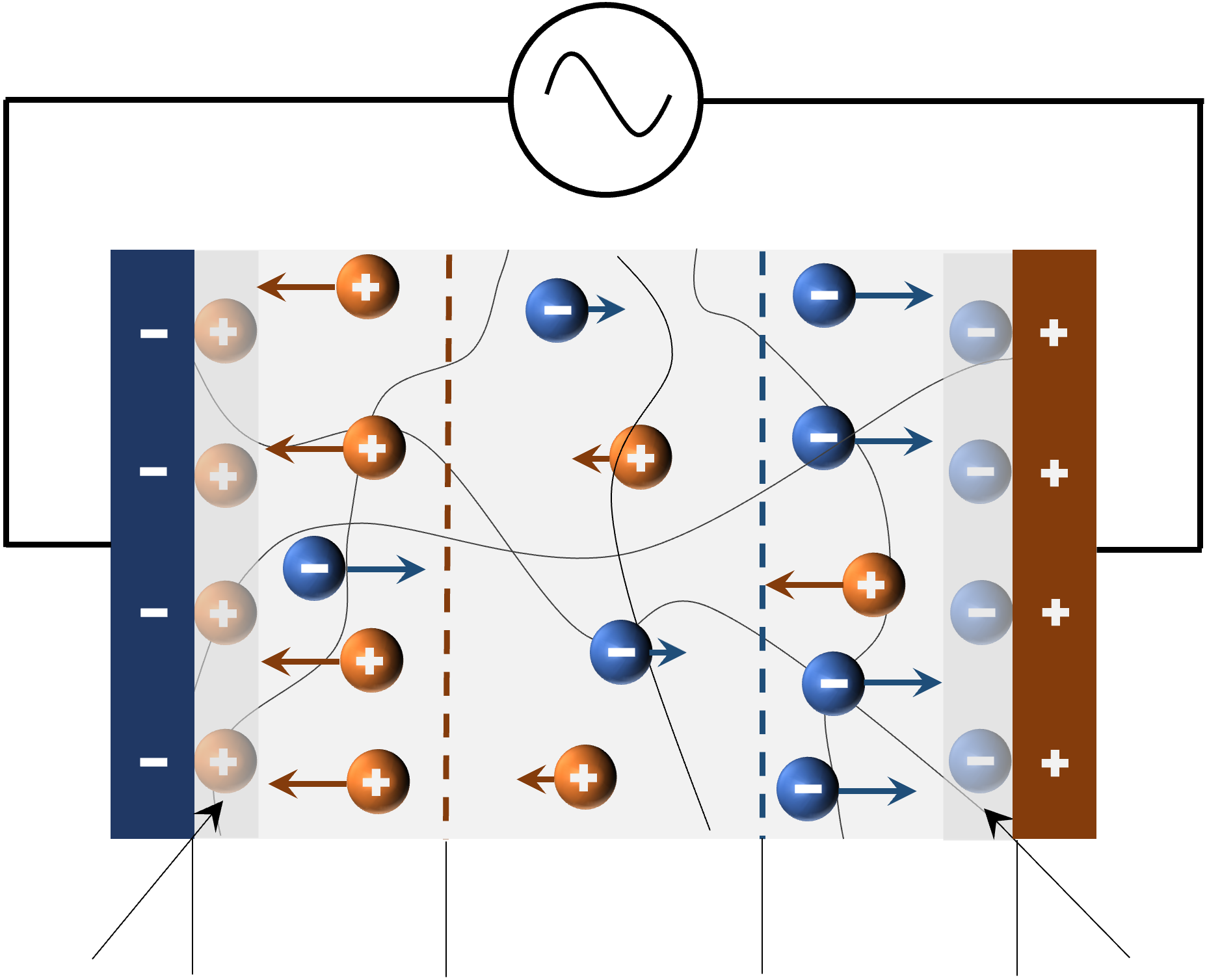}};

\draw (-3.2,-3.1) node {\footnotesize Stern };
\draw (-3.2,-3.6) node {\footnotesize layer};

\draw (-1.7,-3.1) node {\footnotesize diffuse };
\draw (-1.7,-3.6) node {\footnotesize layer};

\draw (0,-3.1) node {\footnotesize bulk };
\draw (0,-3.6) node {\footnotesize electrolyte};

\draw (1.7,-3.1) node {\footnotesize diffuse };
\draw (1.7,-3.6) node {\footnotesize layer};

\draw (3.2,-3.1) node {\footnotesize Stern };
\draw (3.2,-3.6) node {\footnotesize layer};

\draw (0.85,2.75) node {$+$};
\draw (-0.85,2.75) node {$-$};

\end{tikzpicture}\\

a)&b)
\end{tabular}
\caption{Fundamentals of EIS and underlying physics. a) EIS involves applying a small amplitude sinusoidal voltage V and measuring the current signal I.  The two sinusoidal waves are then compared to obtain the impedance magnitude as $|$V$|/|$I$|$, and phase shift angle $\delta$. Here, $\alpha$ denotes the frequency dependent conversion factor given in units s$/^\circ$.  b) A snapshot of the ionic conductor subjected to an external AC voltage source V. The electric field is established and the mobile ions migrate towards the oppositely charged electrode. At the polymer/electrode interface the ions screen the charges on the electrode surface, thus forming a diffuse double layer. The Stern layer represents the first ion layer adsorbed onto the surface, denoted with gray shading (note that we do not explicitly model this layer).}
\label{fig:PhysicsAndEIS}
\end{figure}
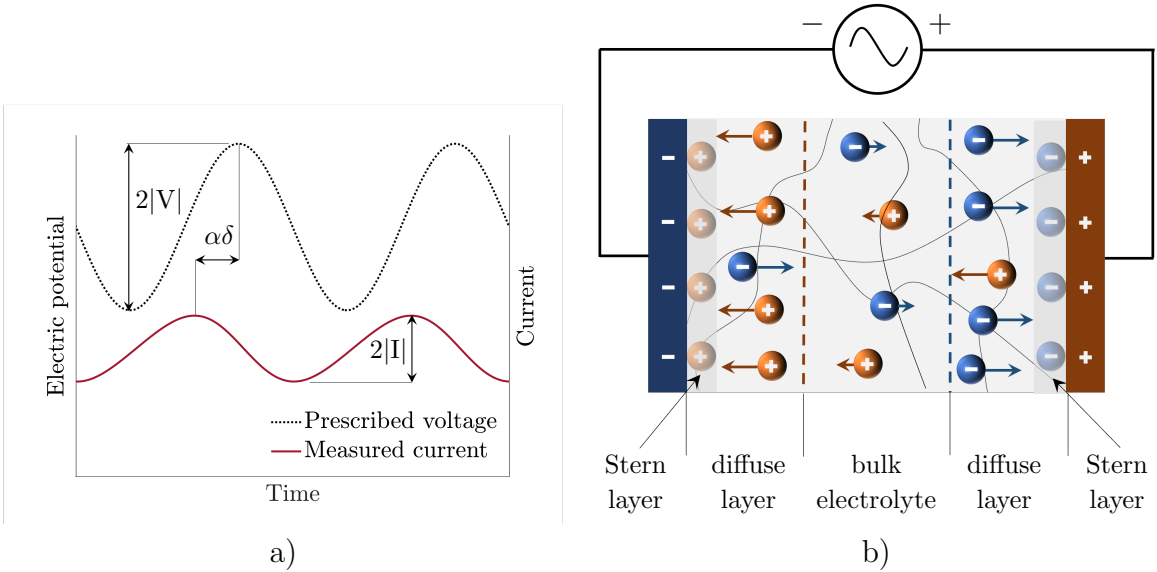

Thorough characterization of the electrochemical response greatly faciliates successful implemention of soft polymers into ionotronic devices. Electrochemical impedance spectroscopy (EIS) is a common experimental method that involves applying a small electric potential perturbation across the specimen while measuring the current signal.  The perturbation is typically applied as a sinusoidal voltage signal with a constant and small amplitude, to ensure the material response is within the linear regime.  The EIS data is then analyzed by comparing the voltage and current waves, as shown in Figure \ref{fig:PhysicsAndEIS}a. The experimental data is commonly presented as Bode plots, with the impedance magnitude $|$Z$|$ computed as the voltage-to-current amplitude ratio $|$V$|\,/\,|$I$|$, and plotted together with the phase shift angle $\delta$. Overall, EIS provides information about the mechanisms of charge conduction, and can be used to evaluate the performance of ionic materials and ionotronic devices.


Despite the rapid increase in the number of potential applications, there is a lack of overarching design tools, hindering the further advance of ionotronic technology.  The main challenge in developing reliable simulation capabilities for ionotronics is the coupled electro-chemo-mechanical behavior.  The most common approach for evaluating ionotronic device and material performance is fitting EIS data to equivalent circuit models. While these models provide information on the response of systems within electrical circuits, there is minimal insight into the relations among the material, device structure, charge transport mechanisms, and observed electrochemical response. EIS data can be fitted to many different equivalent circuits, and understanding the mechanisms driving the electrochemical phenomena is required for selecting an accurate circuit model. Moreover, many soft ionic conductors are expected to operate under large mechanical deformation -- another feature not captured by equivalent circuit models.


Aside from the equivalent circuit modeling, there are many continuum-level models that capture the electrochemical behavior of ionotronic materials.  The seminal works of Walther Nernst \citep{nernst1888kinetik} and Max Planck \citep{planck1890ueber} capture the transport of charged species in an electrolyte. The introduced modification to the conservation of mass included the influence of electric fields on ionic diffusion -- a relation commonly referred to as the Nernst-Planck equation. Additionally, the charge concentration alters the electric field; and the relation between the two is given by one of the four Maxwell equations -- Gauss's law \citep{gauss1877theoria}, commonly represented in the form of Poisson's equation \citep{poisson1826memoire}. These coupled effects of charge transport and electric potential are captured by the Poisson-Nernst-Planck relations. Building upon the seminal theories, there have been many notable works capturing charge transport in the presence of electric fields \citep[cf.,e.g.,][]{buck1969diffuse,mizushina1971electrochemical,kornyshev1981conductivity,corry2000tests,gillespie2002coupling,bazant2004diffuse}. Such models allow for computational investigation of electrochemical systems and open the door for analysis of advanced ionotronic devices through numerical simulations on intricate geometries and under complex boundary conditions. Towards this goal, different numerical methods have been proposed for solving the electrochemically coupled problems, including the finite difference method \citep{brumleve1978numerical,flavell2014conservative,liu2014free}, finite volume method \citep{lopreore2008computational}, and finite element method \citep{lu2010poisson,eisenberg2011mathematical,paz2011modeling}.

The coupling between the electrochemical processes and large mechanical deformation in polyelectrolytes has been modeled previously in the context of environmentally responsive ionic gels and ionic electromechanical transducers. In current literature on electrochemically and mechanically coupled response of environmentally responsive polymers, there are various models accounting for the ionic-driven swelling of polymer gels \citep[cf., e.g.,][]{doi1992deformation,de2000mechanoelectric,wallmersperger2004coupled,keller2011modeling,drozdov2015modeling,leichsenring2017modeling,zhang2020kinetics,narayan2022coupled}.
Recent years also saw notable efforts towards developing multiphysics framework for coupled electrochemistry and mechanics of soft ionic actuators, with most authors focusing on the mechanical response of ionic polymer-metal composites under external electric fields \citep[cf., e.g., ][]{nemat2002micromechanics,toi2005finite,nardinocchi2011thermodynamically,rossi2018thermodynamically,narayan2021coupled}. In addition, similar models have been applied to investigate the perturbation in electrochemical behavior in response to an applied mechanical deformation in sensor-like systems \citep[cf., e.g., ][]{ganser2019finite}. To complement modeling efforts, the numerical approach laid out by \citet{narayan2022coupled} allows for solving the coupled set of electro-chemo-mechanical equations in three-dimensional space under large mechanical deformation. Nonetheless, in current state of the art there is a lack of robust computational capabilities for the influence of multiple mobile ionic species, charged polymeric backbones and large mechanical deformation on ionotronic material and device performance. While out of scope of this research, we also note that literature on modeling of energy storage devices, such as batteries, provides ample electro-chemo-mechanical models. However, the coupled electrochemical and mechanical models developed for batteries are usually focusing on the swelling-induced deformation and electrochemical reactions \citep[cf. e.g.,][]{golmon2009numerical,bucci2014measurement,rejovitzky2015theory,sauerteig2018electrochemical}.

To bridge the gap between the increasing demand for soft ionotronic devices and lack of computational tools, in this work, we introduce a continuum-level multiphysics framework and a constitutive model incorporating the coupled effects of electric field, ion transport, and large mechanical deformation on the overall behavior of soft ionotronic materials and devices. We account for the influence of both neutral and charged polymeric backbones, along with the diffusion of multiple ionic species. Our modeling effort is aimed at simulating the electrochemical characterization of soft ionic conductors, along with the operation of ionotronic devices. The usefulness of such a computational tool is twofold: (i) it provides the guidance for modifying the polymer structure towards target electrochemical properties and (ii) it allows for design evaluation of devices through finite element analysis. In our first step towards this goal, we numerically implement the electrochemical framework (i.e., the framework in the absence of mechanical deformation) as a 1D custom finite element code in Matlab. We showcase the reliability of our computational approach by comparison against analytical solutions and EIS data. Then, we include the effects of large mechanical deformation and  implement the framework as a user element (UEL) subroutine in Abaqus/Standard. We demonstrate the potential of our numerical tool to understand and design ionotronic systems by simulating the operation of devices found in literature. Accordingly,  the remainder of this paper is organized as follows: in Section \ref{sec:Modeling} we develop a continuum-level thermodynamic framework and a constitutive model; in Section \ref{sec:Simulations} we present numerical simulation results; in Section \ref{sec:Conclusion} we provide concluding remarks.


\section{Multiphysics model}\label{sec:Modeling}

In this section we present a framework and constitutive model for electro-chemo-mechanics of ionically conducting polymers. To facilitate understanding, we start with an overview of the underlying physics, including typical transport mechanisms.


\subsection{Overview of the governing physics} \label{sec:GoverningPhysicsAndEIS}

Polymeric materials consist of mutually entangled or chemically cross-linked polymer chains which form a polymer network. Ionically conductive polymers contain mobile charged species, i.e. ions. Subjecting these materials to an electric field causes the mobile ions to migrate towards the oppositely charged electrode.  At the polymer/electrode interface,  ions screen charge at the electrode surface, thus forming a diffuse double layer, as shown in Figure \ref{fig:PhysicsAndEIS}b. In this study we assume there is no charge transfer across the electrode/polymer interface, which is analogous to saying we have blocking electrodes. The layer of ions at the interface is known as the Stern layer and its thickness is on the order of angstroms.  A common way to quantify the distance over which mobile ions screen charge at the electrode is by computing the Debye length, which is found assuming a Boltzmann distribution for the concentration of charges:
\begin{equation}
\lambda_D = \sqrt{\frac{\varepsilon_r \varepsilon_0 RT}{\left(e N_a \right)^2  c^{E} }}\,,
\label{eqn:DebyeLength}
\end{equation}
where $\varepsilon_r$ is the relative permittivity, $\varepsilon_0$ is the dielectric permittivity of vacuum, $R$ is the universal gas constant, $T$ is absolute temperature, $e$ is the elementary charge, and $N_a$ is Avogadro's number. The average concentration of charged species in the electrolyte is obtained as $c^{E} = \sum \left(z^{(i)}\right)^2 c^{(i)}_0\,$, where  $z^{(i)}$ denotes the charge and $c^{(i)}_0$ the initial concentration of $i$-th mobile ionic species.

\subsubsection{Polymer backbone charge}\label{sec:BackboneCharge}
The polymer chains in the network can themselves be either neutral or charged. The degree of charge often depends on pH of the environment, i.e., association/dissociation chemical reactions might occur on groups tethered to the polymer backbone, changing both the degree of fixed charge and concentration of mobile ions. For example, at high pH, mobile protons (H$^+$) can easily dissociate from carboxyl groups RCOOH, leaving the negative carboxylate RCOO$^-$ tethered to the backbone ($\textrm{R}\textrm{COOH}\rightleftharpoons \textrm{R}\textrm{COO}^- + \textrm{H}^+$).  Backbones with these negatively charged groups are termed anionic backbones, and the materials are called polyanions or polyanionic.  On the other hand, at low pH, the hydrogen can associate with certain functional groups, such as a neutral amino group RNH$_2$,  yielding a positively charged and tethered RNH$_3^+$ ($\textrm{R}\textrm{NH}_2 + \textrm{H}^+\rightleftharpoons \textrm{R}\textrm{NH}_3^+$). The positively charged backbones are termed cationic backbones, and the materials are called polycations or polycationic.

\subsubsection{Charge transport mechanisms}\label{sec:TransportMechanism}

There are four major mechanisms for achieving charge transport in ionically conductive polymers; and the effective diffusivity of mobile ions depends on the particular mechanisms dominating the charge transport:
\begin{enumerate}[(1)]
    \item \underline{Diffusion through an electrolyte} is dominant in swollen polymer networks, such as hydrogels, where transport of dissolved and dissociated salts is almost unaffected by the highly hydrated polymer network. The diffusion values are typically  $10^{-9}$m$^2$/s plus or minus an order of magnitude. \citep{lobo2001transport,wu2009effect,schuszter2017determination}
    \item \underline{``Hopping"} occurs along the charged functional groups tethered to the backbone of nearly dry and solvent-free polymers. The diffusion coefficients for this regime vary drastically based on the material and processing (e.g. $\sim 10^{-11}\,$m$^2$/s  \citep{choi2005thermodynamics}, $\sim 10^{-12}\,$m$^2$/s  \citep{ochi2009investigation}, $\sim 7\cdot 10^{-15}\,$m$^2$/s \citep{kim2020ionoelastomer}).
    \item The \underline{Grotthuss mechanism} (transport by hopping along water molecules) is dominant in hydrated proton exchange membranes (PEM), and can be also found in anion exchange membranes (AEM) \citep[cf., e.g.,][]{miyake2015grotthuss}. At a continuum scale, this process yields a relatively high diffusion coefficient, up to $\approx 7\cdot 10^{-9}\,$m$^2$/s \citep{cukierman2006tu,agmon1995grotthuss,paddison2002nature,choi2005thermodynamics}.
    \item \underline{Vehicular or \emph{en masse} transport} involves migration of fully solvated ions together with the solvent, i.e., using solvent molecules as a vehicle for transport. The resulting diffusion coefficients are $\approx 2 \cdot 10^{-9}\,$m$^2$/s  \citep{agmon1995grotthuss,li2001theoretical,choi2005thermodynamics}.
\end{enumerate}
In our continuum-level modeling approach we are not explicitly modeling the above mentioned micromechanics of charge transport. However, we use these as guidelines for selecting a reasonable diffusion coefficient value for each material in our simulations.

It is worth noting that ions diffuse significantly faster than water in the presence of an electric field.  Consequently, the change in water content is negligible for the boundary conditions and timescales under consideration here. We therefore assume the hydration level uniform and constant except when directly coupled to ion motion. In addition, we consider pH to affect only the initial  charge of the polymer chains and the electrolyte strength, while neglecting any pH driven swelling. We also choose to neglect thermal effects in this manuscript since they will be negligible in the devices we examine.


\subsection{Thermodynamic framework} \label{sec:Balance laws}
\renewcommand{\thefootnote}{\fnsymbol{footnote}} 
We commence the modeling procedure by introducing the useful kinematic relations, and consider a body $\calB_\mat$ in its referential configuration undergoing a motion $\bfchi$ to its spatial (i.e., deformed) configuration $\calB$. The deformation gradient and velocity are then given by\footnote[1]{The symbols $\nabla$ and Div denote the gradient and divergence with respect to the material point $\bfx_\mat$ in the referential configuration; grad and div denote these operators with respect to the point $\bfx=\bfchi(\bfx_\mat,t)$ in the spatial configuration. Symbol $\dot{\square}$ denotes the time derivative d${\square}$/d${t}$.} 
\begin{equation}
 \bfF = \nabla \bfchi \quad \textrm{and} \quad \bfv = \dot{\bfchi}
 \label{eqn:Kinematics}
\end{equation}
respectively, along with the volumetric deformation $J=\det\bfF$. In addition to the mechanical deformation, the presence of ionic species induces deformation of the polymer network. Therefore, we make use of an approach commonly found in gel mechanics literature \citep[cf., e.g.,][]{hong2008,chester2010}, and decompose the deformation gradient into a mechanical and swelling part
\begin{equation}
 \bfF = \bfF^m\bfF^s\,,
 \label{eqn:GradientDecomposition}
\end{equation}
where $\bfF^m$ and $\bfF^s$ are the mechanical and swelling components of the deformation gradient respectively. The volumetric deformation can then be decomposed into
\begin{equation}
 J = J^m J^s,
 \label{eqn:VolumetricDecomposition}
\end{equation}
where $J^m$ and $J^s$ represent the mechanical and swelling volume change respectively. We assume the change in volume results directly from the change in mobile species concentration, with corresponding volumetric swelling in the form
\begin{equation}
J^s = \det \bfF^s = 1+ \sum_i \Omega^{(i)} \left(c_\mat^{(i)}- c_{\mat 0}^{(i)}\right)\,,
 \label{eqn:VolumetricDecomposition2}
\end{equation}
where $\Omega^{(i)}$, $c_\mat^{(i)}$, and $c_{\mat 0}^{(i)}$ are the molar volume, current concentration and initial concentration of the $i$-th mobile ionic species in the referential configuration, respectively.

The right Cauchy-Green tensor is given by
\begin{equation}
  \bfC = \bfF^{\,\trans}\bfF \, = \bfC^m \bfC^s, 
\end{equation}
where $\bfC^m$ and $\bfC^s$ are the mechanical and swelling part of the right Cauchy-Green tensor, respectively. In addition, the left Cauchy-Green tensor is given by
\begin{equation}
  \bfB = \bfF\bfF^{\,\trans} \,. 
\end{equation}

Next, to capture the multiphysics behavior of soft ionic conductors, we assume the material is affected by (i) the electric field, (ii) the balance of ionic species, and (iii) mechanical stress. As in much of the previous literature on electrically responsive polymers \citep[cf., eg.][]{wang2016modeling,rossi2018thermodynamically,narayan2021coupled}, we choose the form of Gauss's law that gives the relation between the concentration of charged species and the electric displacement in the referential configuration $\bfd_\mat$:
\begin{equation}
 eN_a \left( \sum_i \left(z^{(i)}c^{(i)}_\mat\right) + z^{(\textrm{fix})}c^{(\textrm{fix})}_\mat \right)=\Div \, \bfd_\mat  \,,
\label{eqn:GaussLaw}
\end{equation}
where $z^{(i)}$ denotes the charge number of $i$-th mobile species, and $z^{\textrm{(fix)}}$ and  $c_\mat^{\textrm{(fix)}}$ the charge number and concentration of chemical groups tethered to the polymer matrix, and consequently fixed in position, in the reference configuration. We also introduce the net charge density in the spatial configuration as
\begin{equation}
    \rho = \sum_i \left(z^{(i)}c^{(i)}\right) + z^{(\textrm{fix})}c^{(\textrm{fix})}\,.
\end{equation}

To capture the change in concentration of mobile ions, we employ balance of mass in the form  
\begin{equation}
\dot{c}_\mat^{(i)} = - \textrm{Div}\, \bfj_\mat^{(i)} = -J \textrm{div}\, \bfj^{(i)}\,,
\label{eqn:MassBalance}
\end{equation}
where $\bfj_\mat^{(i)}$ and $\bfj^{(i)}$ denote the flux for each mobile species in the referential and spatial configuration, respectively.

In the absence of inertial effects, the balance of forces and moments is given by
\begin{equation}
\Div \bfT_\mat + \bfb_\mat = \bf0
\label{eqn:ForceBalance}
\end{equation}
where $\bfT_\mat$ represents the first Piola stress and $\bfb_\mat$ the body forces per unit reference volume. 

Next, we consider an arbitrary part $\calP_\mat$ of the body in its referential configuration $\calB_\mat$. As a consequence of the first two laws of thermodynamics, the temporal increase in free energy cannot be greater than the power expended, and therefore the thermodynamic imbalance in the referential configuration takes the form
\begin{multline}
\int_{\calP_\mat} \dot{\psi_\mat} \, \textrm{dv}_\mat \leq -  \int_{\partial\calP_\mat} \varphi\,\dot{\bfd}_\mat  \bfn_\mat \, \textrm{da}_\mat -   \sum_i \int_{\partial\calP_\mat} \mu^{\textrm{EC}(i)}\cdot \bfj_\mat^{(i)} \bfn_\mat  \,  \textrm{da}_\mat\\ + \int_{\partial\calP_\mat} \bfT_\mat \bfn_\mat \cdot \bfv \,  \textrm{da}_\mat + \int_{\calP_\mat} \bfb_\mat \cdot \bfv \,  \textrm{dv}_\mat \,,
\label{eqn:ThermImb}
\end{multline}
where $\psi_\mat$ denotes the free energy per unit reference volume, $\varphi$ is the electric potential,  and $\mu^{\text{EC}(i)}$ is the electrochemcial potential. The outwards unit normal in referential configuration is given by $\bfn_\mat$, and $\text{v}_\mat$ and $\text{a}_\mat$ are the volume and surface area in the referential configuration, respectively. The first term on the right hand side of equation \eqref{eqn:ThermImb} represents the power due to electric displacement, the second term is the power due to ion transport, and the last two terms account for the power due to mechanical work.

Employing the divergence theorem, the balance laws \eqref{eqn:GaussLaw}, \eqref{eqn:MassBalance} and \eqref{eqn:ForceBalance}, and since $\calP_\mat$
is arbitrary, we can rewrite the thermodynamic imbalance in \eqref{eqn:ThermImb} as
\begin{equation}
\dot{\psi}_\mat - \bfe_\mat\cdot \dot{\bfd}_\mat + \sum_i \left(\varphi eN_a z^{(i)} - \mu^{\textrm{EC}(i)}\right)\dot{c}_\mat^{(i)} + \sum_i \nabla \mu^{\textrm{EC}(i)} \bfj_\mat^{(i)}    \,-\, \bfT_\mat: \dot{\bfF} \leq 0 \,,
\label{eqn:ThermImb2}	
\end{equation}
where $\bfe_\mat = - \nabla \varphi$ is the electric field in the reference configuration. Considering the multiplicative decomposition of the deformation gradient in \eqref{eqn:GradientDecomposition}, along with the relation between the ion concentration and swelling in \eqref{eqn:VolumetricDecomposition2}, the last term on the left hand side of equation \eqref{eqn:ThermImb2} can be further expanded as
\begin{equation}
\bfT_\mat : \dot{\bfF}  = \bfT^m: \dot{\bfF}^{m} + p \sum_i \Omega^{(i)}\dot{c}_\mat^{(i)}\,,
\label{eqn:StressDeformationGrad}
\end{equation}
where we introduce the mechanical Piola stress and a pressure-like component 
\begin{equation}
\bfT^m \Def  \bfT_\mat \bfF^{s} \quad \text{and} \quad p  \Def \frac{1}{3} \Tr \left(\bfF^{m \trans}\bfT_\mat \bfF ^{s}\right)\, .
\label{eqn:StressDeformationGrad2}
\end{equation}

Finally, implementing \eqref{eqn:StressDeformationGrad2} into the thermodynamic imbalance \eqref{eqn:ThermImb2}, we obtain
\begin{multline}
\dot{\psi}_\mat 
 + \dot{\bfd}_\mat \, \nabla \varphi 
+\sum_i \left( \left(   eN_a z^{(i)}  \varphi  - \mu^{\text{EC}(i)} + p\Omega^{(i)} \, \right)  \dot{c}^{(i)}_\mat  \right)\\
+ \sum_i  \bfj_\mat \, \nabla \mu^{\text{EC}(i)}
+ \bfT^m : \dot{\bfF}^m
\leq 0
\label{eqn:ThermImb3}
\end{multline}

To satisfy the thermodynamic imbalance in \eqref{eqn:ThermImb3}, the constitutive relations for the electric field, the electrochemical potential, and the first Piola stress are obtained as 
\begin{equation}
\begin{split}
\bfe_\mat &= \pards{\psi_\mat}{\bfd_\mat}\,,\\
\mu^{\textrm{EC}(i)} &= \pards{\psi_\mat}{c_\mat^{(i)}} + p\Omega^{(i)} + \varphi eN_a z^{(i)} \,,\\
\bfT^m & =    \pards{\psi_\mat}{\bfF^m}\,.\\
\end{split}
\label{eqn:BasicConstitutive}
\end{equation}
In addition,  the ionic flux term in \eqref{eqn:ThermImb2} must obey $\bfj_\mat^{(i)}\nabla\mu^{(i)} \leq 0$. The first two terms in electrochemical potential \eqref{eqn:BasicConstitutive}, are commonly termed the chemical potential $\mu$ \citep{narayan2021coupled,ganser2019finite}. The obtained expression for the electrochemical potential extends from the definition by the International Union of Pure and Applied Chemistry (IUPAC) \citep{mcnaught1997compendium} to include mechanical pressure effects. 





\subsection{Constitutive relations}\label{sec:FreeEn}

For our constitutive models, we choose to consider three distinct contributions to the total free energy: (i) the ideal dielectric nature of the polymer network, (ii) the ideal solution mixing of the ions, and (iii) the non-linear Neo-Hookean mechanical response of the polymer. The total free energy is then comprised of
\begin{multline}
\psi_\mat =  \frac{1}{2\varepsilon}\,\bfd_\mat \bfC  \bfd_\mat + \sum_{i}\,RT c_\mat^{(i)} \left(\ln\frac{c_\mat^{(i)}}{\bar{c}^{(i)}}-1\right)\\ + \frac{1}{2}G \left[ \left(J^s\right)^{2/3} \Tr \bfC^m- 3 - 2\ln J \right] + \frac{1}{2} J^s K\left(\ln J^m\right)^2 \,,
\label{eqn:FreeEn}
\end{multline} 
where $\bar{c}^{(i)}$ is the maximum concentration of $i$-th species, while $G$ and $K$ denote the shear and bulk moduli, respectively.

The soft ionic conductors can undergo large deformation, and therefore, we convert the electric field, the electric displacement, species concentration and stress to the spatial configuration \citep[cf., e.g.,][]{suo2008nonlinear} as:
\begin{equation}
\bfe = \bfF^{-\trans} \bfe_\mat\,, \quad
\bfd = J^{-1}  \bfF \bfd_\mat\,, \quad
c^{(i)} =  J^{-1}\, c^{(i)}_\mat\,, \quad \textrm{and} \quad
\bfT = J^{-1} \bfT^m \bfF^{m \trans}
\,
\label{eqn:BasicConstitutiveSpatial}
\end{equation}
where $\bfe$ and $\bfd$ are the electric field and dielectric displacement in spatial configuration, respectively, and $\bfT$ represents the Cauchy stress. 

Based on the specialized free energy in \eqref{eqn:FreeEn}, the constitutive relations in \eqref{eqn:BasicConstitutive}, and \eqref{eqn:BasicConstitutiveSpatial}, the electric field and the electrochemical potential in spatial configuration, along with the Cauchy stress take the forms
\begin{equation}
\begin{split}
\bfe &= \frac{1}{\varepsilon} \bfd\,,\\
\mu^{\textrm{EC}(i)} &= RT \ln \frac{c^{(i)}}{\bar{c}}\, + p\Omega^{(i)} + z^{(i)}eN_A\varphi,\\
\bfT &=  J^{-1} G\left(\bfB - \id \right) + \left(J^m\right) ^{-1} K\left(\ln J^m \right)\id + \varepsilon J \left(\bfe\otimes\bfe - \frac{1}{2}\left(\bfe\cdot\bfe\right)\id\right)  \,.\\
\end{split}
\label{eqn:Constitutive}
\end{equation}

In this work, the gradient of the electrochemical potential $\mu^{\textrm{EC}(i)}$ is the driving force for ion flux, and thus, to account for the transport of ions through the continuum, we employ the Darcy-type form
\begin{equation}
\bfj^{(i)} = -M^{(i)} \grad \mu^{\textrm{EC}(i)}\,,
\label{eqn:Flux}
\end{equation}
where  the ion mobility follows the Einstein-Stokes relation 
\begin{equation}
M^{(i)} = \frac{D^{(i)} c^{(i)}}{RT} \,,
\label{eqn:Mobility}
\end{equation} 
with $D^{(i)}$ denoting the ion diffusivity. To satisfy the thermodynamic imbalance in \eqref{eqn:ThermImb2}, the mobility has to be non-negative $M^{(i)}\geq 0$  when  \eqref{eqn:Flux} is also taken into account.

In the absence of mechanical deformation, and assuming the material acts as a binary electrolyte combined with an ideal dielectric, the developed multiphysics framework reduces to:
\begin{equation}
\begin{split}
\dot{c}^{(+)}&=-\text{Div}  \left(-D^{(+)} \nabla c^{(+)} - M^{(+)} z^{(+)} eN_a\nabla\varphi\right)\,,\\
\dot{c}^{(-)}&=-\text{Div}  \left(-D^{(-)} \nabla c^{(-)} + M^{(-)} z^{(-)} eN_a\nabla\varphi\right)\,,\\
 \varepsilon \, \Div\left(\nabla\varphi\right)&= eN_a\left(z^{(+)}  c^{(+)}- z^{(-)} c^{(-)}\right)\,.
 \end{split}
\label{eqn:PNP}
\end{equation}
which are also known as the Poisson-Nernst-Planck (PNP) relations.

The above constitutive model is implemented as a UEL in Abaqus. Numerical implementation details can be found in \ref{sec:NumericalImplementation}. A more limited, 1D electrochemical version is implemented in Matlab and Python for greater accessibility.


\section{Numerical results}\label{sec:Simulations}

In this section we demonstrate the computational capabilities of our multiphysics model to simulate electro-chemo-mechanically coupled phenomena in soft ionic conductors. First, we verify the model presented in Section \ref{sec:Modeling}, as well as the numerical approach, against analytical results found in literature. Next, we validate the model using the experimental data obtained through EIS performed on an alginate hydrogel. Finally, we simulate the operation of three electromechanical transducers, compare our results with the reported device performances, and discuss parameter sensitivity.

It is worthwhile noting that we follow the approach proposed by \cite{bazant2004diffuse}, and non-dimensionalize the space and time domain, $\bfX=\bfx/L$ and $\tau=tD/(\lambda_D L)$, respectively, where $\bfx = [\,x_1,\, x_2,\, x_3\,]$ denotes the position vector, $t$ is the time, and $L$ is the length of the system. Additionally, we non-dimensionalize the electrochemical degrees of freedom, i.e.,  the electric potential with the thermal voltage $\Phi=\varphi eN_a/RT$ and  the chemical potential $M^{(i)} = \mu^{(i)}/RT$, as well as the mobile ion concentration field with the average concentration of ionic species in the electrolyte $C^{(i)}=c^{(i)}/c^E$. By doing so, we obtain values that are more amenable numerically for the iterative Newton-Raphson solver (see \ref{sec:NumericalImplementation}).


\subsection{Verification -- DC simulation}\label{sec:Verification}

The electrochemical portion of our model and numerical tool is verified against analytical results from \cite{bazant2004diffuse}. Here, we simulate the diffusion of ions through an $L=0.2\,\mu$m long  binary electrolyte under an applied electric potential of 5mV across the length.  The remaining simulation parameters are $D=10^{-9}$m$^2$/s, T = 293 K, and $\lambda_D = 5$ nm. The finite element mesh consists of 100 1D elements, each $10^{-2}\,L$ long. Under 1D conditions the non-dimensional space vector $\bfX = [\,X_1,\, X_2,\, X_3\,]$  becomes a scalar $X$.  The electric potential is applied at $X=0$ and $X=1$, and the non-dimensionalized initial and boundary conditions are as follows:
\begin{itemize}
\item initial concentration of mobile ions  $C^{(+,-)}(\tau=0) = 1$ applied across the mesh,
\item initial electric potential profile is defined as $\Phi (\tau=0) = -0.01 + 0.02X\,$,
\item electric potential boundary conditions are prescribed as  $\breve{\Phi} (X=0) = -0.01$ and  $\breve{\Phi} (X=1) = 0.01\,$,
\item zero flux condition at the boundaries given by $\breve{J}^{(+,-)}(X=0) = 0$ and $\breve{J}^{(+,-)}(X=1) = 0\,$.
\end{itemize} 


\begin{figure}[h!]
\centering




\begin{tabular}{cc}
\begin{tikzpicture}
\node[inner sep=0] (image) at (0,0){\includegraphics[width=0.45\linewidth]{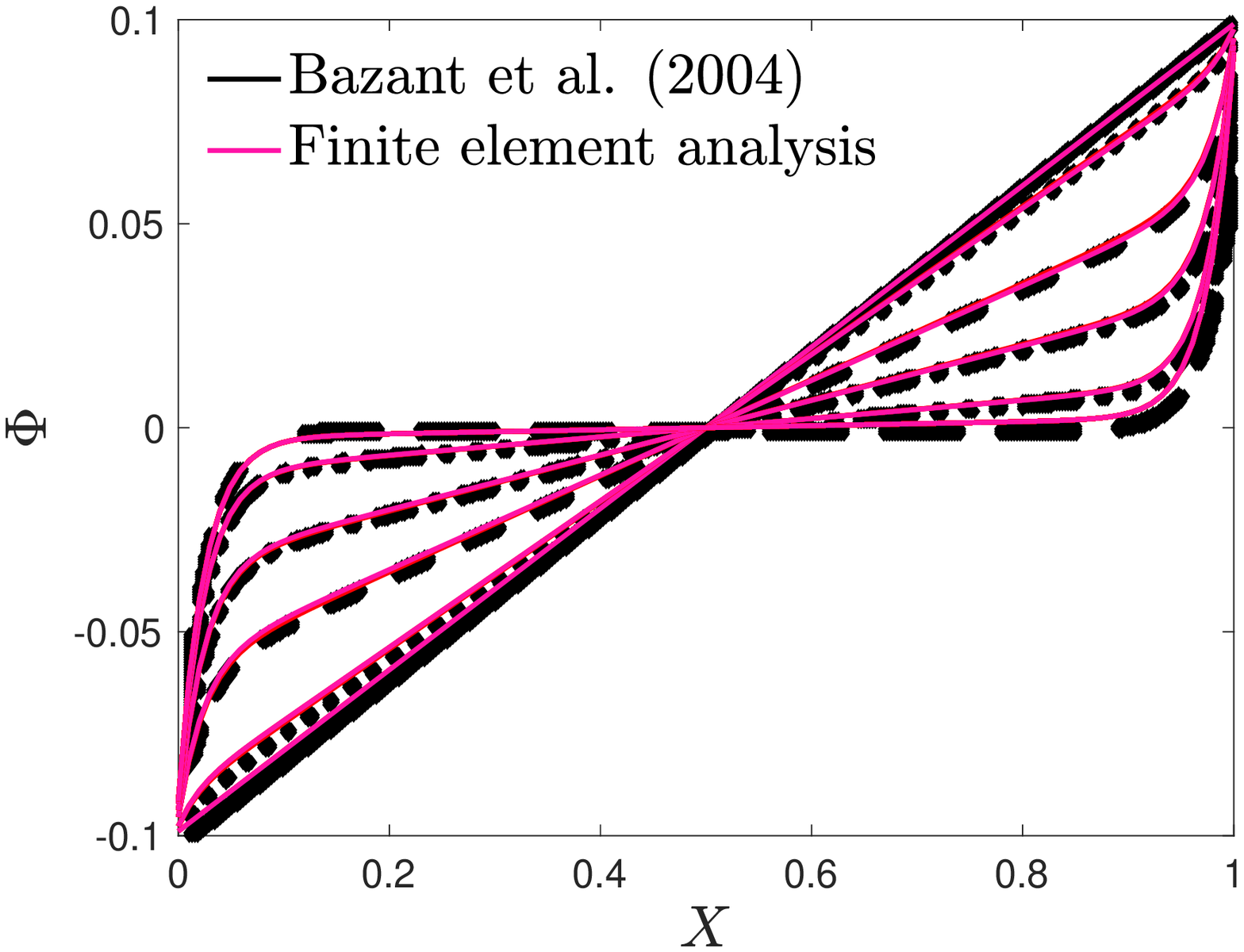}};

\draw [black, thin,-{Stealth[scale=0.75]}] (-1.15,-1.5) -- (-1.15,0.6);
\draw (-1.4,-1.75) node {\footnotesize $\tau = 0$ };
\draw (-1.15,0.85) node {\footnotesize $5$};

\draw [black, thin,-{Stealth[scale=0.75]}] (2.,1.95) -- (2.,-0.5);
\draw (2.,2.15) node {\footnotesize $\tau = 0$};
\draw (2.,-0.75) node {\footnotesize $5$};

\end{tikzpicture}
&
\begin{tikzpicture}
\node[inner sep=0] (image) at (0,0){\includegraphics[width=0.45\linewidth]{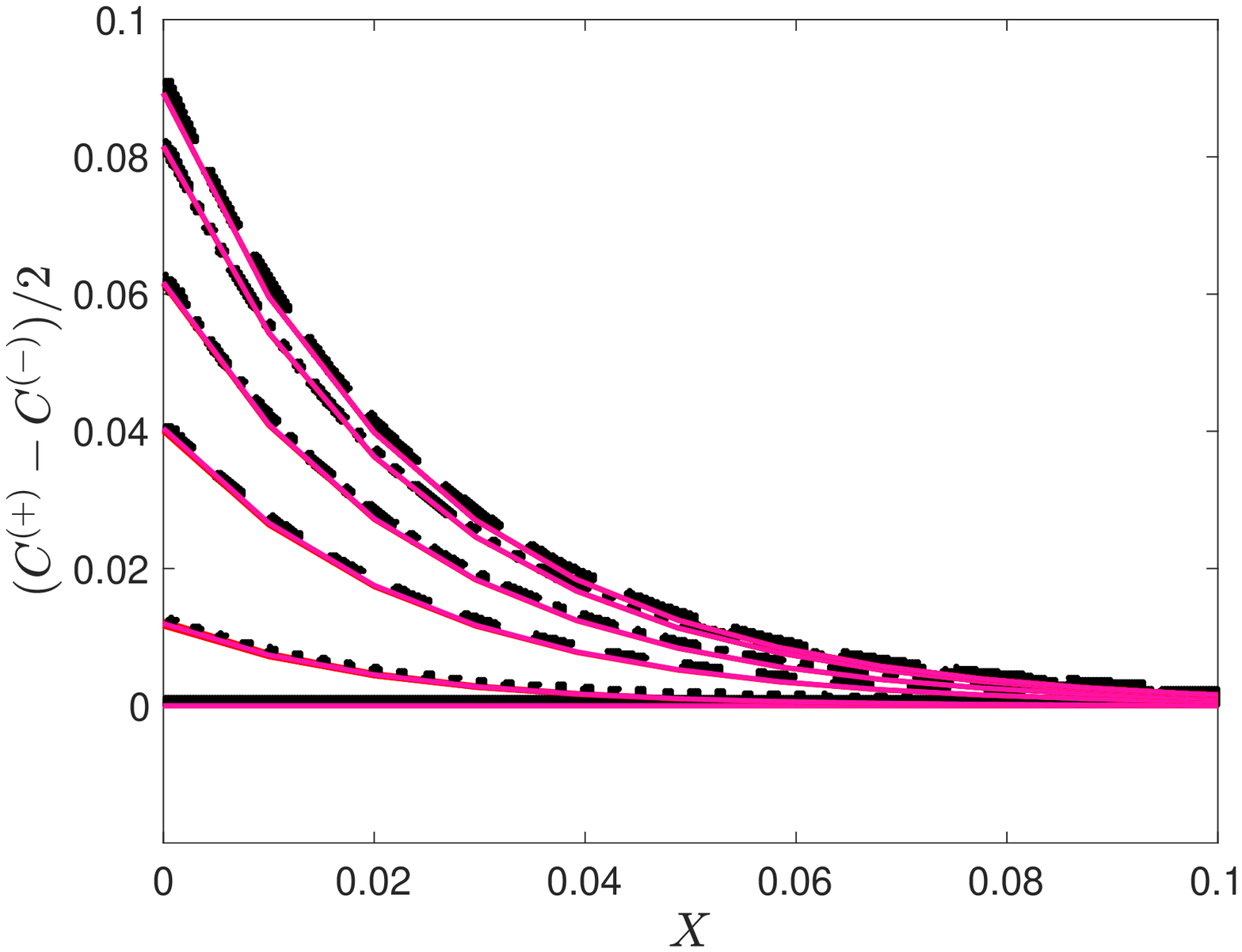}};

\draw [black, thin,-{Stealth[scale=0.75]}] (-1.5,-1.7) -- (-0.45,0.2);
\draw (-2,-1.8) node {\footnotesize $\tau = 0$};
\draw (-0.35,0.4) node {\footnotesize $5$};

\end{tikzpicture}\\
a)&b)
\end{tabular}
\caption{Verification of our computational tool by comparison with the analytical results from \citet{bazant2004diffuse}. 
 a) Non-dimensionalized electric potential $\Phi$ across the entire length. b) Difference in normalized concentration of positive and negative ions $(C^{(+)}-\,C^{(-)})/2$ close to the negative electrode. The black lines represent the analytical results and the solid pink lines represent our simulation results; both are shown at five different non-dimensional time points $\tau=$ 0, 0.05, 0.25, 0.5, 1 and 5. For reference, the non-dimensional time $\tau = 1$ corresponds to $t = 1\,\mu$s. The arrows indicate increasing time. }
\label{fig:Verification}
\end{figure}


In Figure \ref{fig:Verification} we show the comparison between our numerical simulation and the analytical results from \citet{bazant2004diffuse}. As time elapses, the electric potential profile in Figure \ref{fig:Verification}b deviates from the initial linear distribution, and we observe a sharp drop in the electric potential close to the electrodes caused by the migration of ions. The redistribution of ions due to the electric field can be observed in Figure \ref{fig:Verification}c, where we show the difference in concentration between the positive and negative ions close to the negative electrode ($X=0$). As expected, the positive ions are attracted to the negatively charged electrode, while the negative ions are repelled; these observations are symmetric (with the sign flipped) across $X=0.5$. Finally, our finite element analysis predictions are in excellent agreement with the analytical results, thus verifying the electrochemical portion of our computational approach.


\subsection{Validation -- EIS simulation}

We validate the developed numerical tool by simulating EIS performed on an alginate hydrogel. To synthesize the hydrogel, we adapted the procedure proposed by \citet{kaklamani2014mechanical}. We first prepared approximately 1\% by weight Na-alginate solution by placing dry Na-alginate in a beaker, adding water, and stirring for 72 hours at room temperature until the solution was homogeneous. We also prepared a $0.5\,$M solution of CuCl$_2$. Next, we filled a vertical mold, consisting of two glass plates separated by a $2.2\,$mm thick silicone spacer, with Na-alginate solution, and then added the CuCl$_2$ solution. After 72 hours, the Cu-alginate was removed from the mold and soaked in deionized water for 24 hours to remove the excess sodium, copper, and chloride ions. The obtained hydrogel was then cut into a 24x$50\,$mm sample. 

\begin{figure}[h!]
\centering




\begin{tabular}{cc}

\begin{tikzpicture}
\node[inner sep=0] (image) at (0,0){\includegraphics[width=0.45\linewidth]{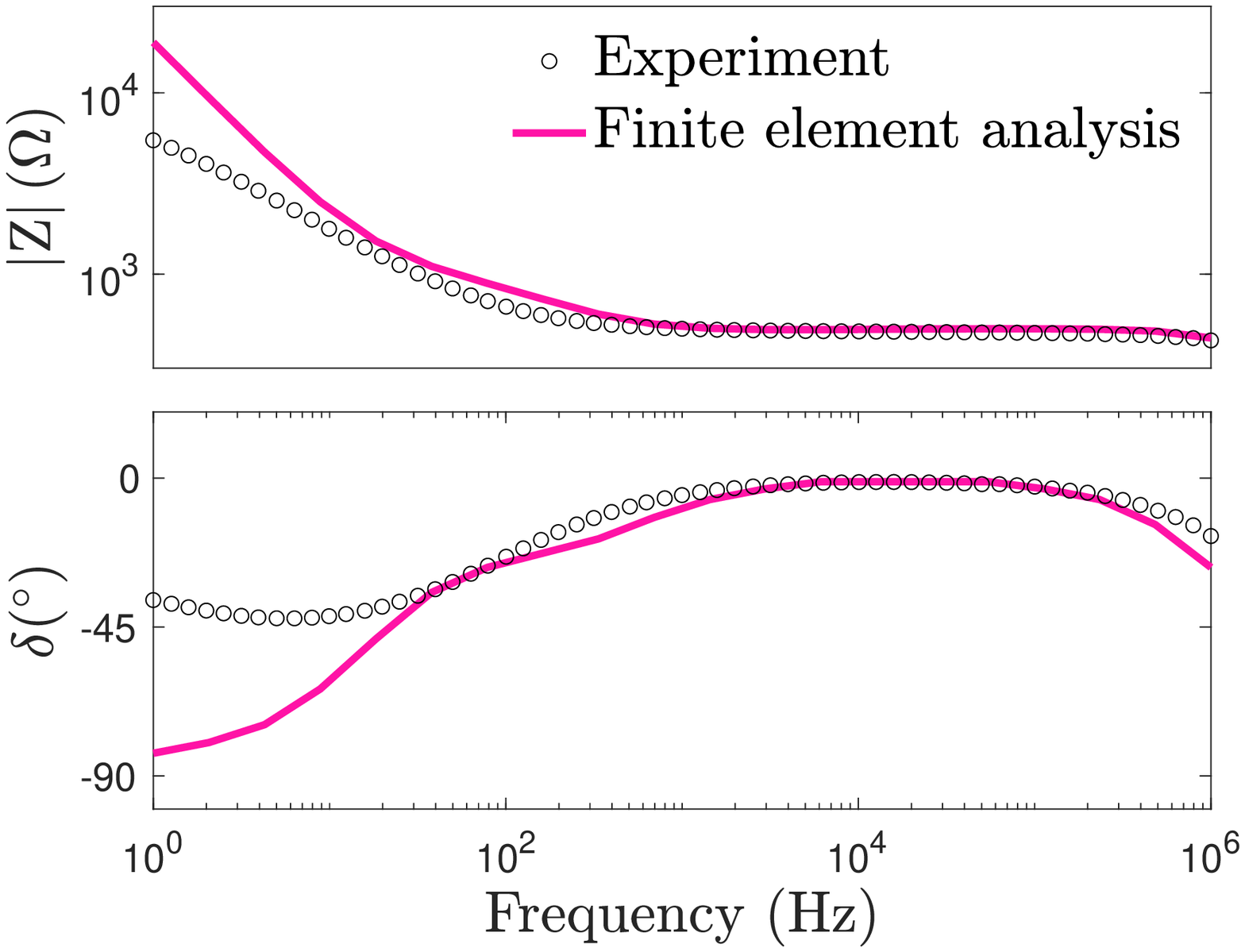}};
\end{tikzpicture}
&
\begin{tikzpicture}
\node[inner sep=0] (image) at (0,0){\includegraphics[width=0.45\linewidth]{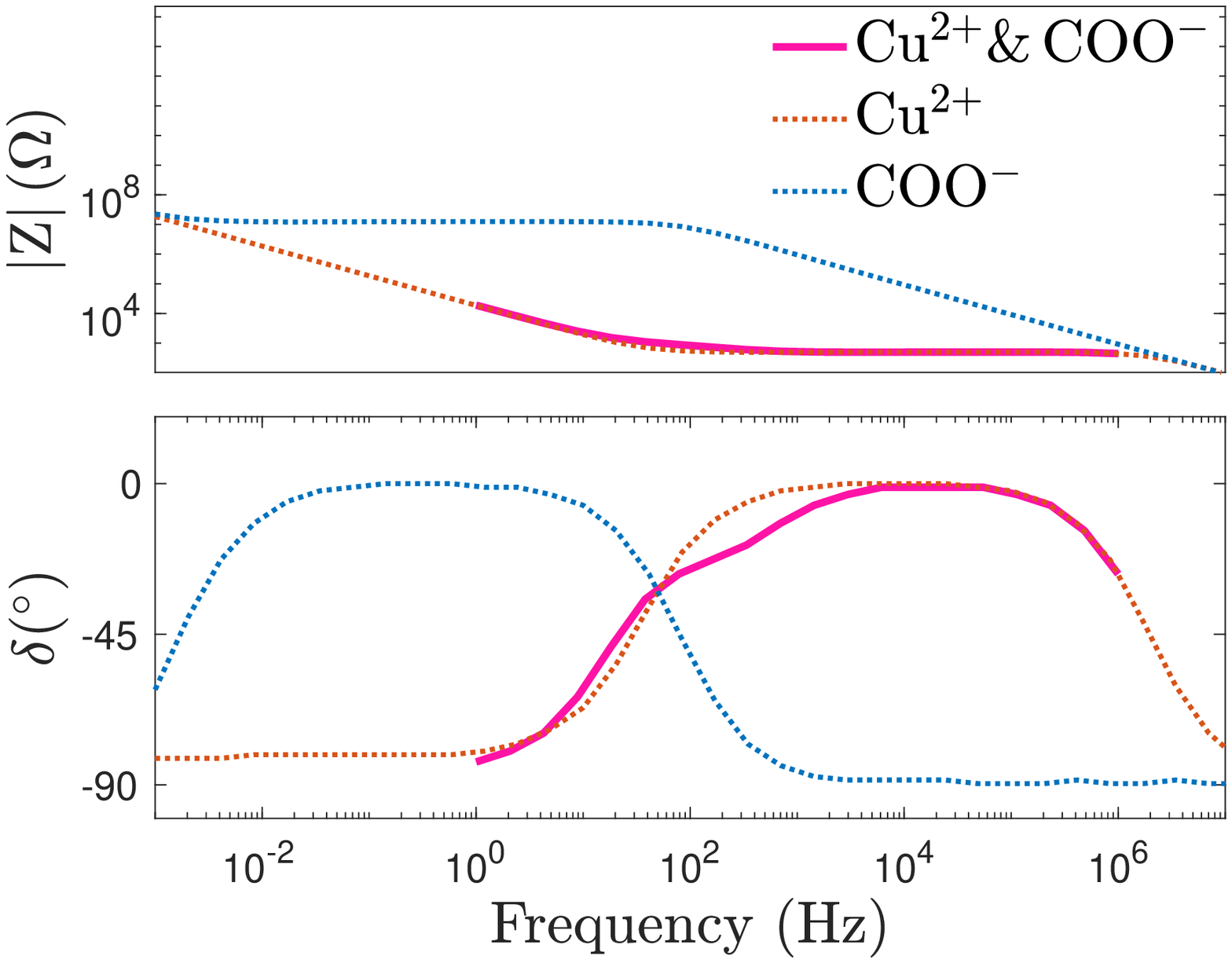}};
\end{tikzpicture}\\
a)&b)

\end{tabular}
\caption{Comparison between EIS experimental data and finite element analysis. The experiment is conducted on an alginate hydrogel synthesized by our group. Finite element analysis is performed under 1D electrochemical boundary conditions. a) Bode plots showing the  comparison between experimentally determined and simulated impedance magnitude $|$Z$|$ and phase shift angle $\delta$. b) Bode plots with simulation results showing the response of faster diffusing Cu$^{2+}$, and slower diffusing COO$^-$. The full simulation result from part a) is replotted as well for comparison.}
\label{fig:EISvalidation}
\end{figure}

EIS experimental data was obtained by through-plane testing at room temperature with a Gamry Reference 3000AE potentiostat. The sample was mounted on a custom testing fixture, placing it between two parallel gold electrodes. The surface area of sample in contact with the electrodes was $1\,$cm$^2$. Voltage was applied as a sinusoidal wave $\left(10\,\right.$mV$\left.\right) \sin(\omega t)$, where $\omega = 2\pi f$ denotes the angular frequency. Testing was performed at  60 different frequencies ranging from $f=1\,$Hz to $f=1\,$MHz. The results are shown as Bode plots in Figure \ref{fig:EISvalidation}a, where we observe three distinct regions: a mixed resistive/capacitive region at low frequency ($\delta \approx -45^\circ$), a resistive region at intermediate frequencies ($\delta \approx 0^\circ$), and a dielectric capacitor type region at high frequencies dominated by material polarization ($\delta \rightarrow -90^\circ$). For a material with single conductive ion species (or alternatively multiple ion species with similar diffusivities), we would expect that the low frequency regime would be dominated by slow EDL charging and discharging ($\delta \rightarrow 0^\circ$). However, the experimental data obtained at low frequencies suggests that, aside from the dominant mobile ion, there is another mobile ionic species with a significantly lower mobility. This can be observed at $10\,$Hz as a slight increase in phase angle with a decrease in frequency, as the current response of the less mobile species becomes in-phase with the applied electric potential.

To simulate EIS, we use a 1D finite element model, with the same electrode-to-electrode distance ($L = 0.8\,$mm) and surface area in contact with electrodes (1 cm$^2$) as in the experiments. A fine mesh is applied close to each electrode, consisting of 10 1D elements, each $\approx10^{-5}\,L$ long. The remaining geometry is meshed with 10 coarser 1D elements, each $\approx 10^{-1}\,L$ long. The initial conditions and parameters are provided in Table \ref{tab:EIS_parameters}. To determine the concentration of copper ions Cu$^{2+}$, we follow the ion exchange capacity approach found in \citet{hagiri2021modification}, and also compute the concentration based on the available binding sites (i.e., carboxyl acid groups COO$^-$) on the alginate network, both resulting in a similar concentration $\approx 0.1\,$M. These ions are loosely bound to the polymer network, and are mobile enough to impart conductivity via a ``hopping" mechanism \citep{schauser2019multivalent}. In addition, the alginate network itself is negatively charged due to the presence of carboxyl acid groups. While these charges are tethered to the backbone, the ionically bound network can still flow slowly in response to the applied voltage; and this response is orders of magnitude slower than the one of more mobile copper, as evidenced by the low frequency phase shift angle in Figure \ref{fig:EISvalidation}a. Therefore, we consider the electrochemical system consisting of two mobile species, with significantly different diffusion coefficients, and we use our computational tool to further examine the material behavior. We also take the dielectric constant to be $\varepsilon_R = \varepsilon_{\text{H}_2\text{O}} = 78.4\,$, which is a reasonable assumption for a highly swollen hydrogel. Based on the relation in \eqref{eqn:DebyeLength}, the Debye length is $\lambda_D = 0.7\,$nm at $T = 293\,$K.

\begin{table}[h!]
    \centering
    \begin{tabular}{lcc}
    Parameter & Unit & Value\\
        \hline
        \hline
           $\varphi_0$ & $\left(\text{V}\right)$ & 0  \\
           $c^{(+)}_0$ & $\left(\text{mol/m}^3\right)$  & 100\\
           $z^{(+)}$     &     & 2\\
           $c^{(-)}_0$ & $\left(\text{mol/m}^3\right)$  & 200\\
           $z^{(-)}$     &     & -1\\
           $c^{(\text{fix})}$ & $\left(\text{mol/m}^3\right)$  & 0 \\
           \hline
    \end{tabular}
    \caption{Initial conditions and material parameters for simulating the EIS experiment on alginate hydrogel.}
    \label{tab:EIS_parameters}
\end{table}

The boundary conditions are prescribed as to reflect the experimental setup:
\begin{itemize}
\item electric potential boundary conditions are prescribed as  $\breve{\varphi} (x=0) = 0$ and $\breve{\varphi} (x=0.8\,$mm$) = \left(10\text{mV}\right) \sin(\omega t)\,$, where $\omega$ is the angular frequency,
\item zero flux conditions are prescribed at the boundaries as $\breve{j}^{(+,-)}(x_1=0) = 0$ and $\breve{j}^{(+,-)}(x_1=0.8$\,mm$) = 0$.
\end{itemize}
For comparison with the experimental data, we perform simulations at 20 different frequencies ranging from $f=1\,$Hz to $f=1\,$MHz. Further, to probe the individual contribution of each ionic species, and to help calibrate our model, we conduct two additional simulation sets between $f=10^{-3}\,$Hz and $f=10\,$MHz. These simulations are conducted on: (1) alginate with only Cu$^{2+}$ as the mobile ions and immobile negative charges (to ensure neutrality), and (2) alginate with only COO$^-$ as the mobile ions, and immobile positive charges. After each simulation, the electric potential and current sinusoidal waves are analyzed as in Figure \ref{fig:PhysicsAndEIS}a.


The frequency range at which a material behaves as a capacitor or a resistor depends on the electrode-to-electrode distance, the Debye length, and diffusion coefficients. The electrode-to-electrode distance is easily measured, and the Debye length can be computed based on ion content and dielectric constant. Therefore, the two calibration parameters we use to fit the model are the diffusion coefficients of Cu$^{2+}$ and COO$^-$. We determined that the combination of $D^{(+)} = 3\cdot10^{-10}\,$m$^2/$s for faster diffusing Cu$^{2+}$, and $D^{(-)} = 1.2\cdot10^{-14}\,$m$^2/$s for the less mobile COO$^{-}$ provides the best fit with the experimental data (Figure \ref{fig:EISvalidation}a).

In Figure \ref{fig:EISvalidation}c, we show the simulated response when considering only Cu$^{2+}$ mobile ions, only COO$^{-}$ mobile ions, and both mobile species. The impedance magnitude plot clearly indicates that the conductivity of this material is almost completely imparted by Cu$^{2+}$. More specifically, in the absence of Cu$^{2+}$, the impedance magnitude is $\sim10^4$ greater in the resistive region (i.e., when $\delta \approx 0^\circ$ and $|$Z$|$ is flat). When both mobile ions are present, the impedance magnitude curve is almost identical to the one with only Cu$^{2+}$. The phase shift curve on the other hand does deviate significantly when both mobile species are present compared to just Cu$^{2+}$. Alginate with slower diffusing COO$^{-}$ exhibits an in-phase response between $\approx 10^{-2}\,$ and $10\,$Hz; for alginate with Cu$^{2+}$ this region is located between $\approx10^{2}\,$ and $10^{5}\,$Hz. As the phase shift angle for alginate with Cu$^{2+}$ begins to decrease towards lower frequencies around $10^2\,$Hz, the phase shift angle for the COO$^-$ case increases. The combined effects of the two mobile species can be observed in this region, as the phase shift angle for alginate with both mobile ions changes its curvature around $200\,$Hz.

The comparison between the simulation results and the experimental data is shown in Figure \ref{fig:EISvalidation}a. At higher frequencies, when the material response is dominated by the faster diffusing Cu$^{2+}$, the numerical results are in a good agreement with the experimental data. The numerical results also suggest the influence of slower diffusing COO$^-$ becomes more noticeable at the intermediate frequencies $10^2<f<10^3\,$Hz, observed as a change in phase shift angle curvature at $f \approx 200\,$Hz. However, the experimental measurements reveal this effect is occurring around $f \approx 10\,$Hz. Due to this disagreement, at low frequencies $f < 10\,$Hz we observe a discrepancy between the numerical results and experimental data. The difference in impedance magnitude between the experimental data and numerical results is less apparent, and as previously mentioned this feature is mostly affected by the faster diffusing ion Cu$^{2+}$. We expect that the overall response could be better captured by modeling the polymer chain diffusion as having a wide spectrum of time scales. The spectrum of time scales would arise not only from diffusion of different sized portions of the chains (analogous to a Rouse model), but also from the dynamic breaking and reforming for the copper-carboxylate bonds acting as temporary crosslinks in the material.

\subsection{Electromechanical transducers}

Finally, we  simulate the operation of three soft ionotronic devices using  the full electro-chemo-mechanically coupled model. For that purpose, we implement our framework and model as a user defined element (UEL) subroutine for use in Abaqus/Standard.


\subsubsection{Resistive sensor}\label{sec:ResistiveSensor}

\begin{figure}[h!]
\centering
\begin{tikzpicture}
\node[inner sep=0] (image) at (0,0){\includegraphics[width=0.5\linewidth]{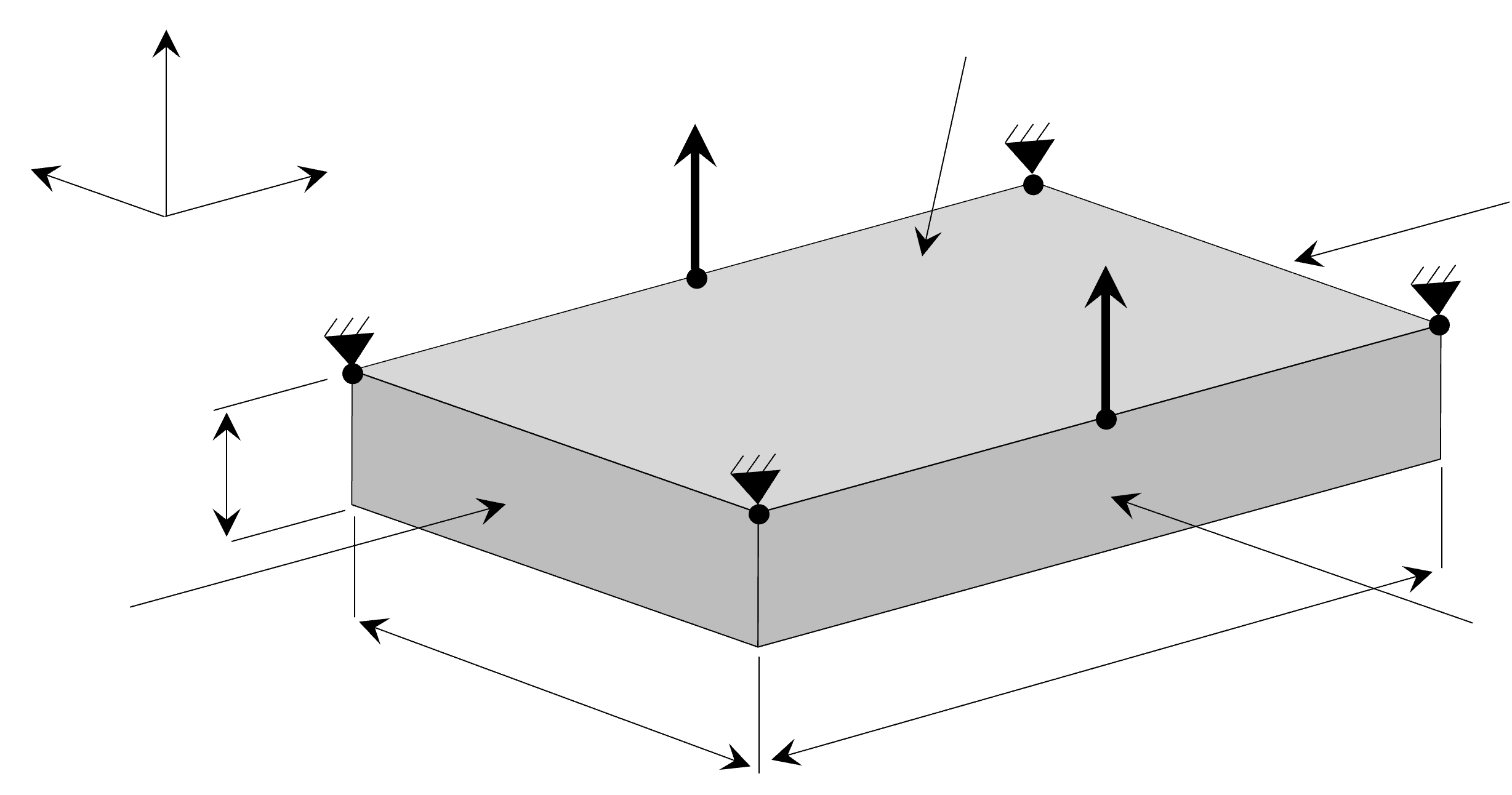}};

\draw (-2.2,1.5) node {\footnotesize 1};
\draw (-4,1.5) node {\footnotesize 2};
\draw (-2.95,2) node {\footnotesize 3};

\draw (-4,-1.1) node {\footnotesize face 1};
\draw (4.75,1.1) node {\footnotesize face 2};
\draw (4.5,-1.25) node {\footnotesize face 3};
\draw (1.1,2.1) node {\footnotesize face 4};

\draw (-3.5,-0.4) node {\footnotesize {$L/10$}};
\draw (1.9,-2.1) node {\footnotesize $L$};
\draw (-1.15,-1.95) node {\footnotesize $L/2$};

\draw (0.,1.2) node {\footnotesize $\bfu$};
\draw (1.6,.65) node {\footnotesize $\bfu$};

\end{tikzpicture}\\
a)\\
\begin{tabular}{cc}
\begin{tikzpicture}
\node[inner sep=0] (image) at (0,0){\includegraphics[width=0.5\linewidth]{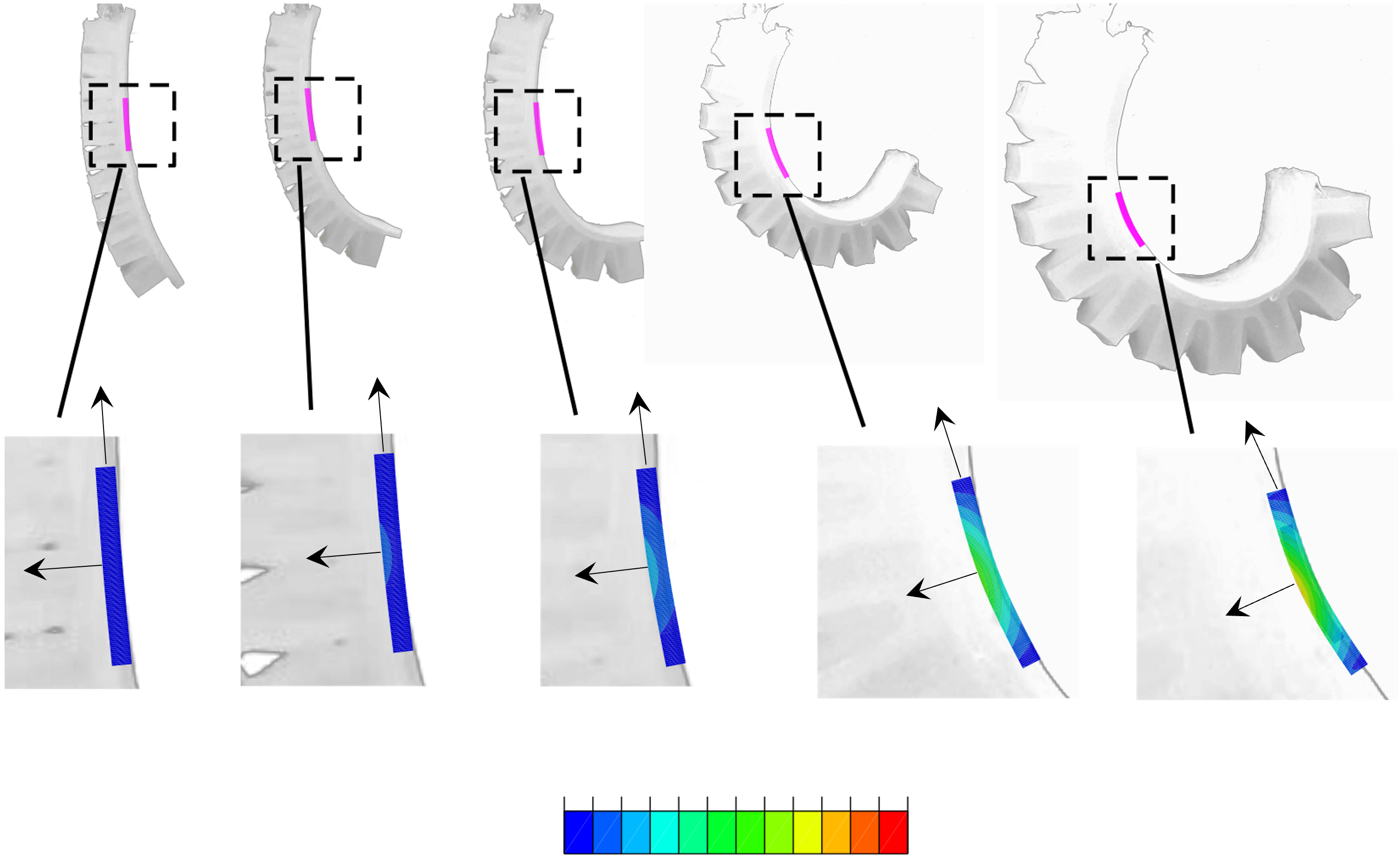}};
\draw (-3.25,2.7) node {\footnotesize $30^\circ$ };
\draw (-2.1,2.7) node {\footnotesize $60^\circ$ };
\draw (-.8,2.7) node {\footnotesize $90^\circ$ };
\draw (0.75,2.7) node {\footnotesize $150^\circ$ };
\draw (2.75,2.7) node {\footnotesize $180^\circ$ };

\draw (-3.3,0.2) node {\footnotesize 1};
\draw (-4,-.6) node {\footnotesize 3};

\draw (-1.85,-2.4) node {\footnotesize true strain};

\draw (-0.8,-2) node {\footnotesize $0$ };
\draw (1.3,-2) node {\footnotesize $0.05$ };

\end{tikzpicture}&
\includegraphics[width=0.5\linewidth]{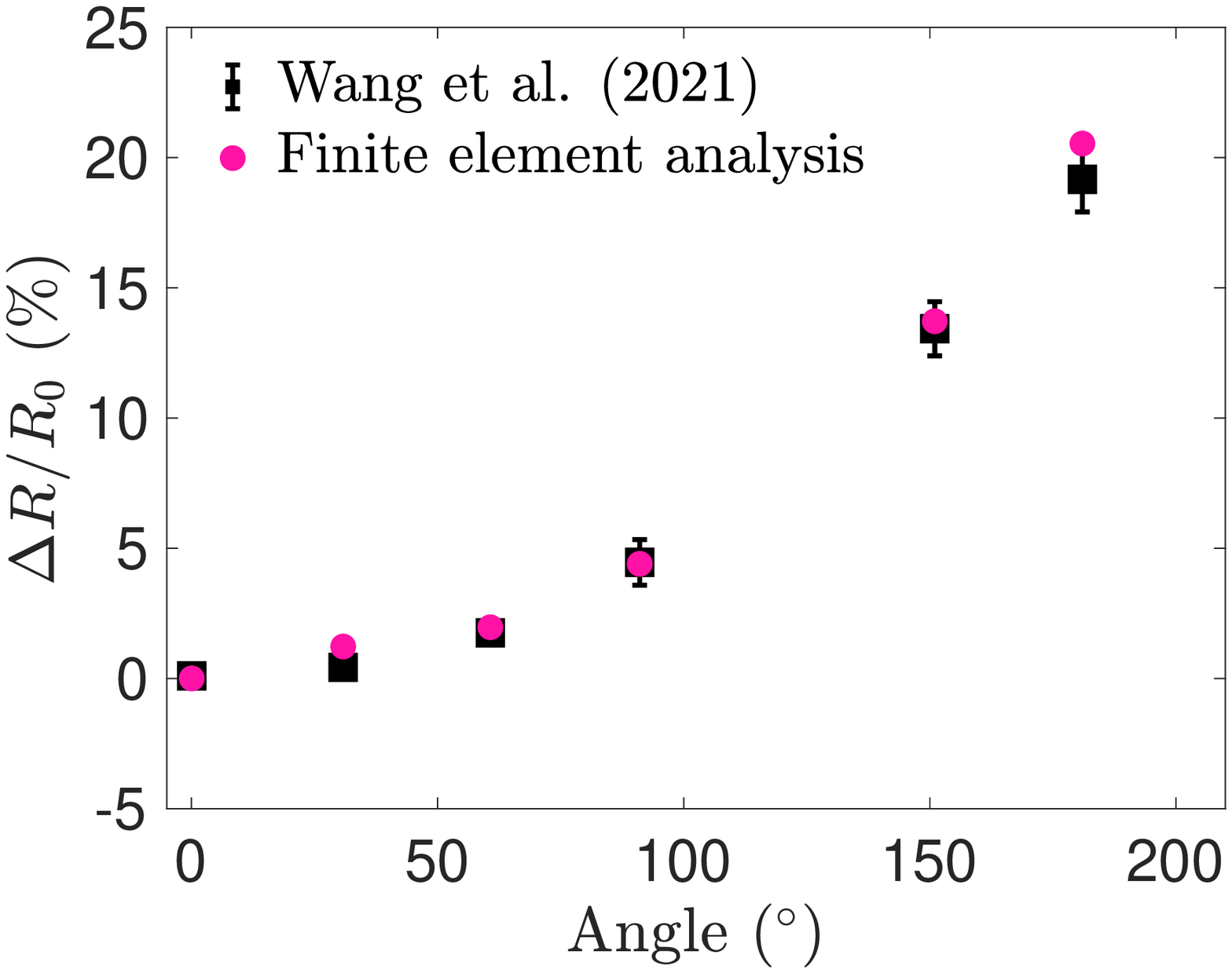}\\
b)&c)
\end{tabular}
\caption{Finite element analysis of a resistive strain sensor mounted on a robotic finger-like actuator from \citet{wang2021highly}. a) Non-dimensional geometry of the device. On face 4 the four outer nodes are pinned, and the two midplane nodes are given matching applied displacement, to drive a three point bending deformation. b) Deformation comparison between the  simulation results and the experimental images. The pink color in the top row denotes the deformed finite element mesh. The closeup views below show the true strain component (${\bf \epsilon}^t_{11}$)  from our finite element analysis. c) Comparison of the relative resistance obtained from finite element analysis and experiments. Here, $R_0$ denotes the initial resistance, i.e., the resistance at 0$^\circ$.}
\label{fig:wang2021}
\end{figure}

The first ionotronic device we will simulate is a flexible resistive sensor published by  \cite{wang2021highly}.  The sensor is made from an ionically conductive hydrogel with two soft copper wires, acting as electrodes, attached at each end. Na$^+$ and Cl$^-$ are the mobile ionic species and the polymer chains are neutral. In their work, the device is mounted on a robotic finger-like actuator, and the resistance $R$ is measured at seven different bending angles, ranging from 0$^\circ$ to 180$^\circ$, with a 30$^\circ$ increment. The stretchable region of this sensor is 10$\,$mm x 10$\,$mm, and we assume a thickness of $1\,$mm based on the experimental images. The distance between the electrodes is $L=10\,$mm. The experimental results from \citet{wang2021highly} (Figures \ref{fig:wang2021}b and \ref{fig:wang2021}c) showcase an increase in resistance with an increase in bending angle. The increase in resistance arises due to the sensor elongation caused by the bending motion; elongation corresponds to an increase in distance between the electrodes and a decrease in the cross-sectional area, both of which drive an increase in resistance.


Due to symmetry, we simulate one half of the geometry, as shown in Figure \ref{fig:wang2021}a. The finite element mesh consists of 300 3D elements, with a single element in the cross-section. A fine mesh, consisting of 100 elements, each $10^{-9} \,L$  long, $0.5\,L$ wide, and $0.1\,L$ thick, is applied adjacent to faces 1 and 2 (see Figure \ref{fig:wang2021}a), to capture the phenomena close to the electrodes. The remaining part of the geometry is meshed with 100 elements, each  $\approx10^{-2}\,L$ long, $0.5\,L$ wide, and $0.1\,L$ thick.

\begin{table}[h!]
    \centering
    \begin{tabular}{lcc}
    Parameter & Unit & Value\\
        \hline
        \hline
           $\varphi_0$ & $\left(\text{V}\right)$ & 0  \\
           $c^{(+)}_0$ & $\left(\text{mol/m}^3\right)$  & 3600\\
           $z^{(+)}$     &     & 1\\
           $c^{(-)}_0$ & $\left(\text{mol/m}^3\right)$  & 3600\\
           $z^{(-)}$     &     & -1\\
           $c^{(\text{fix})}$ & $\left(\text{mol/m}^3\right)$  & 0 \\
           $G$ & $\left(\text{kPa}\right)$ & 30\\
           $K/G$ & & 100\\
           \hline
    \end{tabular}
    \caption{Initial conditions and material parameters for simulating the resistive sensor by \cite{wang2021highly}.}
    \label{tab:wang2021_parameters}
\end{table}

We will proceed to describe this system in terms of electrochemical parameters and dimensions with units for ease of analysis and comparison. Nonetheless, parameters were converted to their non-dimensional values for use in finite element analysis, as in our previous simulations. The diffusivity $D = 10^{-9}$m$^2/$s is commonly found in literature for diffusion of ions through a hydrogel, as discussed in Section \ref{sec:TransportMechanism}. The relative permittivity $\varepsilon_R = \varepsilon_{\text{H}_2\text{O}} = 78.4$ corresponds to a high water content in hydrogels, yielding a Debye length $\lambda_D = 0.16\,$nm at $T=293\,$K. The initial conditions, along with the material parameters are listed in Table \ref{tab:wang2021_parameters}. The initial concentration of both positive and negative ions is based on the reported material preparation, and the shear modulus value is based on uniaxial tensile testing data provided in \citet{wang2021highly}. We also make a reasonable assumption that the hydrogel behaves as a nearly incompressible solid \citep[cf., e.g.,][]{chester2010,bouklas2015nonlinear}. Therefore, we take the bulk modulus to be two orders of magnitude greater than the shear modulus. The boundary conditions in effect throughout the simulation are:
\begin{itemize}
\item electric potential boundary condition $\varphi = 0$ is prescribed to nodes at face 1,
\item zero flux boundary conditions are prescribed on boundary surfaces located at face 1 and face 2, as $\breve{j}^{(+,-)}= 0\,$,
\item symmetry boundary conditions are prescribed to nodes on face 3; the nodes on edges between faces 1 and 4, and 2 and 4, are mechanically fixed, as shown in Figure \ref{fig:wang2021}a.
\end{itemize}

We perform six different simulations, one for each bending angle. We approximate the deformation of this finger-like actuator as a three-point bending deformation by keeping the ends fixed (see Figure \ref{fig:wang2021}a), and applying the mechanical displacement to the central nodes on face 4. Each simulation consists of two steps:
\begin{enumerate}
\item   Applying the displacement $\bfu$ along 1-direction ($u_1$) to the central nodes on face 4, as shown in Figure \ref{fig:wang2021}a. Displacement is applied with six different magnitudes, corresponding to different actuator bending angles:

\begin{table}[H]
\centering
\begin{tabular}{cc}
  $u_1$ (mm) & Bending angle $\left(^\circ\right)$   \\
  \hline
  \hline
  $0$ & 0   \\
  $0.05$ & 30\\
  $0.075$ & 60\\
  $0.15$& 90\\
  $0.4$ & 150\\
  $0.55$  & 180\\
  \hline
\end{tabular}
\end{table}

The electric potential $\varphi = 0$ is prescribed at face 2, and  the step time is $\tau = 2\cdot10^{-3}$.
\item The electric potential $\varphi = 50\,$mV is applied to nodes on face 2, while measuring the current to obtain the resistance.
\end{enumerate}

Comparison between the simulation predictions and the experimental data is presented in Figures \ref{fig:wang2021}b and \ref{fig:wang2021}c.  In Figure \ref{fig:wang2021}b we show the deformed simulated sensor on top of the experimental images; the insets show closeup views of the sensor with true strain component ${\bf\epsilon}^t_{11}$\footnote[2]{We define the true strain as ${\bf\epsilon}^t = \ln \bfV\,$, where  the left stretch tensor is given by $\bfV = \sqrt{\bfF \bfF^\trans}\,$.} obtained from simulations. Mechanical deformation is in good qualitative agreement with the experimentally observed sensor geometry. As in the experiments, we observe a deformation-driven increase in resistance with the increase in bending angle, shown in Figure \ref{fig:wang2021}c. Due to elongation, i.e. the increase in contour path between faces 1 and 2, the electric field and the concentration gradients decrease. Therefore, the ionic flux is reduced for a given applied voltage, and consequently, we observe greater resistance as the sensor undergoes bending. The  resistance obtained from the finite element analysis closely matches the experimental measurements, thus proving the developed model is able to simulate operation of resistive ionotronic sensors.


\subsubsection{Ionic junctions}\label{sec:EnergyHarvesting} 

Next, we demonstrate the modeling and simulation capabilities on energy harvesting devices based on ionic junctions. These ionic junctions involve an interface between two polymer membranes with opposite fixed charge; i.e. one layer is a polyanion and the other layer is a polycation. Each membrane initially contains sufficient mobile counter ions to result in an overall neutral material. A neutral fixed charge region exists between the two layers either through explicit incorporation of a neutral separator or from interdiffusion of the polymer chains from each membrane.  When joined, ions at the interface diffuse across the neutral layer and into the other polymer due to the large initial chemical potential gradient at the junction. This initial migration of mobile ions leaves the backbone charge of the material near the neutral layer unbalanced. Consequently, an electric field is established between the polyanion and polycation. Electrodes are typically placed on each side of the junction, measuring the difference in voltage as an open circuit electric potential. Once the entropically driven migration is balanced by the internal electric field, the ionic junction is at equilibrium, characterized by the equilibrium electric potential $\varphi_{eq}$. 

Before moving on to describing the device operation, to help discuss the performance of these devices, we make certain analogies to electric circuits. Assuming blocking electrodes, these systems can be considered as ideal capacitors; the relation of the bulk capacitance to the electrode-to-electrode distance $d$ and cross-sectional area $A$ is given by
    \begin{equation}
        \calC = \varepsilon \, \frac{A}{d}\,
        \label{eqn:capacitance}
    \end{equation}
     Therefore, both decrease in electrode-to-electrode distance and increase in cross-sectional area correspond to an increase in device capacitance.  In addition, capacitance can be also represented as the ratio of the charge density at the capacitor plates to the electric potential between them.
    \begin{equation}
        \calC = \rho/\varphi,
        \label{eqn:capacitance2}
    \end{equation}

\begin{figure}[h]
    \centering
    \begin{tabular}{cc}
        \begin{tikzpicture}
            \node[inner sep=0] (image) at (0,0){\includegraphics[width=0.42\linewidth]{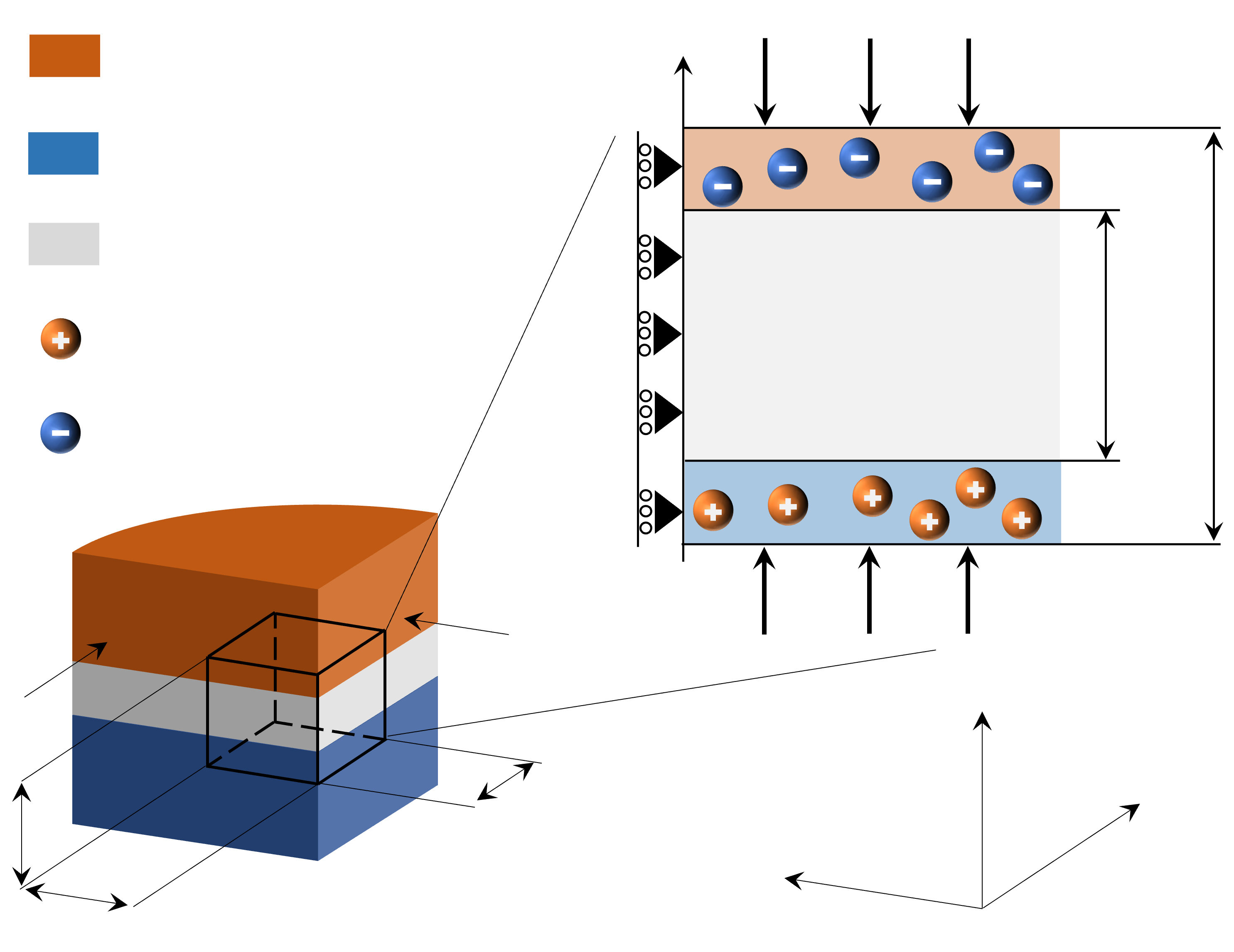}};

            \draw (-2.2,2.35) node {\footnotesize DF25};
            \draw (-2.1,1.85) node {\footnotesize Nafion};
            \draw (-1.85,1.3) node {\footnotesize Separator};
            \draw (-2.35,.8) node {\footnotesize H$^+$};
            \draw (-2.275,0.25) node {\footnotesize OH$^-$};
            
            \draw (1.85,-1.4) node {\footnotesize 1};
            \draw (2.7,-1.7) node {\footnotesize 2};
            \draw (1.,-2) node {\footnotesize 3};
            
            \draw (-3.6,-2.) node {\footnotesize {$L$}};
            \draw (-3.1,-2.65) node {\footnotesize {$W$}};	
            \draw (-0.2,-1.9) node {\footnotesize {$W$}};
            
            \draw (2.3,-.75) node {\footnotesize $\bfu$};
            \draw (2.3,2.2) node {\footnotesize $\bfu$};
            
            \draw (0.2,2.3) node {\footnotesize 1};
            \draw (3.6,0.65) node {\footnotesize $L$};
            \draw (3.05,0.65) node {\footnotesize $t_n$};
            
            \draw (-3.8,-1.3) node {\footnotesize face 1};
            \draw (0.,-.9) node {\footnotesize face 2}; 
        
        \end{tikzpicture}
        &
         \begin{tikzpicture}
            \node[inner sep=0] (image) at (0,0){\includegraphics[width=0.42\linewidth]{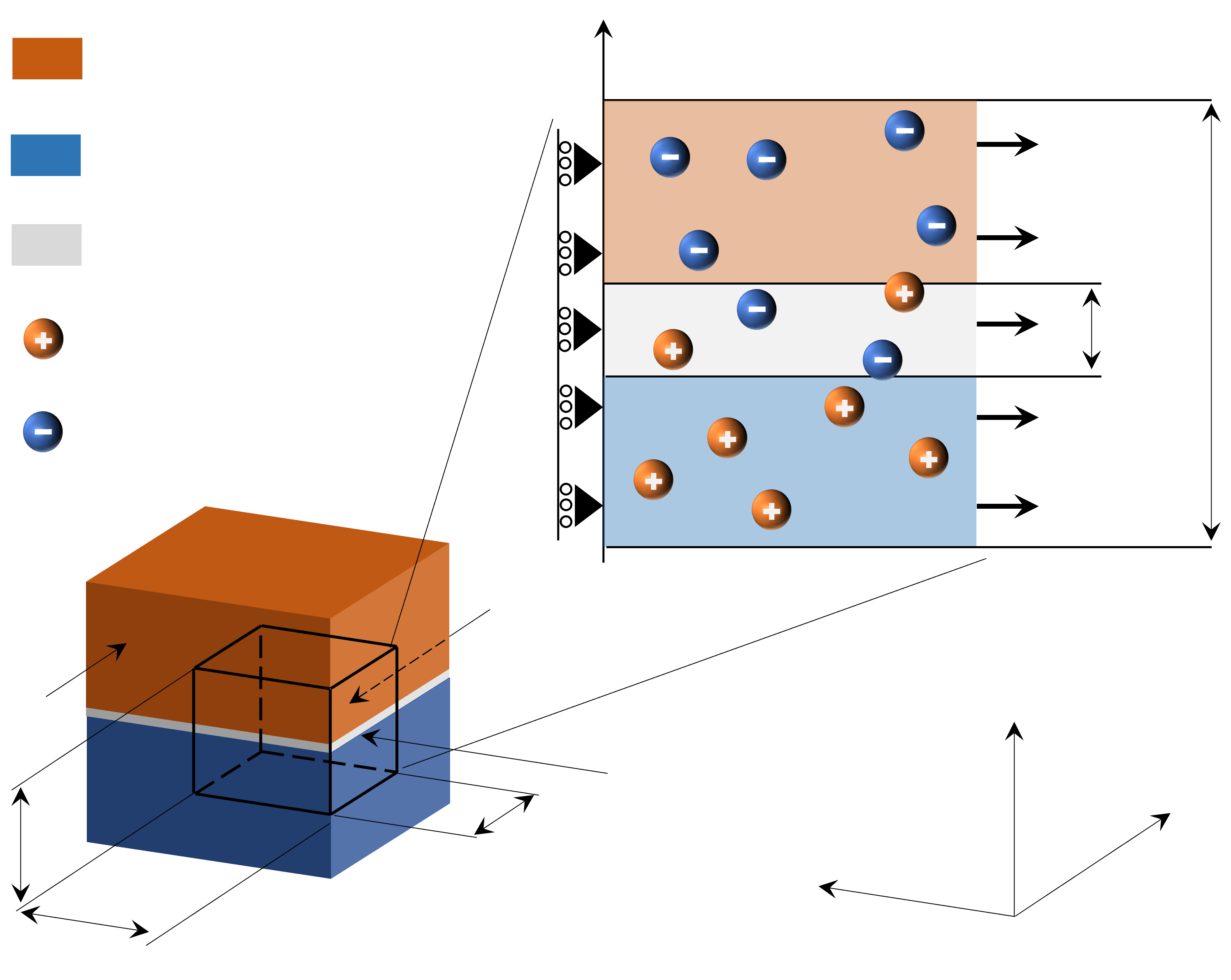}};
            
            \draw (-2.55,2.4) node {\footnotesize AT};
            \draw (-2.55,1.9) node {\footnotesize ES};
            \draw (-1.8,1.35) node {\footnotesize Neutral layer};
            \draw (-2.25,.8) node {\footnotesize EMIM$^+$};
            \draw (-2.25,0.35) node {\footnotesize TFSI$^-$};
            
            \draw (2,-1.25) node {\footnotesize 1};
            \draw (2.9,-1.6) node {\footnotesize 2};
            \draw (1.2,-1.95) node {\footnotesize 3};	
            
            \draw (-3.6,-2) node {\footnotesize {$L$}};
            \draw (-3,-2.7) node {\footnotesize {$W$}};
            \draw (-0.3,-2) node {\footnotesize {$W$}};
            
            \draw (2.5,1.55) node {\footnotesize $\bfu$};
            
            \draw (-0.3,2.65) node {\footnotesize 1};
            \draw (3.6,0.85) node {\footnotesize $L$};
            \draw (3.,0.85) node {\footnotesize $t_n$};
            
            \draw (-3.7,-1.2) node {\footnotesize face 1};
            \draw (0.5,-1.55) node {\footnotesize face 2}; 
            \draw (-0.15,-0.65) node {\footnotesize face 3}; 
            
        \end{tikzpicture}\\
        a) & b)   \\
        \includegraphics[width=0.45\linewidth]{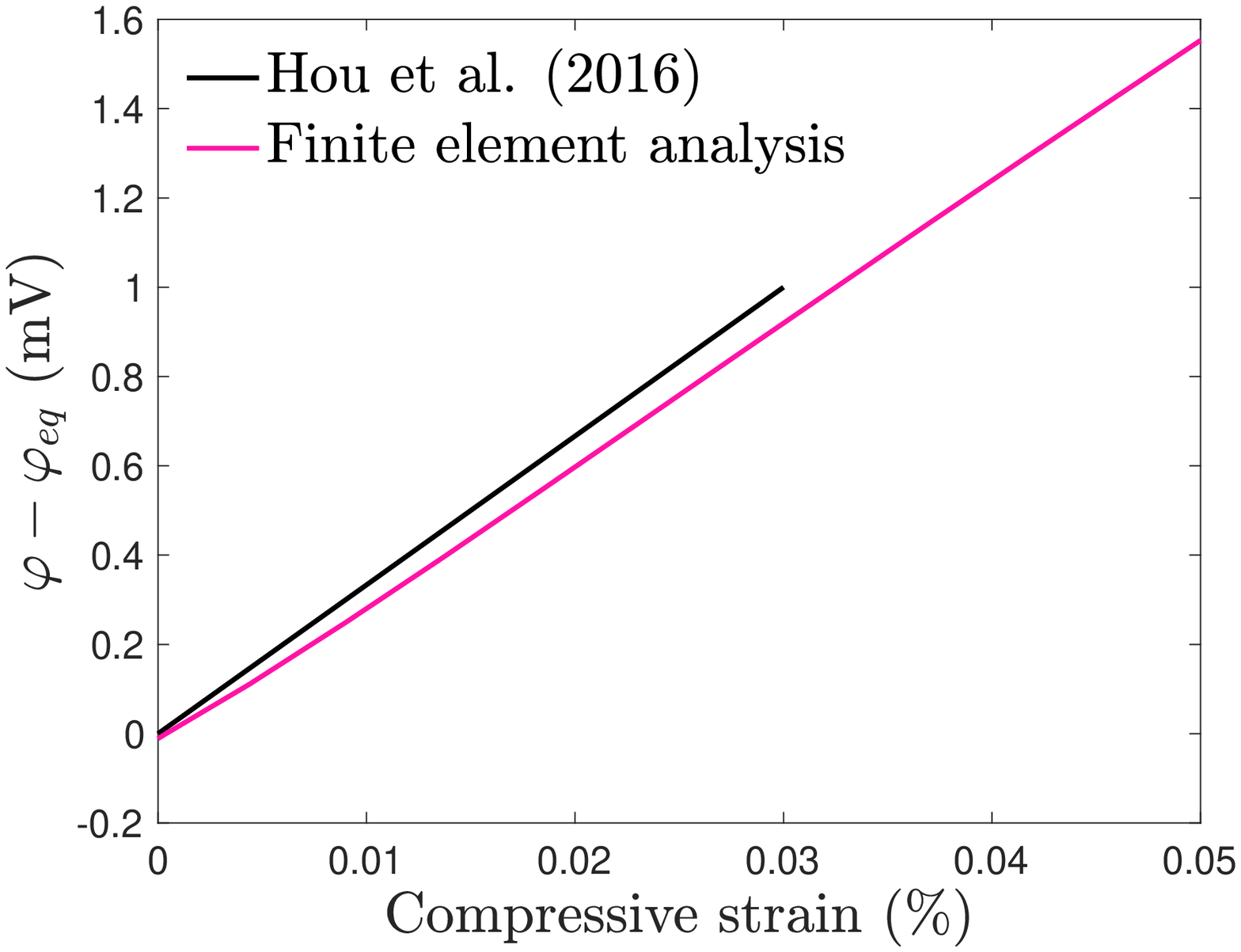}
        &
        \includegraphics[width=0.45\linewidth]{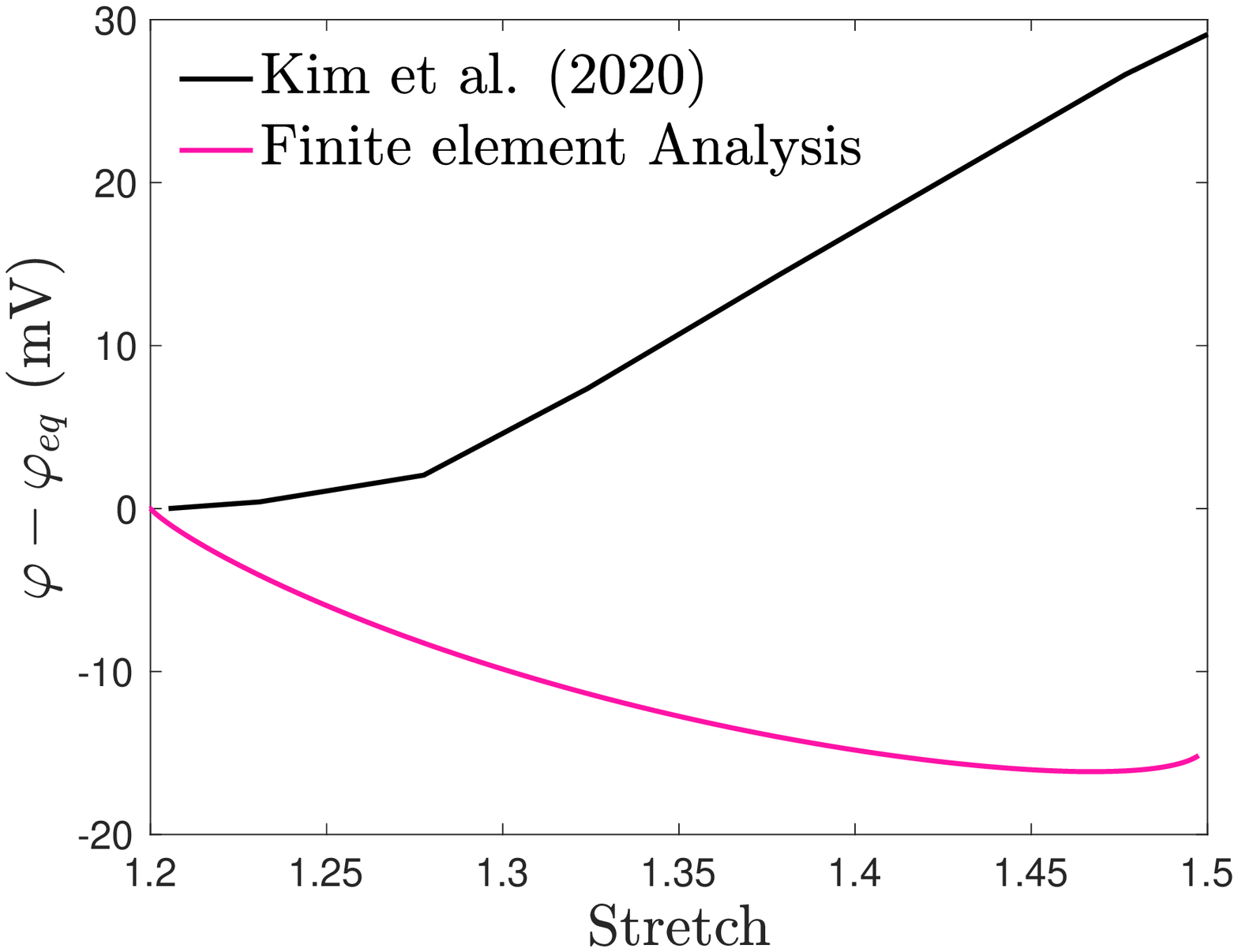}\\
        c) & d)
    \end{tabular}
    \caption{Ionic device summary. Device geometry for a) Nafion/DF25 ionic junction \cite{hou2017flexible}, and b) ES/AT ionic junction \cite{kim2020ionoelastomer}. The blue and orange colors denote regions with negatively and positively charged polymer backbone, respectively. The gray color denotes regions with neutral backbone charge, and circles denote the mobile ions at $t=0$. The meshed region is indicated with thick black lines. Note: geometry is not drawn to proportion. Electric potential across the junction  (i.e., between the top and bottom face) under mechanical loading and open circuit condition: c) in Nafion/DF25 for compression in the 1-direction, and d) in ES/AT for elongation in the 2-direction. Experimental data is plotted alongside simulation results for comparison. }
    \label{fig:EnergyHarvestingDevices}
\end{figure}

In this section we will discuss operation of two devices (Figure \ref{fig:EnergyHarvestingDevices}) relying on ionic junctions to transduce mechanical deformation and stress into electric potential (measured as open circuit potential): (i) a device developed by \cite{hou2017flexible} (the version without carbon nanotube enhancement), and (ii) a device published by. \cite{kim2020ionoelastomer}. The operation of both devices involves a change in the electrode-to-electrode distance (alternating decreasing and increasing) and the cross-sectional area (alternating increasing and decreasing). The first operates via through-plane compression while the latter operates via in-plane extension. Two other key differences are the material properties and the neutral layer composition. The first system uses leathery polymers (developed as fuel cell membranes) with a porous polycarbonate separator and the second system uses ionogels without any separator. Both systems have relatively large counter ions. Because of the material property differences and choices of the authors, the Hou et al. system is characterized with on/off steps in stress every $70\,$s corresponding to $\sim0.03\%$ strain whereas the Kim et al. system is characterized with a pre-stretch of 1.2 and a superposed sinusoid of amplitude 0.15 and frequency of 1 Hz. In addition to predicting performance of these two devices with our best approximation of each system, we will use these cases to study sensitivity to our assumptions and understand the drivers of good device performance. We are especially interested in the seemingly simple question - what drives the open circuit voltage to increase or decrease under mechanical deformation?

\clearpage

\subsection{Simulation results for \citet{hou2017flexible} junction}

The device consists of two oppositely charged membranes, Nafion and DF25 (poly(2,6-dimethyl- 1,4-phenylene oxide (PPO))  with quaternary ammonium bromide side chains, each $12\,\mu$m thick, separated by a $6\,\mu$m thick layer of porous polycarbonate. In our simulation, we consider a region involving a $6\,\mu$m separator, and $2\,\mu$m regions of the ion exchange membranes on each side of the separator, thus the total length of the meshed region is $L=10\,\mu$m (further discussion on the size of the region under consideration is provided in \ref{sec:ParameterSensitivity}). The meshed region is denoted with thick black lines in Figure \ref{fig:EnergyHarvestingDevices}a, and consists of 300 3D elements, with a single element in 1-direction (in \ref{sec:ParameterSensitivity} we provide justification for this choice). Each element is $0.33\cdot10^{-2}\,L$ long, $L$ wide and thick.

As with the previous device, we will describe this system using dimensional parameters. The open circuit potential is computed as the difference between the electric potential at nodes initially at $x_1=-5\,\mu$m and $x_1=5\,\mu$m. We consider low mobile ion concentration on the order of mM, as suggested in \citet{zhou2017biocompatible}.The diffusivity $D = 10^{-9}$m$^2$/s is typical for such ion exchange membranes. The charge numbers for positive and negative ions are $z^{(+)} = 1\,$ and $z^{(-)} = -1\,$, respectively. The dielectric constant $\varepsilon_R = 20$ \citep{paddison1998high} corresponds to a Debye length $\lambda_D = 6\,$nm at $T = 293\,$K. We take the volume of both positive and negative ions $\Omega^{(+,-)} = \Omega^{(\text{H}_2\text{0})} = 1.8\cdot 10^{-5}\,$m$^3$/mol.   The shear modulus for all three materials $G = 350\,$MPa is based on the uniaxial compression data performed on the entire device. We consider both membranes equally compressible, with a bulk modulus $K = G$, which is equivalent to a Poisson's ratio of 1/3. We also take the porous separator to have a bulk modulus of $K = G/3$, corresponding to a Poisson's ratio of 0.  The initial conditions are based on the reported material structure, and are given in Table \ref{tab:Nafion/DF25}. The boundary conditions in effect throughout the simulation are prescribed as follows:  
%


\begin{itemize}

\item electric potential boundary condition $\varphi = 0$ is prescribed at the interface ($x_1=0$); i.e. choosing to ground the device at this location.
\item zero flux boundary conditions prescribed on top and bottom surface (initially at $x_1=5\,\mu$m and $-5\,\mu$m, respectively) as $\breve{j}^{(+,-)}= 0\,$,
\item mechanical symmetry boundary conditions prescribed to nodes on face 1 (1-3 plane) and face 2 (1-2 plane), and the displacement of nodes at the interface ($x_1 = 0$) is constrained in 1-direction.
\end{itemize}

The simulation involves three distinct steps:
\begin{enumerate}
\item Allowing the electric potential across the junction to equilibrate, while keeping the above mentioned boundary conditions.
\item Prescribing mechanical displacement in the 1-direction to nodes at the top and bottom surface to compress the device by 5\%.
\item Allowing the electric potential to equilibrate again in the compressed configuration.
\end{enumerate}

\begin{table}[h!]
    \centering
    \begin{tabular}{lcccc}
        Parameter & Unit & Nafion & DF25 & Separator\\
        \hline
        \hline
           $\varphi_0$ & $\left(\text{V}\right)$ & 0 & 0 & 0  \\
           $c^{(+)}_0$ & $\left(\text{mol/m}^3\right)$  & 1 & 0\footnotemark[3] & 0\\
           $c^{(-)}_0$ & $\left(\text{mol/m}^3\right)$  & 0 & 1& 0\\
           $c^{(\text{fix})}_0$ &$\left(\text{mol/m}^3\right)$  & 1 & 1 & 0 \\
           $z^{(\text{fix})}$ &  &  -1 & 1 & 0 \\
           $G$ & $\left(\text{MPa}\right)$ & 350 & 350 & 350\\
           $K/G$ & & 1 & 1 & 1/3\\
        \hline
    \end{tabular}
    \caption{Initial conditions and material parameters for simulating Nafion/DF25 ionic junction by \citet{hou2017flexible}.}
    \label{tab:Nafion/DF25}

\end{table}

\begin{figure}[!hb]
    \centering
    \begin{tabular}{cc}
        \begin{tikzpicture}
            \node[inner sep=0] (image) at (0,0){\includegraphics[width=0.45\linewidth]{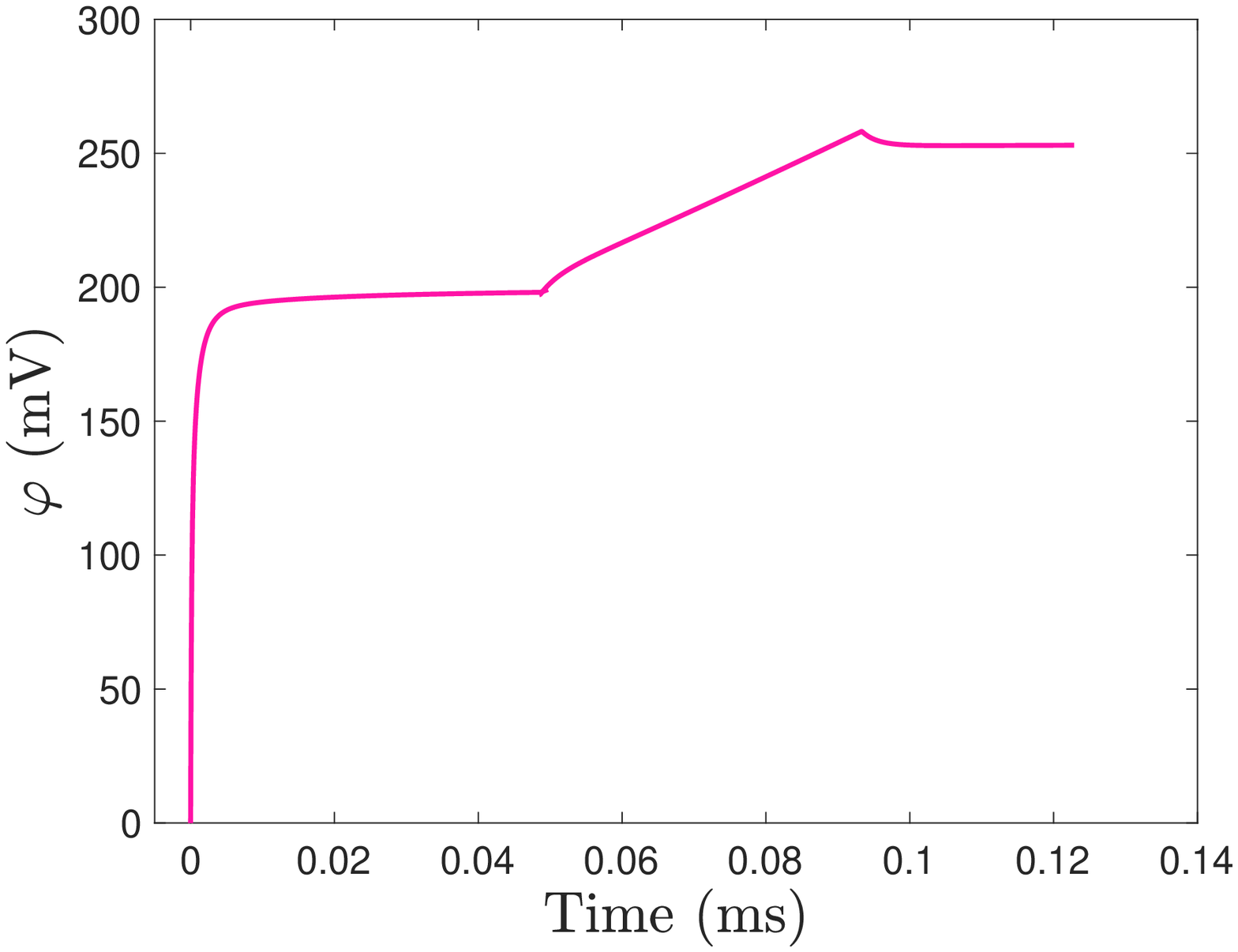}};
           
            \draw [black, thin] (-.6,-2.05) -- (-.6,2.35);
    
            \draw [black, thin] (1.15,-2.05) -- (1.15,2.35);
            
            \draw (-1.6,0) node {\tiny 1.};
            \draw (-1.6,-0.25) node {\tiny equilibrating};
            
            \draw (0.25,0) node {\tiny 2.};
            \draw (0.25,-0.25) node {\tiny compressing};
            
            \draw (2.15,0) node {\tiny 3.};
            \draw (2.15,-0.25) node {\tiny equilibrating};           
        \end{tikzpicture}
       &
       \begin{tikzpicture}
            \node[inner sep=0] (image) at (0,0){\includegraphics[width=0.45\linewidth]{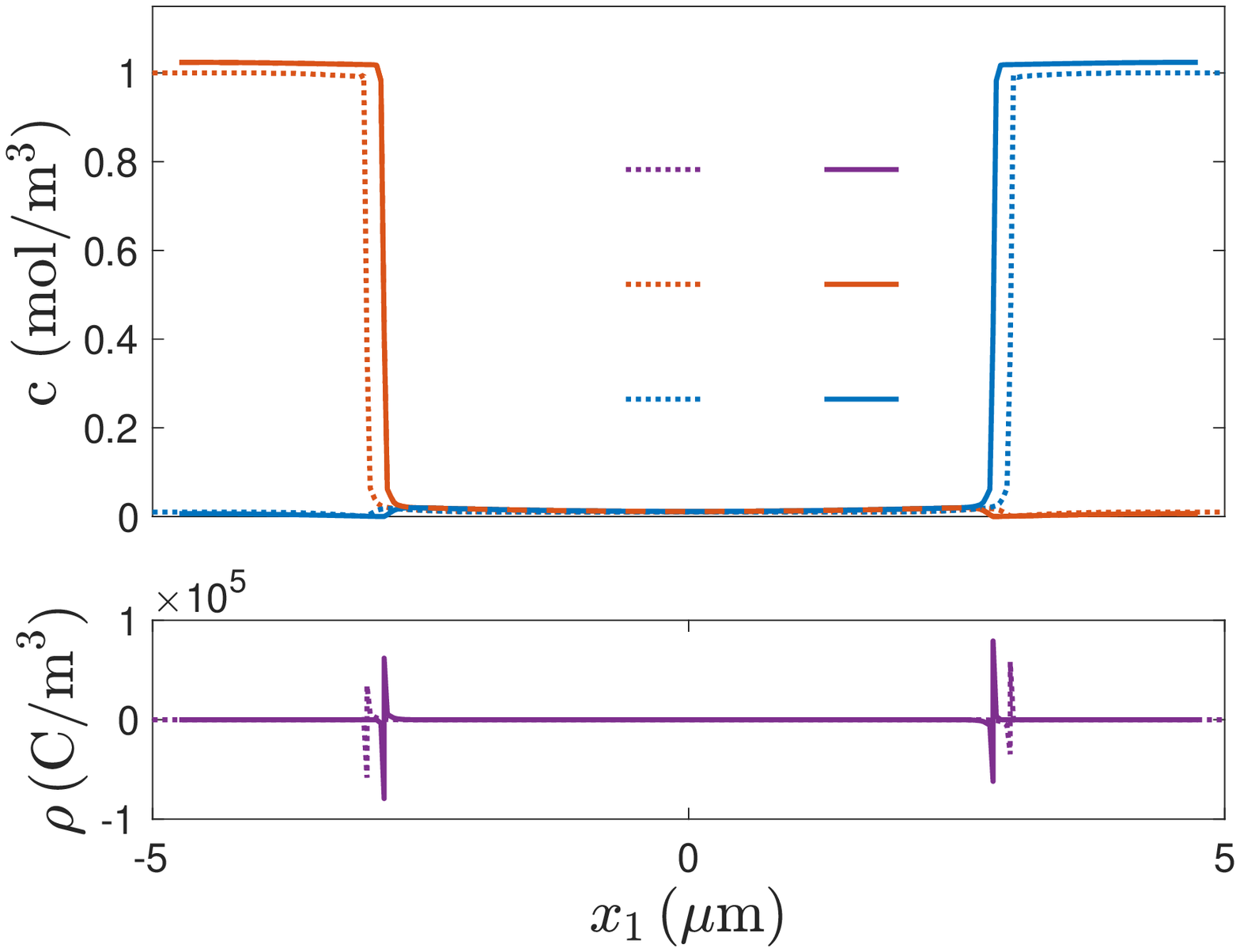}};
           
            \draw (0.,2) node {\footnotesize undef.};
            \draw (1,2) node {\footnotesize def.};

            \draw (-0.8,1.6) node {\footnotesize $\rho$};
            \draw (-0.7,1) node {\footnotesize $c^{(+)}$};
            \draw (-0.7,.4) node {\footnotesize $c^{(-)}$};
            
            
            
        \end{tikzpicture}\\
        a) & b)\\
    \end{tabular}
    \begin{tabular}{c}
        \begin{tikzpicture}
            \node[inner sep=0] (image) at (0,0){\includegraphics[width=0.45\linewidth]{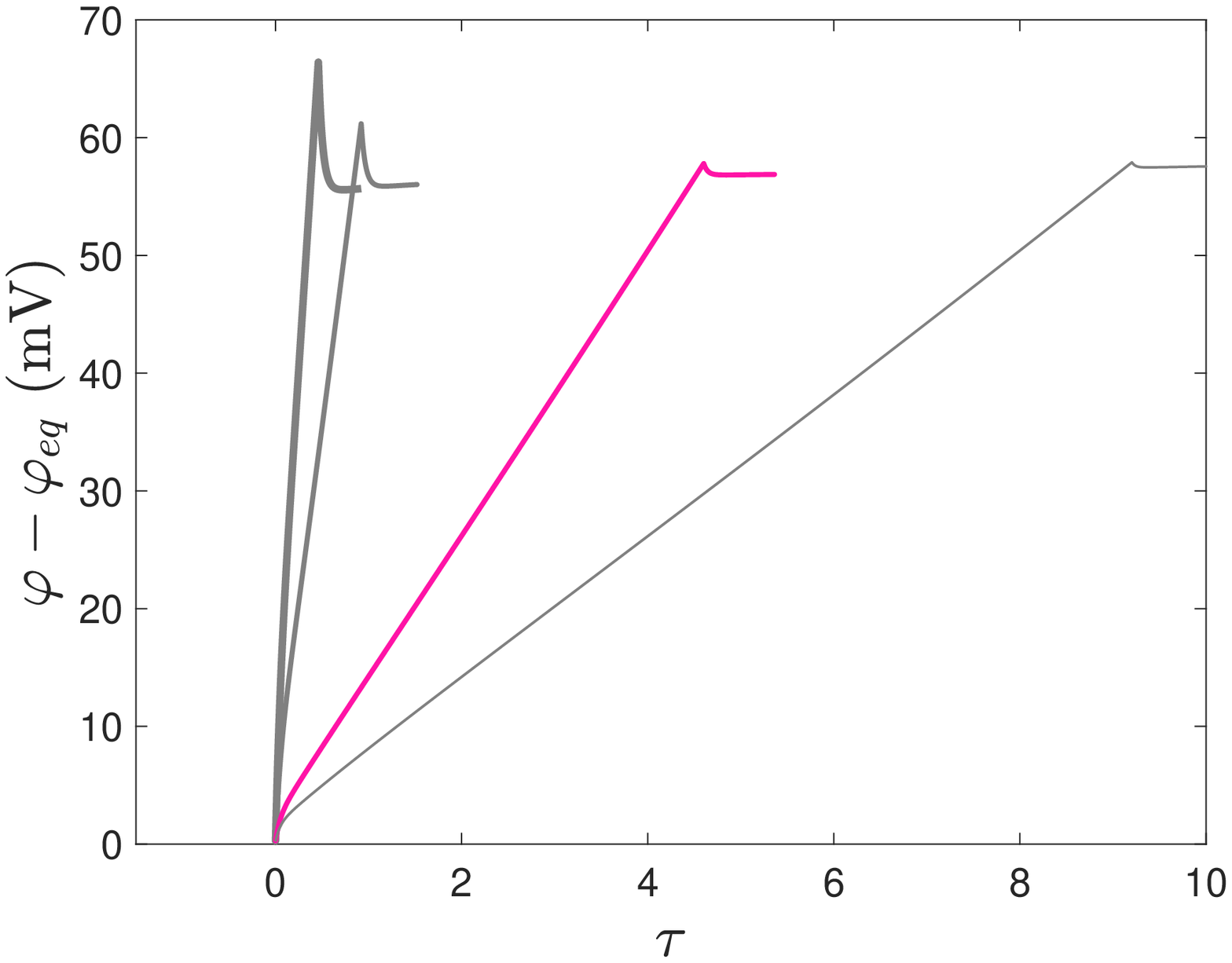}};
            
            \draw [black, thin,{Stealth[scale=0.75]}-] (-2.2,1.7) -- (1,-1);
            
             \draw (1.5,-.7) node {\footnotesize d$\lambda$/d$\tau$};
            \draw (1.65,-1.15) node {\footnotesize  $5\cdot 10^{-3}$};
            \draw (-2.25,1.9) node {\footnotesize  $10^{-1}$};
        \end{tikzpicture}\\ 
        c)     
    \end{tabular}
    \caption{Simulation results for Nafion/DF25. a) Electric potential across the junction under open circuit condition during equilibration step and while compressing. b) Concentration of net charge and mobile ions across the junction when undeformed and at equilibrium and when compressed by 5\% and at the new equilibrium. c) Influence of mechanical compression rate on the electrochemical behavior. Simulations are performed at d$\lambda$/d$\tau = 1\cdot10^{-1}\,$, $5\cdot10^{-2}$, $1\cdot10^{-2}$, and $5\cdot10^{-3}\,$. The pink line represents the same deformation rate as in a). Arrow indicates increase in the compression rate.}
    \label{fig:hou2016}
\end{figure}

Simulation results in Figure \ref{fig:hou2016}a show that the initial, entropically driven, diffusion of ionic species stops after $t\approx0.05\,$ms. The established electric field prevents further ion migration, and the electric potential between the top and bottom face reaches equilibrium at $\varphi_{eq} \approx 195\,$mV. We also observe that the mechanical deformation during the second step induces an increase in electric potential across the junction.  In Figure \ref{fig:hou2016}b, we show the net charge density $\rho$ and ion concentration when undeformed and when compressed by 5\%. The depletion of mobile ions close to the polyelectrolyte/separator interface is caused by the migration of ions into the separator during underformed equilibration. Consequently, the fixed charges in the depletion region are no longer balanced by the counterions, increasing the net charge density at the interface, and thus generating an electric potential across the junction. When compressed, gradients of the electric potential and of the chemical potential are equally increased across the junction through affine deformation of the fixed and mobile charges.  Since the two gradients, which are the driving force for the diffusion based on the flux relation in \eqref{eqn:Flux}, have equal and opposite effect on the movement of mobile ions, no further ion migration takes place.  As expected, once compressed, the overall increase in ion concentration yields a higher electric potential across the junction; and a similar observation was also made by \citet{zhou2017biocompatible}. While we observe the same voltage trend as in the experiment, the increase in voltage at $0.03\%$ compressive strain is $\approx 10$\% less than the experimentally observed one (see Figure \ref{fig:EnergyHarvestingDevices}c).

The particular device design also imposes a multiphysics problem with significantly different time scales between the electrochemical phenomena ($\sim\,$ms) and the applied mechanical load ($\sim 10\,$s). As previously mentioned, we non-dimensionalize the simulation time domain based on the electrochemical parameters: diffusivity coefficient, the length of the system under consideration, and the Debye length. For this system, the time normalization factor $\approx 2\cdot10^4$s$^{-1}$ maps a non-dimensional time increment of $\tau = 1$  to $t = 0.05\,$ms. Consequently, our simulation environment is more conveniently set to capture the electrochemical processes, while the much slower mechanical loading rate requires an extensively long simulation time. Therefore, we ran a set of simulations to investigate the effects of mechanical loading rate on the electrochemical performance. We conducted a set of simulations at four different mechanical compression rates  d$\lambda$/d$\tau = 1\cdot10^{-1}\,$, $5\cdot10^{-2}$, $1\cdot10^{-2}$, and $5\cdot10^{-3}$,  to 5\% strain, while keeping the remaining parameters and boundary conditions the same as in the initial simulation.

In Figure \ref{fig:hou2016}c, we show the device response to different mechanical compression rates as the difference between electric potential in current configuration and in undeformed configuration at equilibrium. The loading rate in our initial simulation is d$\lambda$/d$\tau = 1\cdot10^{-2}$. While the steady-state open circuit voltage has negligible dependence on the loading rate, the approach to that value does depend on the loading rate. The faster the loading rate, the larger the overshoot, and the larger the undershoot (though this is much less in magnitude than the overshoot). For all rates, there is a gradual increase to the final value.   Similar peaks, followed by a new equilibrium state, have also been observed through experimental characterization of this system by \cite{hou2017flexible}.



\footnotetext[3]{To ensure numerical stability while preserving the physical features, we take the concentration of ions that should be 0 to be $10^2$ lower than the counter ion.}

\clearpage

\subsection{Simulation results for \citet{kim2020ionoelastomer} junction}
The reported device consists of two $250\,\mu$m thick sheets of ES (poly(1-ethyl-3-methyl imidazolium (3-sulfopropyl) acrylate)) with EMIM$^+$ ( 1-ethyl-3-methylimidazolium) mobile ions and AT (poly(1-[2-acryloyloxyethyl]-3-butylimidazolium bis(trifluoromethane) sulfonimide)) with TFSI$^-$ (bis(trifluoromethane) sulfonimides) mobile ions. To simulate the operation of this ionotronic device, we mesh $L = 1\,\mu$m ($0.5\,\mu$m on each side of the interface); further discussion on the effects of $L$ under consideration can be found in \ref{sec:ParameterSensitivity}. The finite element mesh consists of 2000 3D elements in 1-direction, each $5\cdot10^{-4}\,L$ long, and  $10\,L$ wide and thick, with a single element in the cross-section.



As in the experiments, the ES region has negative fixed charge combined with mobile positive ions, while the AT region contains positive fixed charge and negative mobile ions.  The initial conditions are based on the reported preparation procedure and experimental setup, and are provided in Table \ref{tab:ES/AT}. As suggested in \cite{kim2020ionoelastomer}, we assume low degree of ionization, and take concentration of ions on the order of \,mM, and a diffusion coefficient $D = 10^{-14}$m$^2$/s based on the reported conductivity. The dielectric constant  $\varepsilon_R = 15$ is common for ionic liquids \citep[cf. e.g.,][]{weingaertner2014static}, and yields a Debye length of $\lambda_D \approx 6\,$nm at $T = 293\,$K. The molar volume of these ionic liquids is $\Omega^{(+)} = 1.4\cdot10^{-4}$m$^3$/mol and $\Omega^{(-)} = 1.9\cdot10^{-4}$m$^3$/mol. The shear modulus for both materials $G = 38\,$kPa is based on the uniaxial tensile testing performed on ES/AT junction. In our initial simulation,  we assume that the materials are compressible, more specifically $K = G$. We also set the thickness of the neutral interface between the two ionogels to be $t_n = 10\,$nm, and assume it has the same mechanical properties as the pure polyanion and polycation regions. The boundary conditions set throughout the simulation are as follows:

%

\begin{itemize}
\item electric potential boundary condition $\varphi = 0$ is prescribed at the interface ($x_1=0\,$); i.e. choosing to ground voltage at this position,
\item zero flux boundary conditions prescribed on the top and bottom surfaces (initially at $x_1=0.5\,\mu$m and $x_1=-0.5\,\mu$m), as $\breve{j}^{(+,-)}= 0\,$,
\item symmetry boundary conditions are prescribed at nodes on the face 1 (1-3 plane) and face 2 (1-2 plane),  and the displacements of nodes at the interface ($x_1 = 0$) are constrained in the 1-direction.
\end{itemize}

\begin{table}[b!]
    \centering
    \begin{tabular}{lcccc}
        Parameter & Unit & ES & AT & Neutral layer\\
        \hline
        \hline
           $\varphi_0$  & $\left(\text{V}\right)$ & 0 & 0 & 0  \\
           $c^{(+)}_0$  & $\left(\text{mol/m}^3\right)$  & 1 & 0& 0.5\\
           $c^{(-)}_0$  & $\left(\text{mol/m}^3\right)$  & 0 & 1& 0.5\\
           $c^{(\text{fix})}$ & $\left(\text{mol/m}^3\right)$  & 1 & 1 & 0 \\
           $z^{(\text{fix})}$ & &  -1 & 1 & 0 \\
           $G$ &$\left(\text{MPa}\right)$ & 38 & 38 & 38\\
           $K/G$ & & 1 & 1 & 1\\
        \hline
    \end{tabular}
    \caption{Initial conditions and material parameters for simulating ES/AT junction by \citet{kim2020ionoelastomer}.}
    \label{tab:ES/AT}
\end{table}

The simulation involves three distinct steps:
\begin{enumerate}
\item Allowing the electric potential across the junction to equilibrate, while keeping the above mentioned boundary conditions.
\item Prescribing the mechanical displacement along the 2-direction to nodes on face 3 (located opposite of the face 1) to reach the stretch $\lambda = 1.2$, and allowing the electric potential to equilibrate again.
\item Prescribing the mechanical displacement to nodes at face 3 to match the experimentally applied stretch $\lambda_{\,\text{sin}}$. The displacement is applied along the 2-direction in the form of a sinusoidal wave.
\end{enumerate}

\begin{figure}[h!]
\centering
\begin{tabular}{cc}
        \begin{tikzpicture}
            \node[inner sep=0] (image) at (0,0){\includegraphics[width=0.45\linewidth]{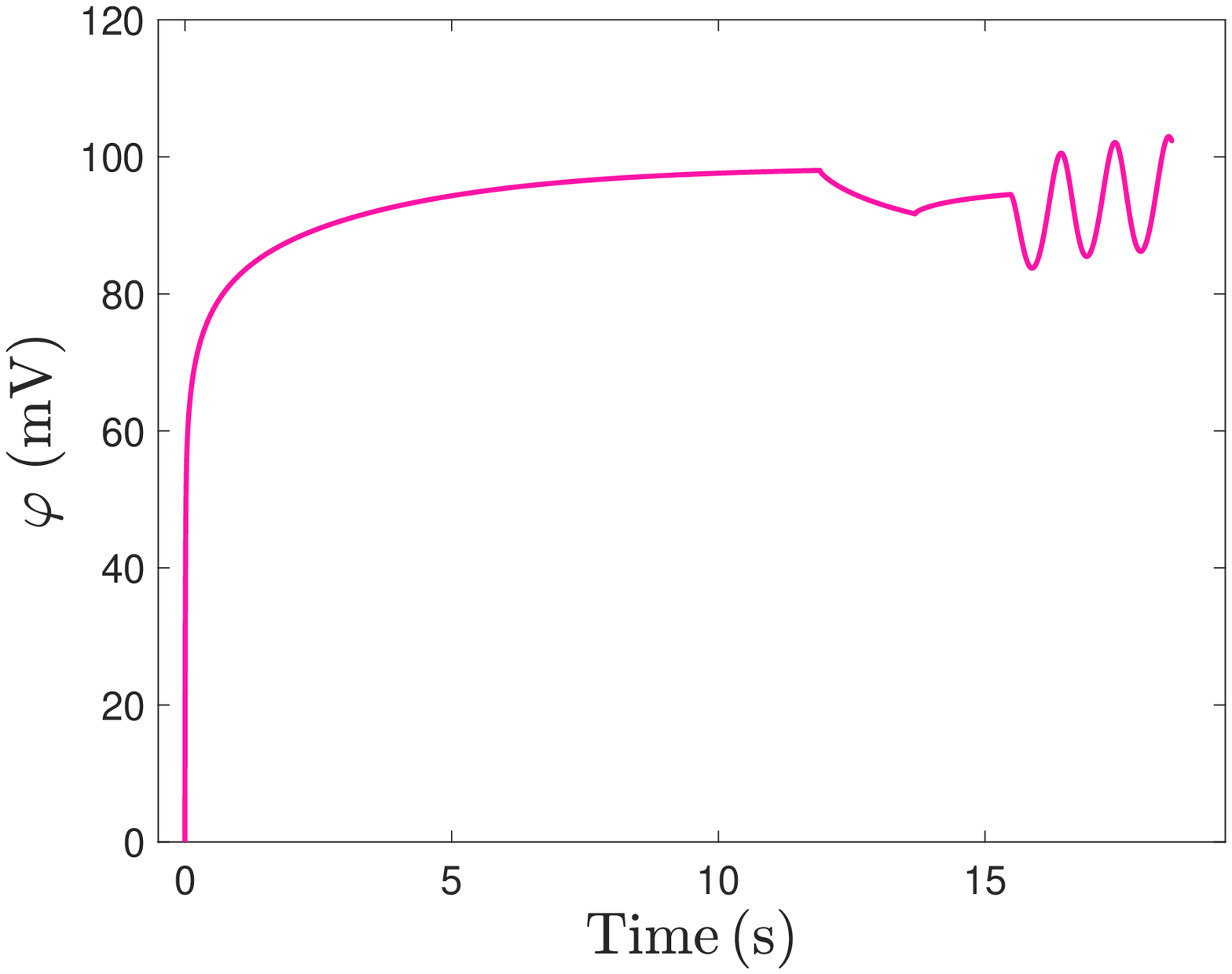}};
           
            \draw [black, thin] (.8,-2.05) -- (.8,2.35);
            \draw [black, thin] (1.85,-2.05) -- (1.85,2.35);
            
            \draw (-0.5,0) node {\tiny 1.};
            \draw (-0.5,-0.2) node {\tiny equilibrating};
            
            \draw (1.3,0) node {\tiny 2.};
            \draw (1.3,-0.2) node {\tiny stretch.};
            \draw (1.3,-0.4) node {\tiny \&};
            \draw (1.3,-0.6) node {\tiny equil.};
            
            \draw (2.4,0) node {\tiny 3.};
            \draw (2.4,-0.2) node {\tiny cyclic};   
            \draw (2.4,-0.4) node {\tiny stretch.};    
        \end{tikzpicture}
&
\begin{tikzpicture}
            \node[inner sep=0] (image) at (0,0){\includegraphics[width=0.45\linewidth]{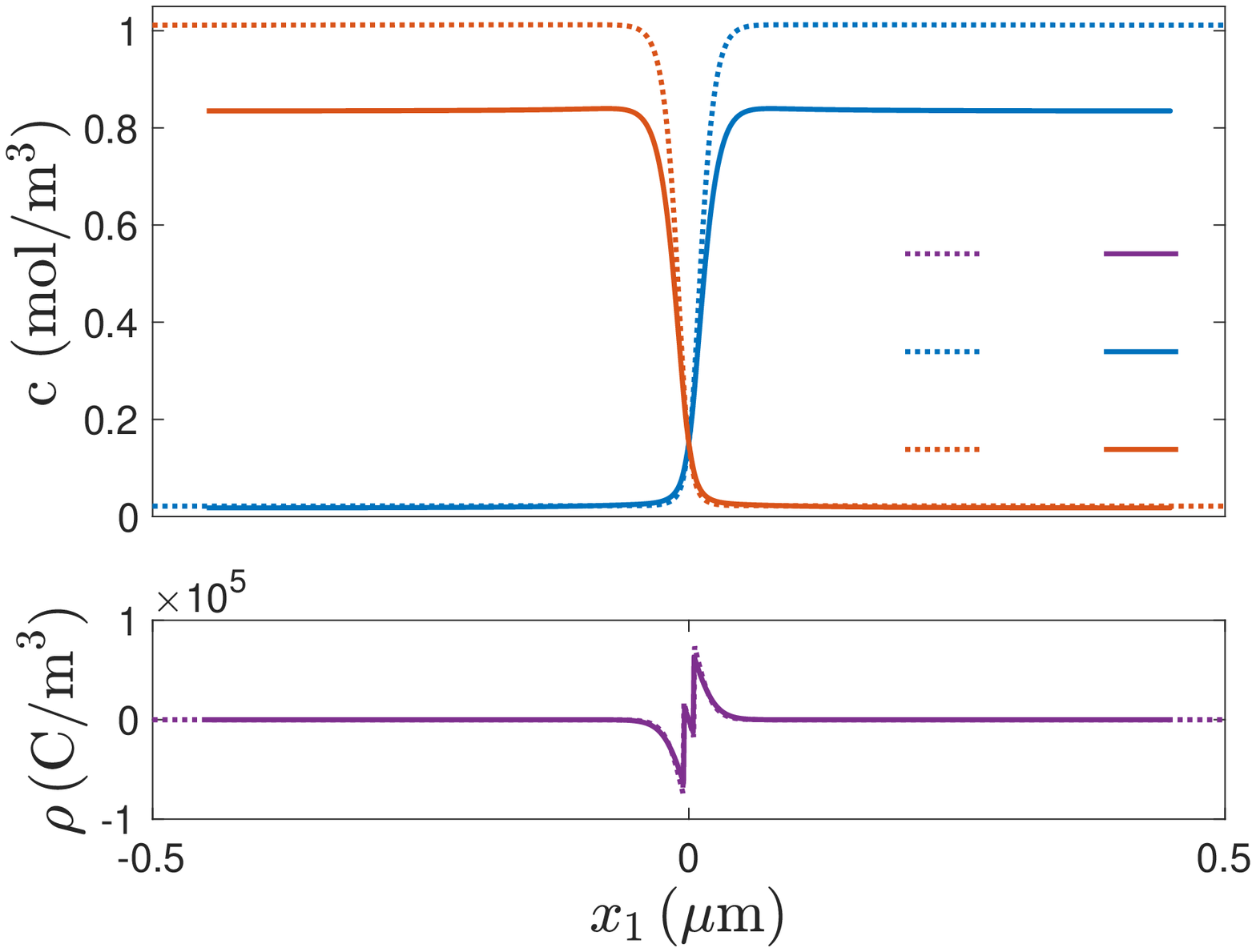}};
           
            \draw (1.5,1.6) node {\footnotesize undef.};
            \draw (2.5,1.6) node {\footnotesize def.};

            \draw (.7,1.15) node {\footnotesize $\rho$};
            \draw (.8,.65) node {\footnotesize $c^{(+)}$};
            \draw (.8,.15) node {\footnotesize $c^{(-)}$};
            
            
            
        \end{tikzpicture}\\
a)& b)\\
\end{tabular}
\includegraphics[width=0.45\linewidth]{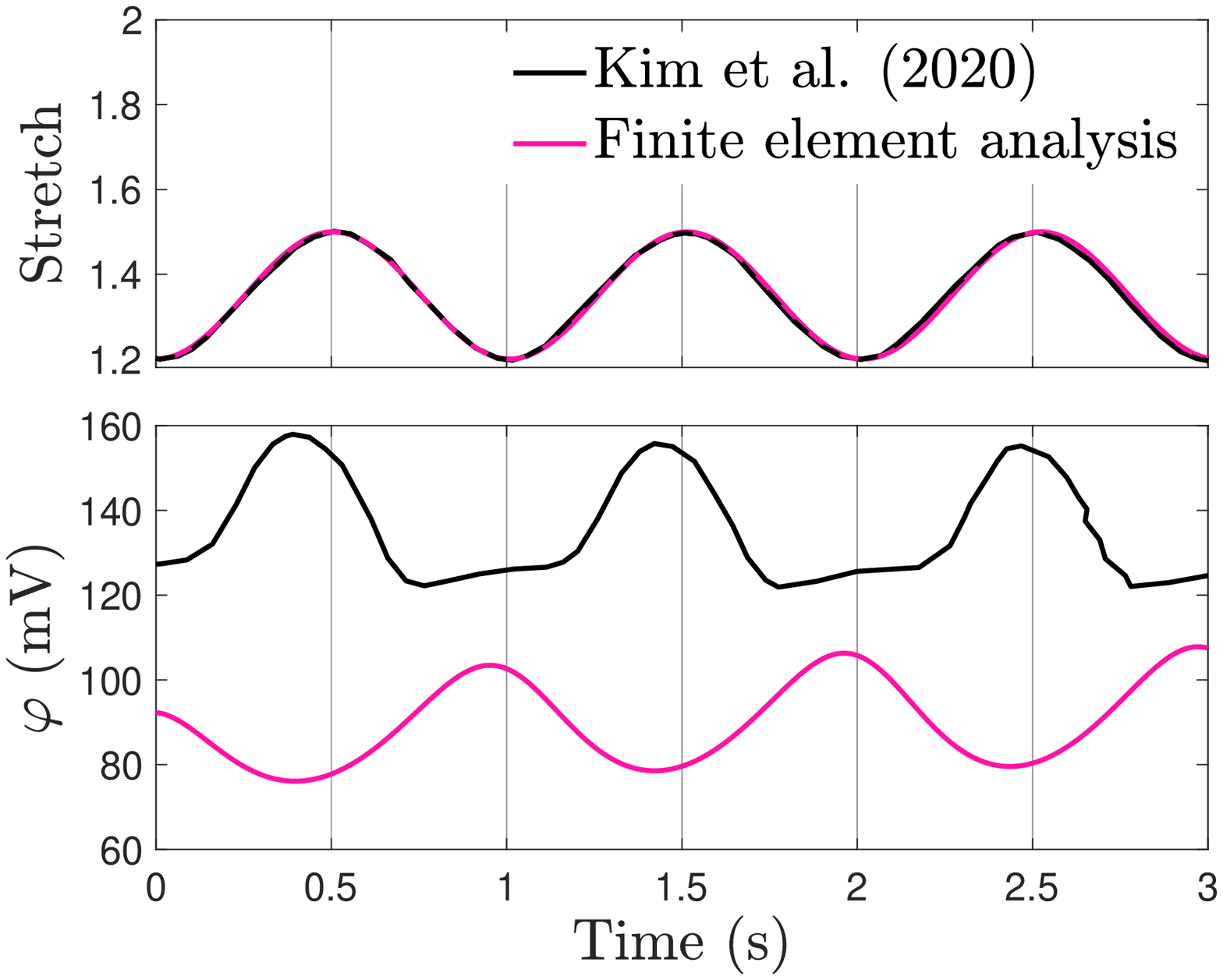}\\
c)
\caption{Finite element analysis of energy harvesting device developed by \cite{kim2020ionoelastomer}. a) Electric potential across the junction under open circuit condition during equilibration, mechanical loading to $\lambda = 1.2$ and second equilibration, and while applying a sinusoidal stretch $\lambda_{\text{sin}}$. b) Concentration of net charge and mobile ions across the junction when undeformed and at equilibrium, and when stretched to 1.5 in the third step.  c) Comparison with the experimental data while applying $\lambda_{\text{sin}}$. Grid lines are added to guide the eye.}
\label{fig:Kim2020_Results}
\end{figure}

Simulation results in Figure \ref{fig:Kim2020_Results}a show that the equilibrium electric potential $\varphi_{eq} = 80\,$mV in the undeformed configuration is established after approximately $10\,$s. We observe a decrease in  electric potential while the device is stretching to $\lambda = 1.2$ in the 2-direction; and this trend extends onto the next step, when the device is subjected to a cycling stretch $\lambda_{\text{sin}}$. On the other hand, reported device performance follows the opposite trend, i.e., an increase in electric potential with elongation in the 2-direction (see Figure \ref{fig:EnergyHarvestingDevices}d).

The decrease in interface thickness brings the polyanion and polycation networks closer, corresponding to an increase in the electric field (i.e., electric potential is acting over shorter distance). However, the net charge concentration $\rho$ slightly decreases when elongated to $\lambda=1.5$, as shown in Figure \ref{fig:Kim2020_Results}b; this effect can be attributed to compressibility as suggested by the drop in ion concentration (this is opposite the device \cite{hou2017flexible} for which compression was applied). Following Gauss's law \eqref{eqn:GaussLaw}, the electric potential needs to decrease to satisfy both the decrease in thickness and the decrease in net charge density. While the computational predictions provide a reasonable and expected dependence between the elongation and electric potential, the experimental data clearly exhibits the opposite trend, as observed in Figure \ref{fig:EnergyHarvestingDevices}d. Moreover, in Figure \ref{fig:Kim2020_Results}c we show the comparison between the simulation results and experimental data while the device is undergoing cyclic loading $\lambda_{\text{sin}}\,$, where we also observe the same/non-matching trends in electrochemical response to an applied mechanical stretch.

In addition to the numerical approach, an analogy between this ionic junction and a capacitor in an electrical circuit provides more insights on the expected device performance. More specifically, elongation in the 2-direction, corresponds to an increase in the cross-sectional area while the electrode-to-electrode distance, i.e.,  the distance between the top and bottom face, is decreasing. Both geometrical effects yield an increase in capacitance based on equation \eqref{eqn:capacitance}; and the increase in capacitance is also documented by  \cite{kim2020ionoelastomer} in their experimental investigation. Therefore, a decrease in the electric potential across the junction is expected based on equation \eqref{eqn:capacitance2} when $\rho$ is constant or decreasing. Nonetheless, the experimental data shows the opposite trend, i.e., the electric potential is increasing with elongation.

One possible source for the experimentally observed increase in the electric potential with increasing stretch is a changing neutral layer thickness. As the interpenetrated neutral region is stretched, the tension could tear the polycation and polyanion polymers back apart, decreasing the size of the neutral layer. In order to determine the effect this thickness change might have on the junction potential, we perform an additional set of five simulations with neutral layer thicknesses of $10\,$nm, $20\,$nm, $30\,$nm, $40\,$nm, and $50\,$nm. We are interested mainly in the equilibrium junction potential, and so simulate only the equilibration step. 

\begin{figure}[h!]
\centering
\begin{tabular}{cc}
\includegraphics[width=0.45\linewidth]{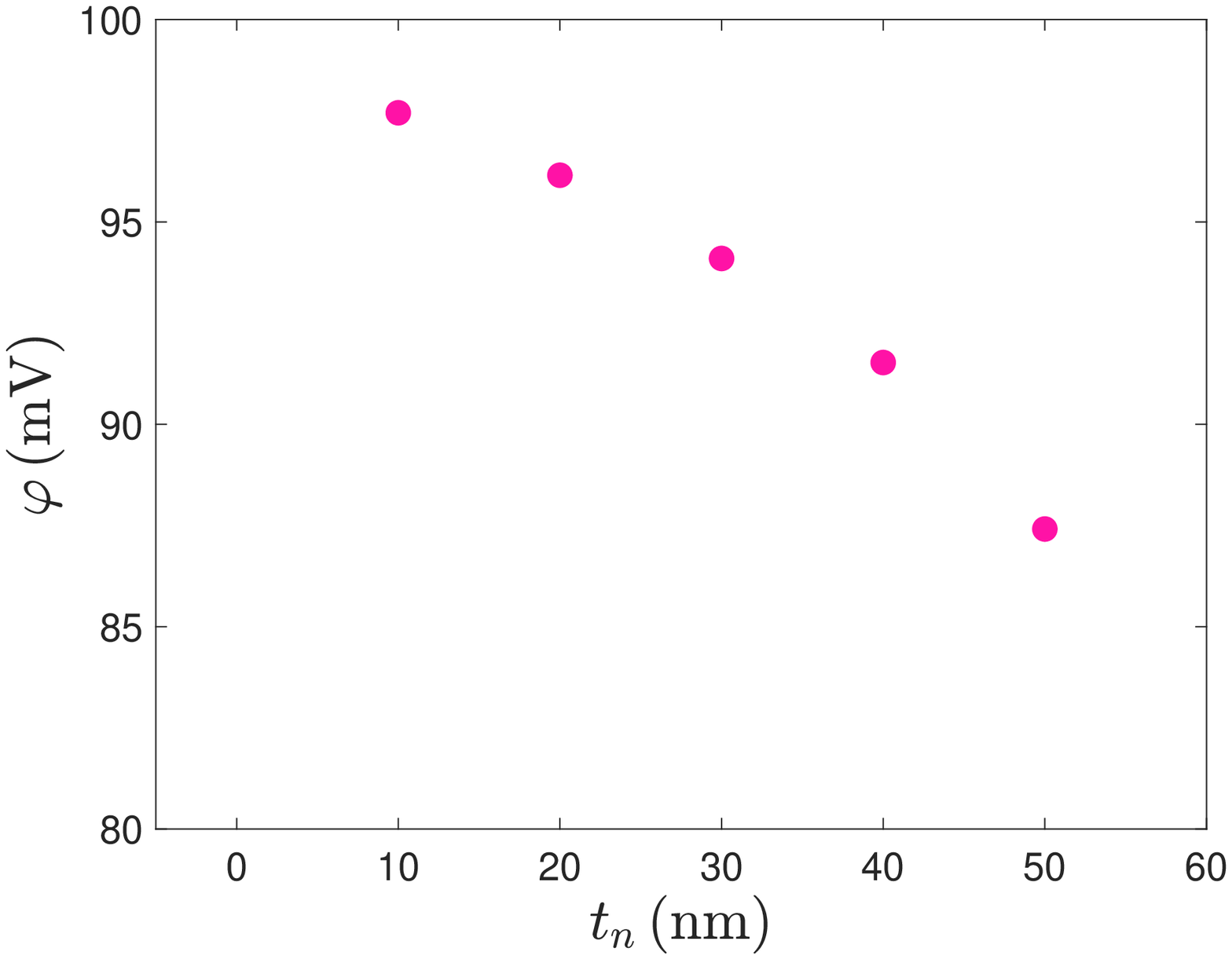}
&
\begin{tikzpicture}
            \node[inner sep=0] (image) at (0,0){\includegraphics[width=0.45\linewidth]{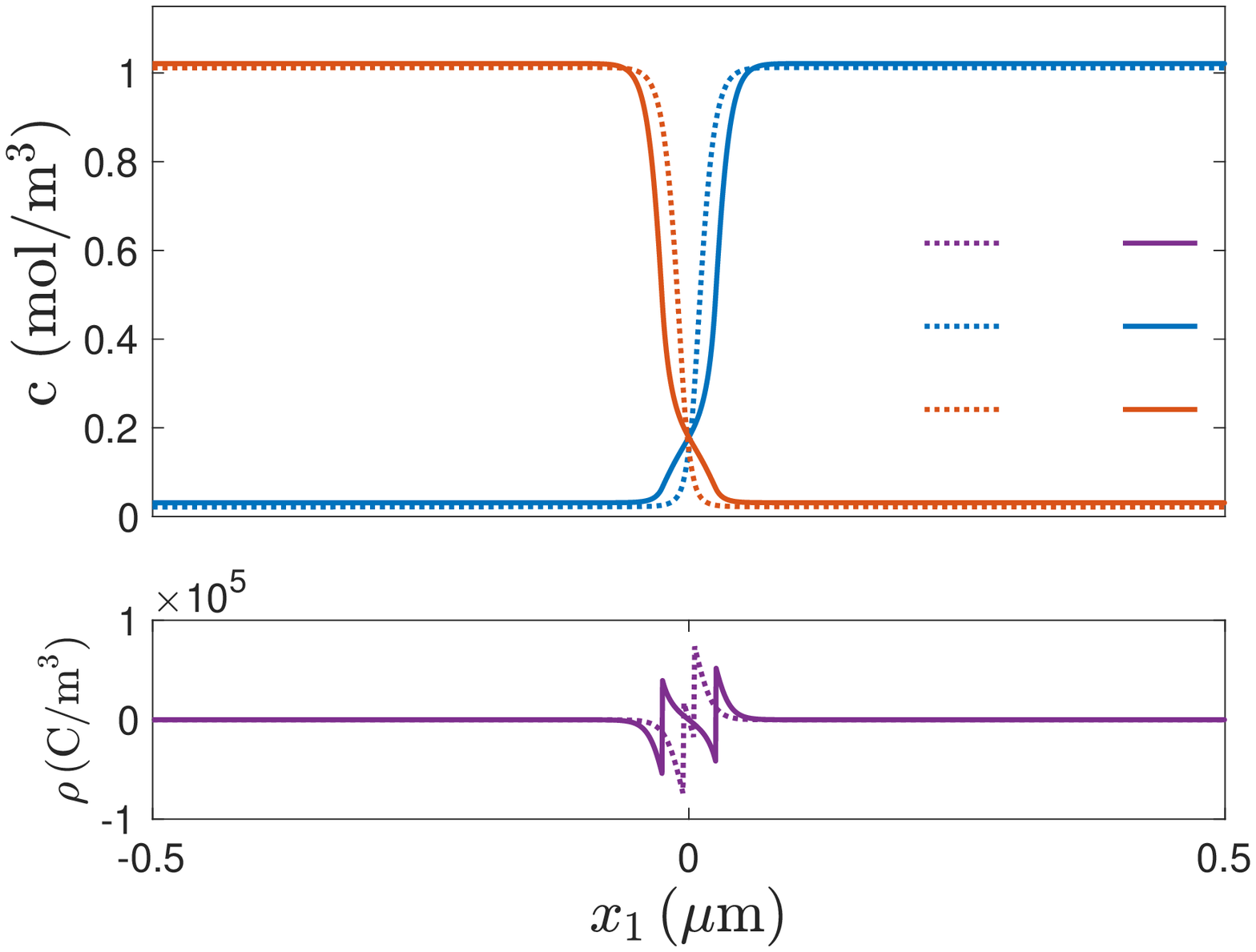}};
           
            \draw (1.6,1.5) node {\footnotesize $10\,$};
            \draw (2.65,1.5) node {\footnotesize $50\,$};

            \draw (.9,1.2) node {\footnotesize $\rho$};
            \draw (1,.85) node {\footnotesize $c^{(+)}$};
            \draw (1,.25) node {\footnotesize $c^{(-)}$};
            
            
            
        \end{tikzpicture}\\
a)& b)\\
\end{tabular}
\caption{Effect of neutral layer thickness on the equilibrium potential for the energy harvesting device developed by \cite{kim2020ionoelastomer}. a) Equilibrium electric potential across the junction obtained at five different neutral layer thickness values $t_n = 10\,,$ 20, 30, 40, and $50\,$nm. b) Net charge density and concentration of mobile species at the end of equilibration step for $t_n = 10\,$ and $50\,$nm.}
\label{fig:Kim2020_NeutralLayer}
\end{figure}

In Figure \ref{fig:Kim2020_NeutralLayer}a we observe an increase in the equilibrium voltage with a decrease in thickness of neutral region between the two oppositely charged polymer networks. This trend holds for all five simulations, and in Figure \ref{fig:Kim2020_NeutralLayer}b,  we show the net charge and ion concentration profiles across the junction when the neutral layer thickness is $t_n = 10$ and $50\,$nm. When the neutral region is large compared to the Debye length, the system can be conceptualized as two separate junctions in series. One junction is formed between the polyanion and the neutral region, and the other junction is formed between the neutral region and polycation. These two junctions are the source of the two discontinuities in the slope $\rho$ in Figure \ref{fig:Kim2020_NeutralLayer}. Across each junction, the concentration of the dominant mobile species drops by half from the polyion side to the neutral side.  For smaller neutral regions, these two junctions begin to overlap, looking more like one junction across which the concentration of each mobile species drops to nearly zero. Because the PNP equations are nonlinear in concentration, the two junction profiles do not superpose. Instead overlapping the junctions with a small neutral layer results in slightly more than double the voltage drop of the individual polyion--neutral junctions. Thus, as the neutral region shrinks, the equilibrium junction voltage increases. Therefore, our computational study shows that a decrease in neutral layer thickness could be responsible for the experimentally measured change in the electric potential.


Next, we probe the effects of material compressibility on the overall electrochemical behavior of the junction. To investigate the influence of this feature, we perform a set of simulations at different bulk-to-shear modulus ratios $K/G$. More specifically, we simulate the response of an incompressible  material  ($K/G = 100$), and then compare the simulation results with the initial case ($K/G = 1$). The remaining material parameters, as well as boundary conditions, are kept the same as in the initial simulation, and we repeat all three simulation steps.

\begin{figure}
\centering
\begin{tabular}{cc}
\begin{tikzpicture}
            \node[inner sep=0] (image) at (0,0){\includegraphics[width=0.45\linewidth]{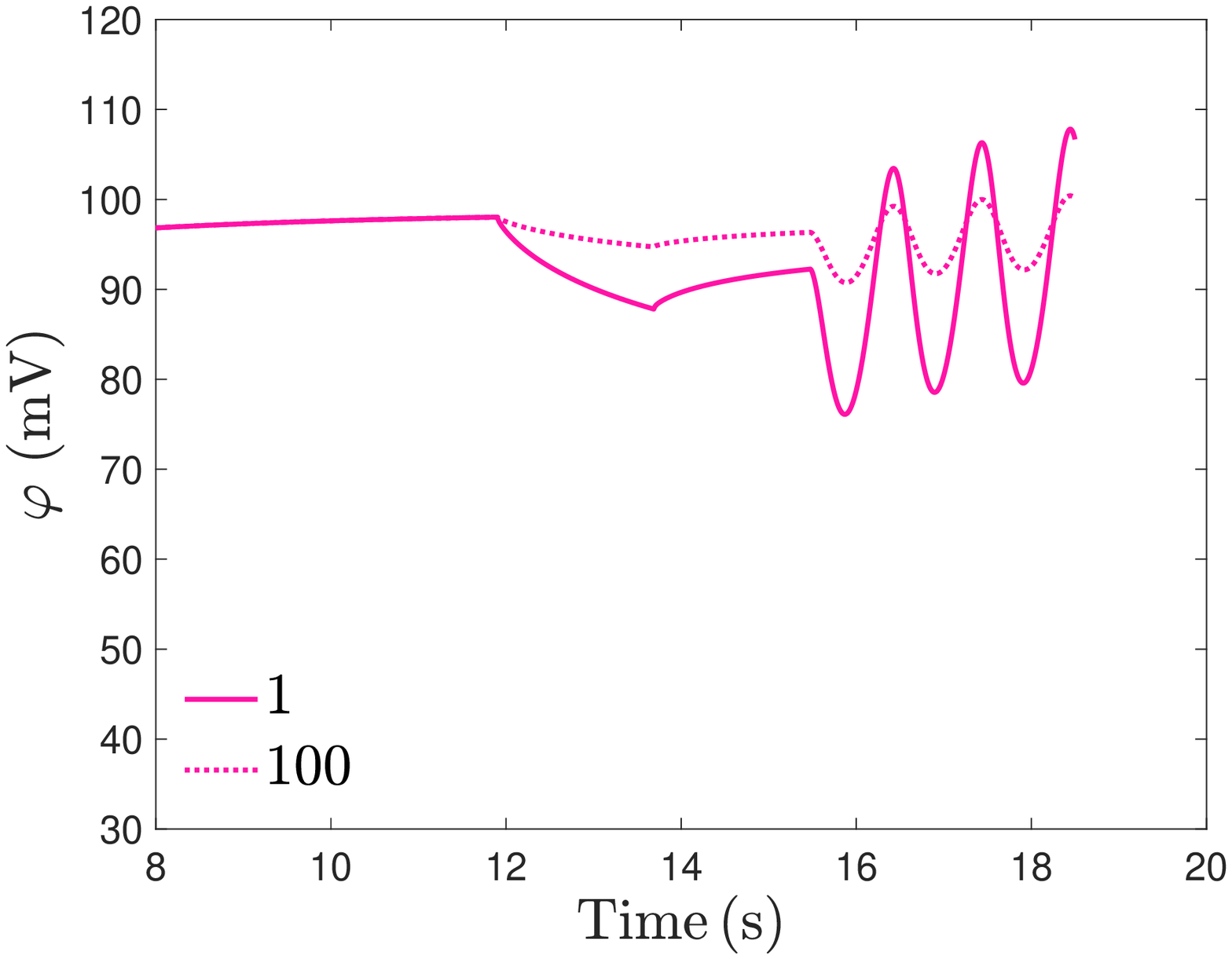}};

            \draw [black, thin] (-.9,-2.05) -- (-.9,2.35);
            \draw [black, thin] (.85,-2.05) -- (.85,2.35);

            \draw (-1.8,0) node {\tiny 1.};
            \draw (-1.8,-0.2) node {\tiny equilibrating};
            
            \draw (0,0) node {\tiny 2.};
            \draw (0,-0.2) node {\tiny stretch.};
            \draw (0,-0.4) node {\tiny \&};
            \draw (0,-0.6) node {\tiny equil.};
            
            \draw (1.7,0) node {\tiny 3.};
            \draw (1.7,-0.2) node {\tiny cyclic};   
            \draw (1.7,-0.4) node {\tiny stretch.};  
            
            \draw (-2,-.9) node {\footnotesize $K/G$};

        \end{tikzpicture}
&
\begin{tikzpicture}
            \node[inner sep=0] (image) at (0,0){\includegraphics[width=0.45\linewidth]{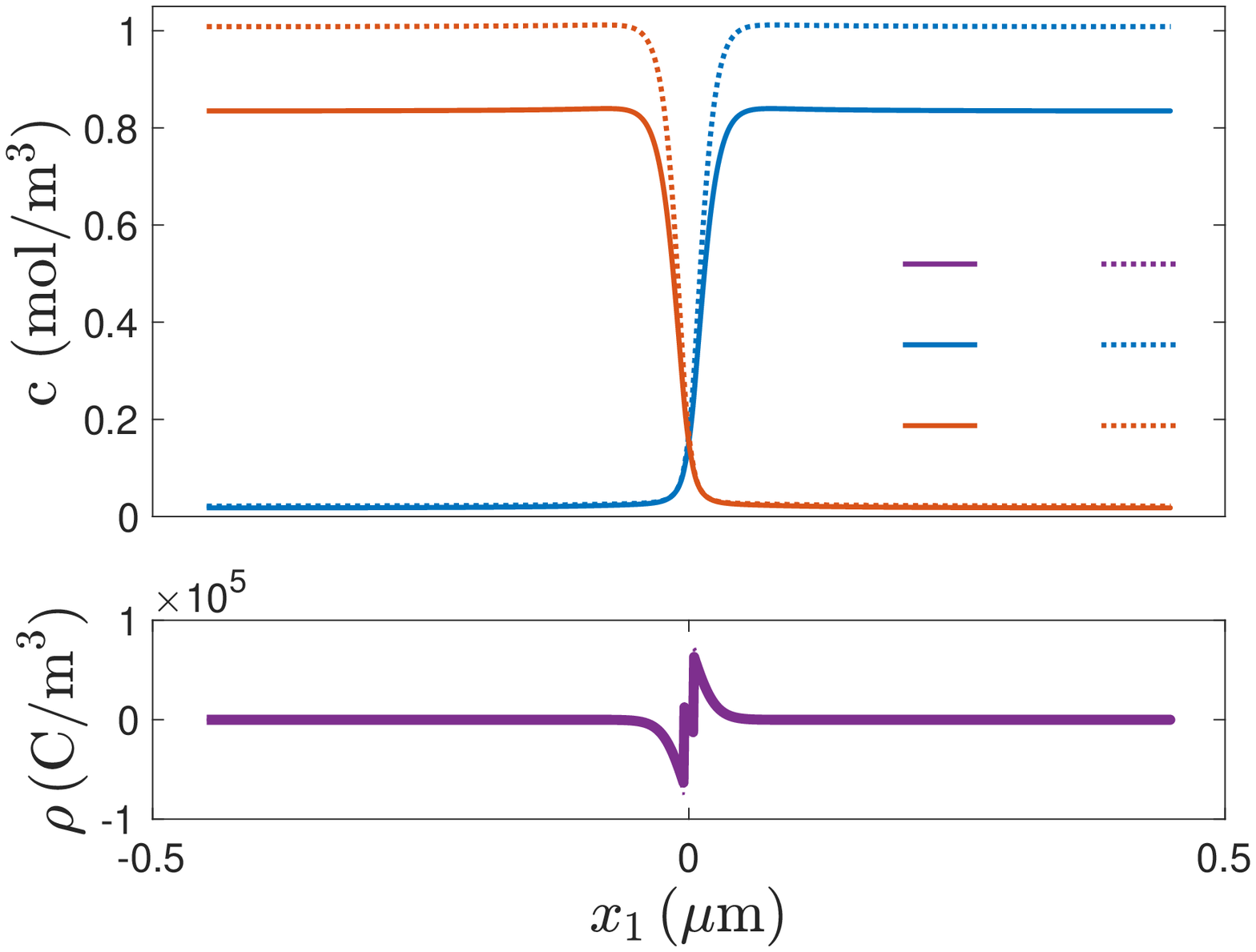}};
           
            \draw (1.425,1.4) node {\footnotesize 1};
            \draw (2.5,1.45) node {\footnotesize 100};
            
            \draw (.9,1.1) node {\footnotesize $\rho$};
            \draw (.85,.75) node {\footnotesize $c^{(+)}$};
            \draw (.85,.25) node {\footnotesize $c^{(-)}$};
       
        \end{tikzpicture}\\
\\
a)& b)\\
\end{tabular}
\caption{Sensitivity of the device performance to compressibility for the energy harvesting device developed by  \cite{kim2020ionoelastomer}. a) Electric potential across the junction under open circuit condition during equilibration step, mechanical loading to $\lambda = 1.2$, and while applying a sinusoidal stretch $\lambda_{\text{sin}}$, at two different bulk-to-shear modulus ratios: 1 and 100. b) Net charge density and concentration of mobile species when stretched to $\lambda = 1.5$ for both bulk-to-shear modulus ratios.}
\label{fig:Kim2020_Compressibility}
\end{figure}

Figure \ref{fig:Kim2020_Compressibility}a shows the device performance at two different shear-to-bulk-modulus ratios during the equilibration step, while loading to $\lambda = 1.2$, and under a cyclic load $\lambda_{\text{sin}}$. Compressibility significantly increases the coupling between the applied mechanical deformation and electric potential, observed as a steeper decrease in potential while stretching, and, similarly, a larger amplitude under cyclic load. As expected, the increase in volume due to a tensile load causes the charge density to decrease, thus yielding a lower electric potential across the junction. This is further corroborated by the net charge and ion concentration profiles when stretched to $\lambda = 1.5$,  shown in Figure \ref{fig:Kim2020_Compressibility}b. The net charge and ion concentration are decreasing with elongation when considering the material as a compressible solid, and remain almost constant when nearly incompressible. The decrease in concentration, coupled with a decrease in distance between the polyanion and polycation layer, yields a lower electric potential across the junction.

One aspect of the device performance we have not investigated in this work is the effect of the dispersed carbon nanotube (CNT) electrodes on the overall electro-chemo-mechanical response. Films of randomly dispersed CNTs are embedded on both sides of ES/AT junctions, acting as high surface area electrodes. Because of the rigidity of individual CNTs relative to the CNT network, the CNT electrode area is independent of ionoelastomer deformation. A major implication of such electrode behavior is that the surface area at the CNT/ES and CNT/AT interface remains constant, while ES/AT interface area is increasing. Consequently, the experimentally observed increase in junction capacitance with in-plane extension is mainly attributed to increasing ES/AT interface area, and also to decreasing distance between the electrodes. In current literature, the effect of high surface area electrodes on the electrochemical response of ionic polymers is captured either by increasing diffusion coefficient and dielectric permittivity values close to the electrodes \citep[cf., e.g.,][]{wallmersperger2008electrochemical}, or by modeling small particles within the polymer matrix \citep[cf., e.g.,][]{akle2011high}. However, current models do not capture the coupled influence of mechanical deformation and the electrochemical response of CNT network and polymer matrix.

Lastly, we note that unlike the previous Nafion/DF25 based device, the difference in timescales between the electrochemical phenomena and the applied mechanical deformation is significantly smaller for ES/AT ionic junction. This is mainly due to three orders of magnitude lower diffusion coefficient; and is also caused by a significantly faster mechanical excitation of $1\,$Hz. Here, the time normalization factor is $\approx 170\,$s$^{-1}$, mapping each non-dimensional time increment $\tau = 1$ to $t=6\,$ms. 

\section{Conclusion}\label{sec:Conclusion}

We have developed a continuum-level multiphysics framework and a constitutive model for the electro-chemo-mechanically coupled behavior of soft ionic conductors. Our modeling approach takes into account the influence of electric fields, the diffusion of multiple ionic species, the influence of fixed charge groups, and mechanical deformation. Both the framework and the model are numerically implemented for use in finite element analysis. We developed two custom finite element codes: (i) a 1D electrochemical version in Matlab and (ii) a coupled electro-chemo-mechanical version, implemented as a user element (UEL) Abaqus subroutine for use in general 3D boundary value problems. We verified  and validated our numerical approach by comparing the finite element analysis with the analytical results and EIS data, respectively. Next, we showcased the usefulness of our tool to accurately capture the change in resistance due to applied mechanical deformation in a hydrogel-based sensor. Finally, we utilized the developed computational tool to analyze the response of two devices involving ionic junctions between polyanionic and polycationic polymer networks. Simulation results helped build understanding of the influence of deformation and mechanical properties on the behavior of these electrochemical systems. More specifically, we investigated the influence of mechanical loading rate, separation between the charged networks, and compressibility, on the performance of ionotronic devices involving junctions between oppositely charged networks. Towards the future, we aim to extend our model to account for the effects of electrode nanostructure, such as CNT networks, on the overall response of the device, as well as the diffusion and interpenetration of polymer chains at the ionic interface between oppositely charged networks. As we continue our experimental characterization of ionotronic materials, we are looking to improve our model to accurately capture the influence of multiple mobile species. Finally, the developed computational tool can be used for design of novel ionotronic devices.

\subsection*{Acknowledgments}
The authors acknowledge the support through the Defense Advanced Research Project Agency Young Faculty Award (DARPA YFA; HR00112010004). The authors would also like to acknowledge Prof. Hyeong Jun Kim at Sogang University for sharing his perspectives on the behavior of ionoelastomer junctions.

\clearpage

\appendix

\section{Numerical implementation}\label{sec:NumericalImplementation}

We make use of the finite element method to solve the governing relations, represented by a set of coupled partial differential equations in \eqref{eqn:GaussLaw}, \eqref{eqn:MassBalance} and \eqref{eqn:ForceBalance}, following the approach laid out in \citet{chester2015}. To find the solutions for this multiphysics probelem, we utilize the Newton-Raphson iterative method and specify a set of element-level residuals and tangents. The developed numerical method is implemented in the form of custom finite element codes in:  (i)  Matlab and Python as a 1D electrochemical code (i.e. in the absence of mechanical deformation), and  (ii) in Abaqus/Standard as a user element (UEL) subroutine. 

To numerically implement the developed framework, first, we define the strong forms, consisting of balance laws and the corresponding boundary conditions. The strong form for Gauss's law is given by:
\begin{equation}
\left. 
\begin{aligned}
 \Div \bfd  &=  eN_a \rho \quad &\text{in} \quad \calB , \\
\bfd \cdot \bfn  &= \breve{q} \quad &\text{on} \quad \calS_q,\\
\varphi &= \breve{\varphi} \quad &\text{on} \quad \calS_\varphi.
\end{aligned}
\right\}
\label{eqn:StrongFormGauss}
\end{equation} 
where $\bfn $ is the outward unit normal, and $\breve{q}$ and $\breve{\varphi}$ represent the prescribed surface charge and electric potential, respectively, on two complementary surfaces $\calS_\Omega$ and $ \calS_\varphi$.

The balance of mass, along with boundary conditions, takes the form:
\begin{equation}
\left. 
\begin{aligned}
\dot{c} ^{(i)}  &= -\Div \bfj ^{(i)} \quad &\text{in} \quad \calB ,\\
\bfj ^{(i)} \cdot \bfn  &= \breve{j}^{(i)} \quad &\text{on} \quad \calS_\bfj,\\
\mu ^{(i)} &= \breve{\mu} ^{(i)} \quad &\text{on} \quad \calS_\mu.
\end{aligned}
\right\}
\label{eqn:StrongFormMass}
\end{equation}
with  $\breve{j}$ and $\breve{\mu}$ denoting the prescribed ion flux and chemical potential, respectively, on two complementary surfaces $\calS_ \bfj$ and $\calS_ \mu$.

Finally, the strong form for balance of forces is:
\begin{equation}
\left. 
\begin{aligned}
\Div\bfT  + \bfb  &= \bf0 \quad &\text{in} \quad \calB , \\
\bfT  \bfn  &= \breve{\bft} \quad &\text{on} \quad \calS_\bft,\\
\bfu &= \breve{\bfu} \quad &\text{on} \quad \calS_\bfu.
\end{aligned}
\right\}
\label{eqn:StrongFormForce}
\end{equation}
where $\bfu$ is the displacement, and $\breve{\bft}$ and $\breve{\bfu}$ are the prescribed surface traction and displacement, respectively, on two complementary surfaces $\calS_ \bft$ and $\calS_ \bfu$.

The body $\calB$ is then approximated with finite elements, such that $\calB  = \cup \calB^e$. Further, we consider the electric potential $\varphi$,  the chemical potential $\mu ^{(i)}$ and the displacement $\bfu$ as the nodal degrees of freedom (DOFs). The nodal DOFs are evaluated within each element using the shape functions $N^A$
\begin{equation}
\varphi = \sum \varphi^A N^A\,, \quad \mu ^{(i)} = \sum \left(
\mu^{(i)} \right)^A N^A \quad \text{and} \quad \bfu = \sum \bfu^A N^A\,,
\end{equation}
 where $A$ denotes the element nodes.

Based on the strong forms in \eqref{eqn:StrongFormGauss}, \eqref{eqn:StrongFormMass} and \eqref{eqn:StrongFormForce}, and using the divergence theorem, we develop weak forms and obtain the element-level residuals for each of the nodal DOF's:
\begin{equation}
\begin{split}
\left( R_\varphi\right) ^A &=  \int_{\calB^e} N^A  \rho\, \text{dv} + \int_{\calB^e} \nabla N^A \, \bfd  \, \text{dv} -  \int_{{\calS_{q^e}}}N^A \breve{q}\,\text{da} \,,\\
\left( R_{\mu(i)} \right)^A &= \int_{\calB^e} N^A \dot{c}_\mat ^{(i)} \,\text{dv} + \int_{\calB^e} \nabla N^A J \bfj ^{(i)} \,\text{dv} - \int_{{\calS_{\bfj^e}}} N^A \breve{j_\mat}^{(i)}\,\text{da}\,,\\
\left( \bfR_\bfu \right)^A &= - \int_{\calB^e} \nabla N^A \bfT  \,\text{dv}  +\int_{{\calS_{\bft^e}}}N^A \, \breve{t}\, \text{da} \,.
\end{split}
\label{eqn:Residuals}
\end{equation}

To provide a convenient set of numbers to the Newton-Raphson solver, we non-dimensionalize the electrochemical quantities, and introduce the non-dimensional electric potential, chemical potential, and charge concentration
\begin{equation}
\Phi = \varphi \frac{eN_a}{RT}\,, \quad \textrm{M} = \frac{\mu}{RT}\, \textrm{, and} \quad C = \frac{c}{c^E}\,,   
\label{eqn:NormalizedDOFs}
\end{equation}
respectively. Considering \eqref{eqn:Residuals} and \eqref{eqn:NormalizedDOFs}, we obtain the non-dimensionalized electrochemical residuals as
\begin{equation}
\begin{split}
\left( R_\Phi\right) ^A &=  \int_{\calB^e} N^A  P\, \text{dv} - \int_{\calB^e} \frac{\partial N^A}{\partial X} \lambda_D^2 L^{-2} \frac{\partial \Phi}{\partial X}   \, \text{dv} - \int_{{\calS_{q^e}}}N^A \breve{Q}\,\text{da} \,,\\
\left( R_{M(i)} \right)^A &= \int_{\calB^e} N^A \frac{\partial C_\mat^{(i)}}{\partial \tau}  \,\text{dv} - \int_{\calB^e} \frac{\partial N^A}{\partial X}  J C^{(i)} \lambda_D L^{-1} \left( \frac{\partial M^{(i)}}{\partial X}+z^{(i)}\frac{\partial \Phi}{\partial X}\right) \,\text{dv}\\
&\quad - \int_{{\calS_{\bfj^e}}} N^A \breve{J}^{(i)}\,\text{da}\,,
\end{split}
\label{eqn:NormResiduals}
\end{equation}
where $P = \sum_i \left( z^{(i)} C^{(i)}\right) + z^{\text{(fix)}} C^{\text{(fix)}}$ is the non-dimensional net charge density.

In addition to non-dimensionalized element-level residuals, we provide a set of non-dimensionalized element-level tangents required to orient the Newton-Raphson solver. The tangents for non-dimensionalized electric potential are obtained as
\begin{equation}
\begin{split}
\left( K_{\Phi\Phi}\right) ^{AB} &= - \pards{\left(R_\Phi\right)^A}{\Phi^B} =   \int_{\calB^e} \frac{\partial N^A}{\partial X} \lambda_D^2 L^{-2}   \frac{\partial N^B}{\partial X} \, \text{dv}\,,\\
\end{split}
\end{equation}
and
\begin{equation}
\begin{split}
\left( K_{\Phi M(i)} \right)^{AB} &= - \pards{\left(R_\Phi\right)^A}{\left(M^{(i)}\right)^B} = - \int_{\calB^e} N^A N^B z^{(i)} \frac{\partial C_\mat^{(i)}}{\partial M^{(i)}}\,\text{dv}.\\
\end{split}
\end{equation}
Similarly, the non-dimensionalized chemical potential tangents are given by
\begin{equation}
\begin{split}
\left( K_{M(i)M(i)}\right) ^{AB} &= - \pards{\left(R_{M(i)}\right)^A}{\left(M^{(i)}\right)^B}\\
&= - \int_{\calB^e} N^A N^B \pards{\left(\partial C_\mat^{(i)} / \partial\tau \right)}{M^{(i)}} \, \text{dv}\\
&\quad+ \int_{\calB^e} J \lambda_D L^{-1} \pards{N^A}{X} \left( N^B \pards{C^{(i)}}{M^{(i)}} \left(\pards{M^{(i)}}{X} + z^{(i)}\pards{\Phi}{X}\right) +  C^{(i)} \pards{N^B}{X} \right)\,\text{dv}\,,
\end{split}
\end{equation}
and
\begin{equation}
\begin{split}
\left( K_{M(i)\Phi}\right) ^{AB} &= - \pards{\left(R_{M(i)}\right)^A}{\Phi^B}\\
&= \int_{\calB^e} J \lambda_D L^{-1} \pards{N^A}{X}  z^{(i)} C^{(i)} \pards{N^B}{X}\,\text{dv}\,.\\
\end{split}
\label{eqn:Tangents}
\end{equation}

\section{Influence of geometrical features and finite element mesh on simulation results}\label{sec:ParameterSensitivity}

In this section we discuss the sensitivity of simulation results to geometrical features and finite element mesh. Throughout this section we consider a region consisting of two oppositely charged polymeric backbones, with geometrical parameters labeled as the length $L$, width $W$, and neutral layer thickness $t_n$, as shown in Figure \ref{fig:Geometry_sensitivity}. The geometry is meshed with 1000 3D elements, each $1\,$nm long, and $1\,\mu$m thick and wide, unless otherwise specified. We take molar volume of both species $\Omega^{(i)} = 10^{-4}\,$mol/m$^3$, and the Debye length $\lambda_D = 10\,$nm. The remaining material parameters and initial conditions are provided in Table \ref{tab:Sensitivity}. Simulations consist of equilibration step over a non-dimensionalized time interval $\tau = 8\,$, and we focus our attention on the electric potential between the top ($x_1=L/2$) and the bottom ($x_1=-L/2$) face. The boundary conditions are as follows:
\begin{itemize}
\item electric potential boundary condition $\varphi = 0$ is prescribed at the interface ($x_1=0\,$); i.e. choosing to ground voltage at this position,
\item zero flux boundary conditions prescribed on the top and bottom surfaces (at $x_1=-L/2$ and $x_1=L/2$), as $\breve{j}^{(+,-)}= 0\,$,
\item symmetry boundary conditions are prescribed to nodes at face 1 (1-3 plane) and face 2 (1-2 plane),  and the displacements of nodes at the interface ($x_1 = 0$) are constrained in the 1-direction.
\end{itemize}
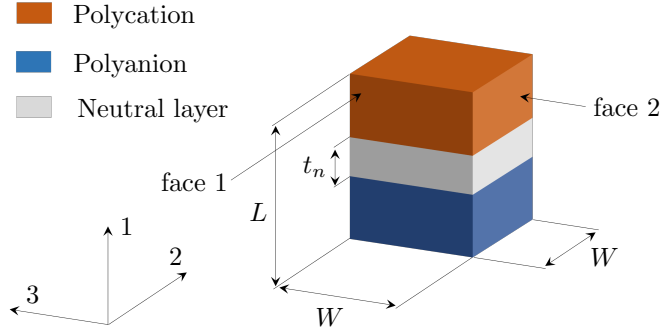
\begin{figure}[t!]
    \centering
         \begin{tikzpicture}
            \node[inner sep=0] (image) at (0,0){\includegraphics[width=0.49\linewidth]{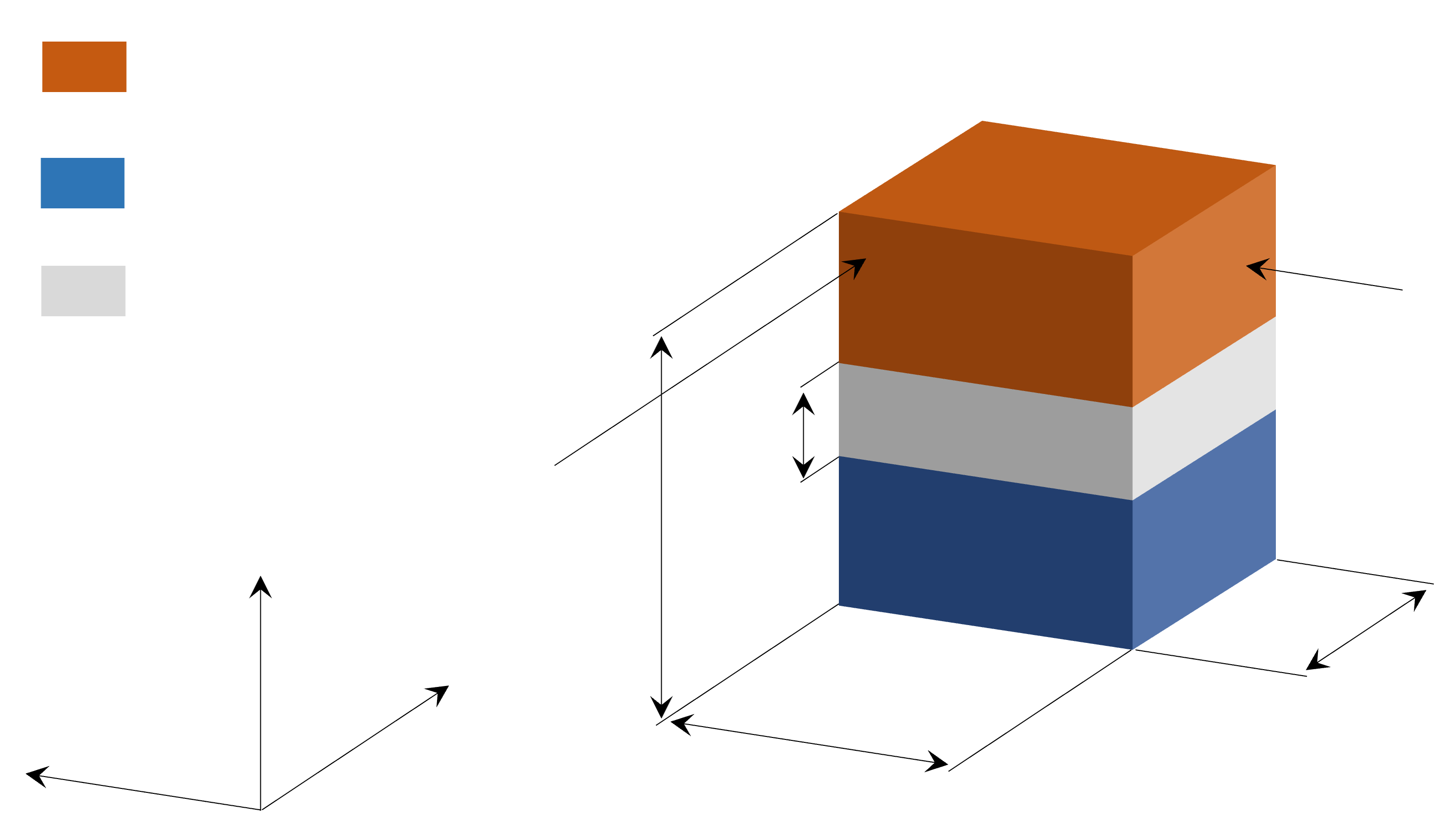}};
            \draw (-2.25,1.9) node {\footnotesize Polycation};
            \draw (-2.3,1.25) node {\footnotesize Polyanion};
            \draw (-2.,0.65) node {\footnotesize Neutral layer};

            \draw (-2.35,-.9) node {\footnotesize 1};
            \draw (-1.7,-1.3) node {\footnotesize 2};
            \draw (-3.6,-1.8) node {\footnotesize 3};	
            
            \draw (-0.6,-0.7) node {\footnotesize {$L$}};
            \draw (0.15,-0.1) node {\footnotesize {$t_n$}};
            \draw (0.35,-2.1) node {\footnotesize {$W$}};
            \draw (4.,-1.3) node {\footnotesize {$W$}};
            
            \draw (-1.45,-0.3) node {\footnotesize {face 1}};
            \draw (4.3,0.7) node {\footnotesize {face 2}};
            
            \end{tikzpicture}
    \caption{Geometry for investigating the effects of geometrical features and finite element mesh.}
    \label{fig:Geometry_sensitivity}
\end{figure}
\begin{table}[h!]
    \centering
    \begin{tabular}{lcccc}
        Parameter & Unit & Polyanion & Polycation & Neutral layer\\
        \hline
        \hline
           $\varphi_0$ & $\left(\text{V}\right)$ & 0 & 0 & 0  \\
           $c^{(+)}_0$ & $\left(\text{mol/m}^3\right)$  & 1 & 0 & 0\\
           $c^{(-)}_0$ & $\left(\text{mol/m}^3\right)$  & 0 & 1& 0\\
           $c^{(\text{fix})}_0$ &$\left(\text{mol/m}^3\right)$  & 1 & 1 & 0 \\
           $z^{(\text{fix})}$ &  &  -1 & 1 & 0 \\
           $G$ & $\left(\text{MPa}\right)$ & 100 & 100 & 100\\
           $K/G$ & & 1 & 1 & 1\\
        \hline
    \end{tabular}
    \caption{Initial conditions and material parameters for investigation of parameter sensitivity.}
    \label{tab:Sensitivity}
\end{table}

We first probe the influence of geometrical features: (1) length-to-Debye length ratio $L/\lambda_D\,$, and (2) the length-to-width ratio $L/W\,$. Then we compare the simulation results when considering a single element in the cross section against the multiple elements in (3).
\begin{enumerate}[(1)]

    \item \underline{Sensitivity to $L/\lambda_D$}\\
     To investigate the sensitivity to $L/\lambda_D$, we conduct a set of simulations at a fixed $\lambda_D = 10\,$nm, and vary the overall length of the geometry $L = 100\,$nm, $200\,$nm, $500\,$nm, $1\,\mu$m, and $2\,\mu$m. In Figure \ref{fig:Sensitivity_Length}a we show the equilibration step results. Here, the electric potential is normalized with $\varphi_{max} = \varphi \left(L/\lambda_D = 200, \tau = 8\right)$, and we observe an increase in electric potential with an increase in $L/\lambda_D$. However, once $L/\lambda_D >80$, there is no noticeable increase in the electric potential (less than 5\% difference), as shown in Figure \ref{fig:Sensitivity_Length}b. Therefore, we take $L/\lambda_D= 100$ as a region sufficiently long to ensure accuracy of our finite element analysis; and the simulation results for this particular ratio are denoted with pink color in \ref{fig:Sensitivity_Length}.
    \begin{figure}[h!]
    \centering
        \begin{tabular}{cc}
            \begin{tikzpicture}
                \node[inner sep=0] (image) at (0,0){\includegraphics[width = 0.45\linewidth]{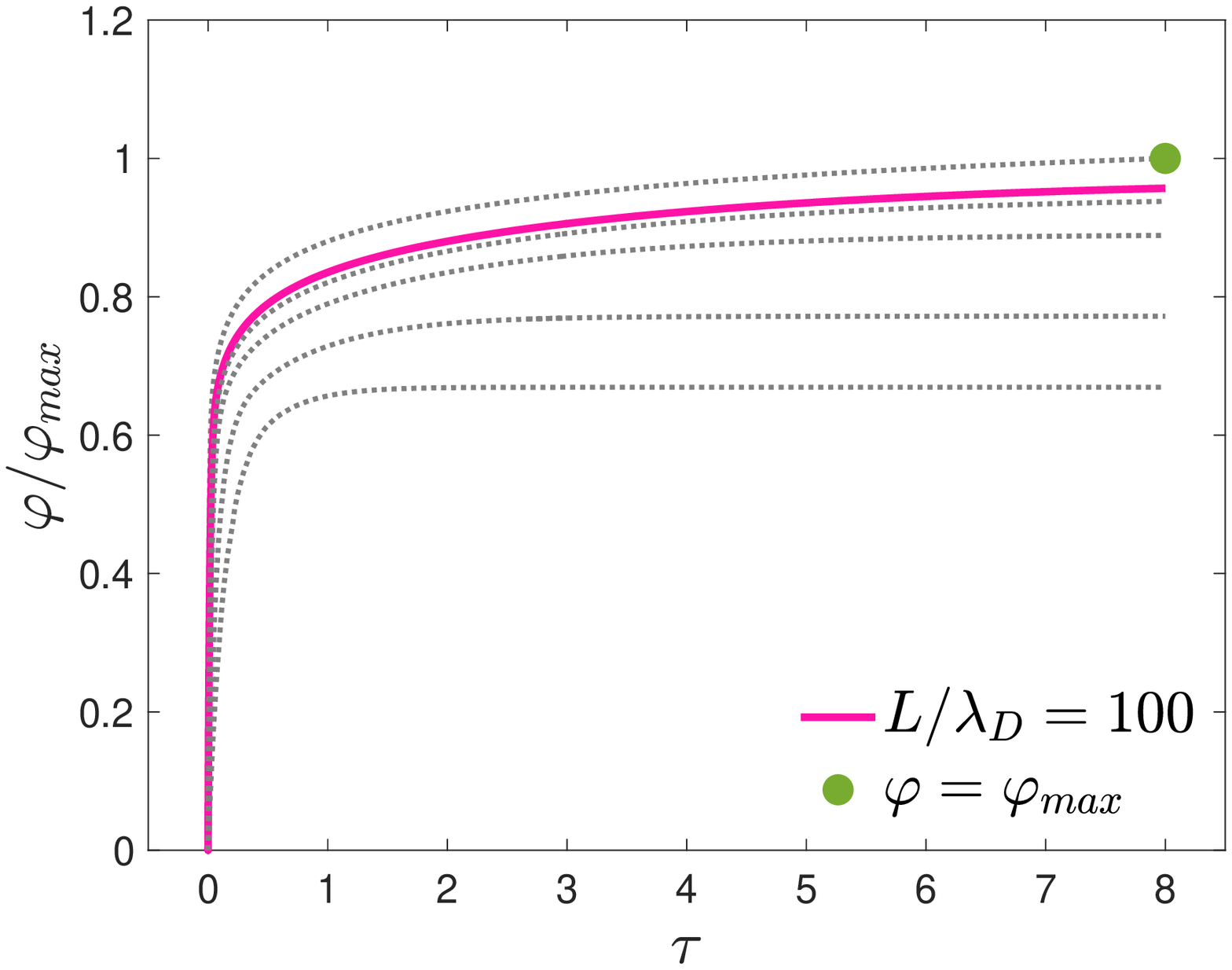}};
                
                \draw [black, thin,-{Stealth[scale=0.75]}] (0.0,0.0) -- (1.45,1.95);
                \draw (-1,-0.1) node {\footnotesize $L/\lambda_D = 10$};
                \draw (1.9,2.05) node {\footnotesize $200$};
            \end{tikzpicture}
            &
                        \begin{tikzpicture}
                \node[inner sep=0] (image) at (0,0){\includegraphics[width = 0.45\linewidth]{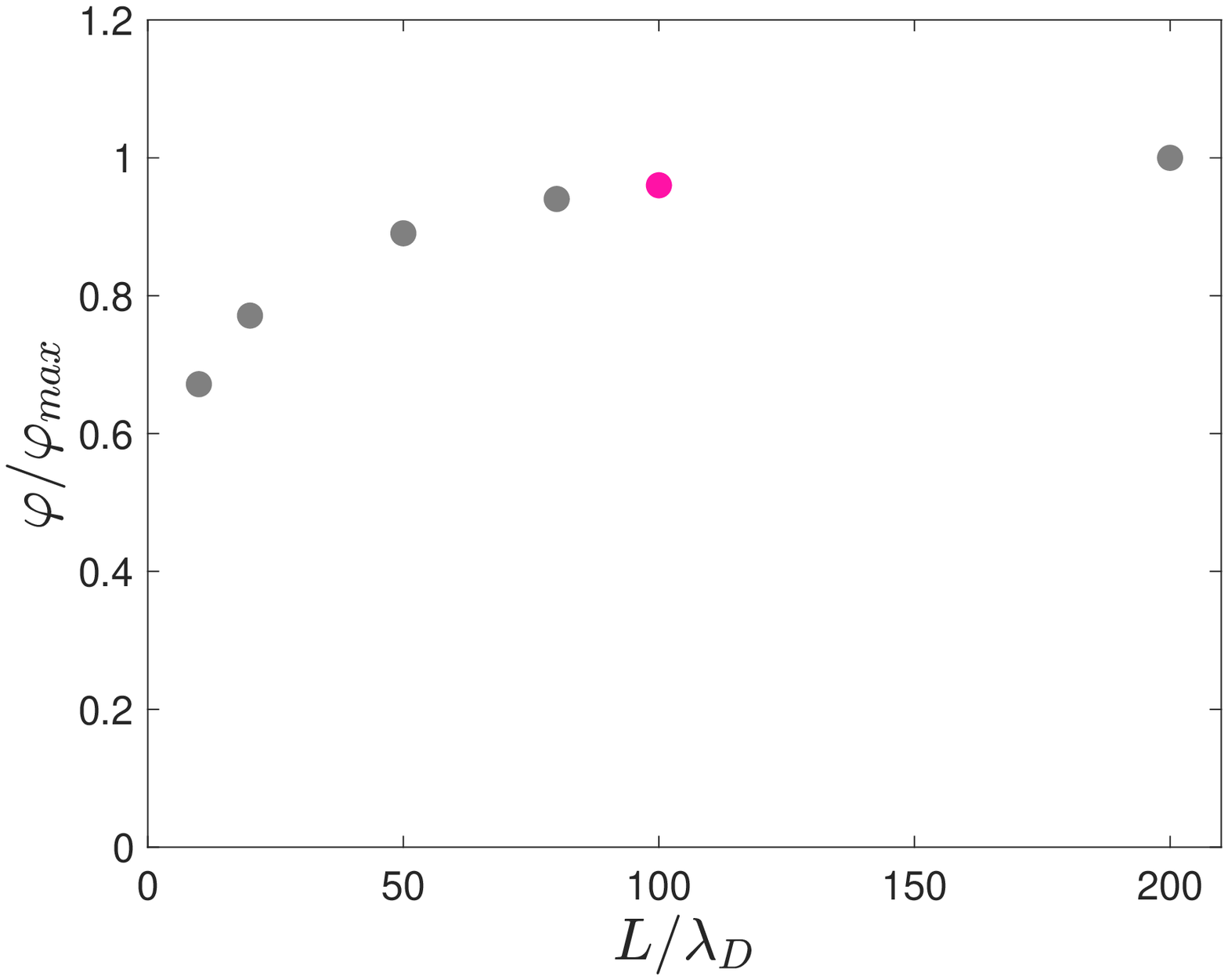}};
    
                \draw (1.95,-1.5) node {\footnotesize at $\tau = 8$};
            \end{tikzpicture}\\
            a)&b)
        \end{tabular}
        \caption{Electric potential across the junction normalized with $\varphi_{max} \left(L/\lambda_D = 200,\tau = 8\right)$ for $L/\lambda_D = 10$ to $200\,$. Pink line denotes the results obtained at $L/\lambda_D = 100$, and the green circle represents the maximum electric potential used for normalization. b) Electric potential across the junction at $\tau = 8$ as a function of $L/\lambda_D$. Arrow indicates increase in $L/\lambda_D$.}
        \label{fig:Sensitivity_Length}
    \end{figure}
    %


     \item \underline{Sensitivity to $L/W$}\\
    Here, we probe the effects of different $W$ values (i.e., the dimensions in 2-direction and 3-direction, with reference to Figure \ref{fig:Geometry_sensitivity}). We keep the length $L=1\,\mu$m constant, and perform three simulations with: $W = 10\,$nm, $1\,\mu$m, and $10\,\mu$m, corresponding to $L/W = 0.1$, $1$, and $10$, respectively. In Figure \ref{fig:Sensitivity_L_to_W}a we show the electric potential normalized with $\varphi_{max} = \varphi \left(L/W = 1, \tau = 8\right)$ during equilibration step, and we observe identical response regardless of $L/W$. The same conclusion can be reached when comparing the normalized electric potential at $\tau = 8$ in Figure \ref{fig:Sensitivity_L_to_W}b.  
  \begin{figure}
    \centering
        \begin{tabular}{cc}
            \begin{tikzpicture}
                \node[inner sep=0] (image) at (0,0){\includegraphics[width = 0.45\linewidth]{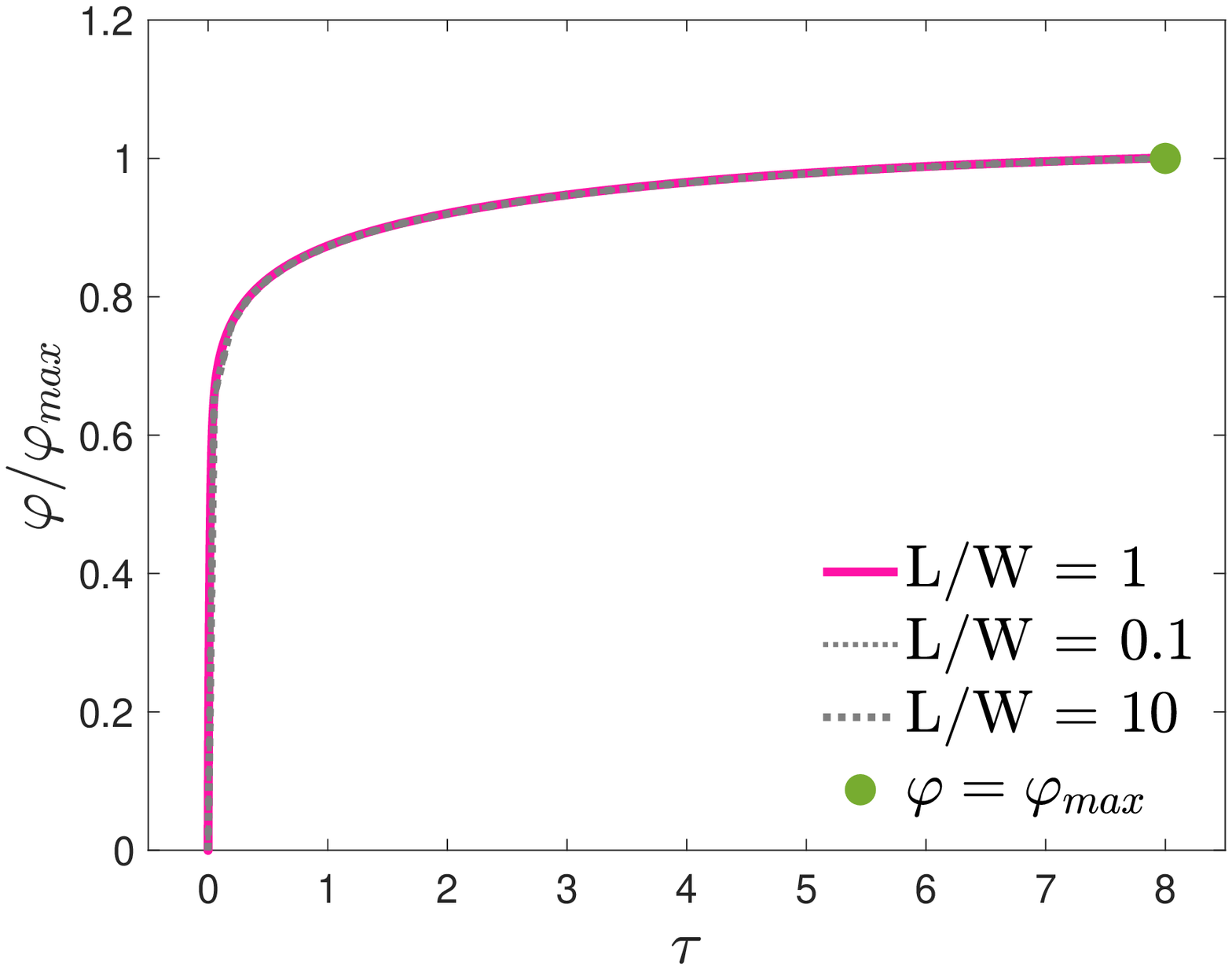}};
                

            \end{tikzpicture}
            &
            \begin{tikzpicture}
                \node[inner sep=0] (image) at (0,0){\includegraphics[width = 0.45\linewidth]{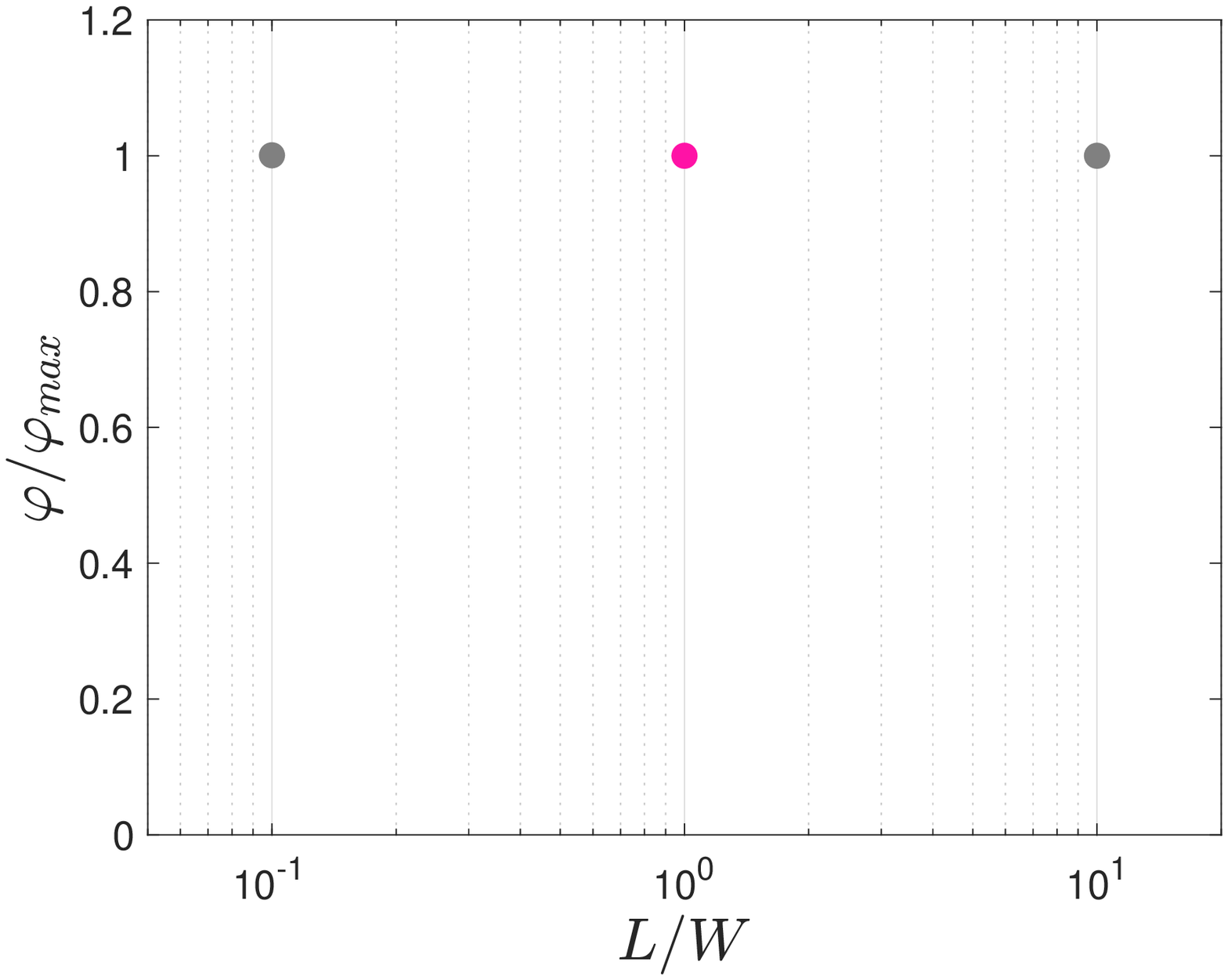}};
    
                \draw (1.95,-1.5) node {\footnotesize at $\tau = 8$};
            \end{tikzpicture}\\
            a)&b)
        \end{tabular}
        \caption{Electric potential across the junction normalized with $\varphi_{max} \left(L/W= 1,\tau = 8\right)$ for $L/W = 1\,,$ $0.1\,,$ and $10\,$. The green circle represents the maximum electric potential used for normalization. b) Electric potential across the junction normalized with $\varphi_{max}$ at $\tau = 8$ as a function of $L/W$.  Grid lines are added to guide the eye.}
        \label{fig:Sensitivity_L_to_W}
    \end{figure}


    \item \underline{Sensitivity to the number of finite elements in cross-section}\\
    To ensure we are accurately capturing the device performance with a single finite element in cross-section, we conduct a set of simulations using 4 and 9 finite elements in the cross-section. The results in \ref{fig:Sensitivity_numElem}a show the electric potential normalized with $\varphi_{max} = \varphi \left(1\text{ element}, \tau = 8\right)$ during equilibration step, and the results for different number of elements in cross-section are indistinguishable. Further, in Figure \ref{fig:Sensitivity_numElem}b we show the normalized electric potential after completion of equilibration step ($\tau = 8$), and we can conclude that having a single element in cross-section is accurately capturing the behavior of electrochemical systems under consideration.
    
    \begin{figure}[!t]
    \centering
        \begin{tabular}{cc}
            \begin{tikzpicture}
                \node[inner sep=0] (image) at (0,0){\includegraphics[width = 0.45\linewidth]{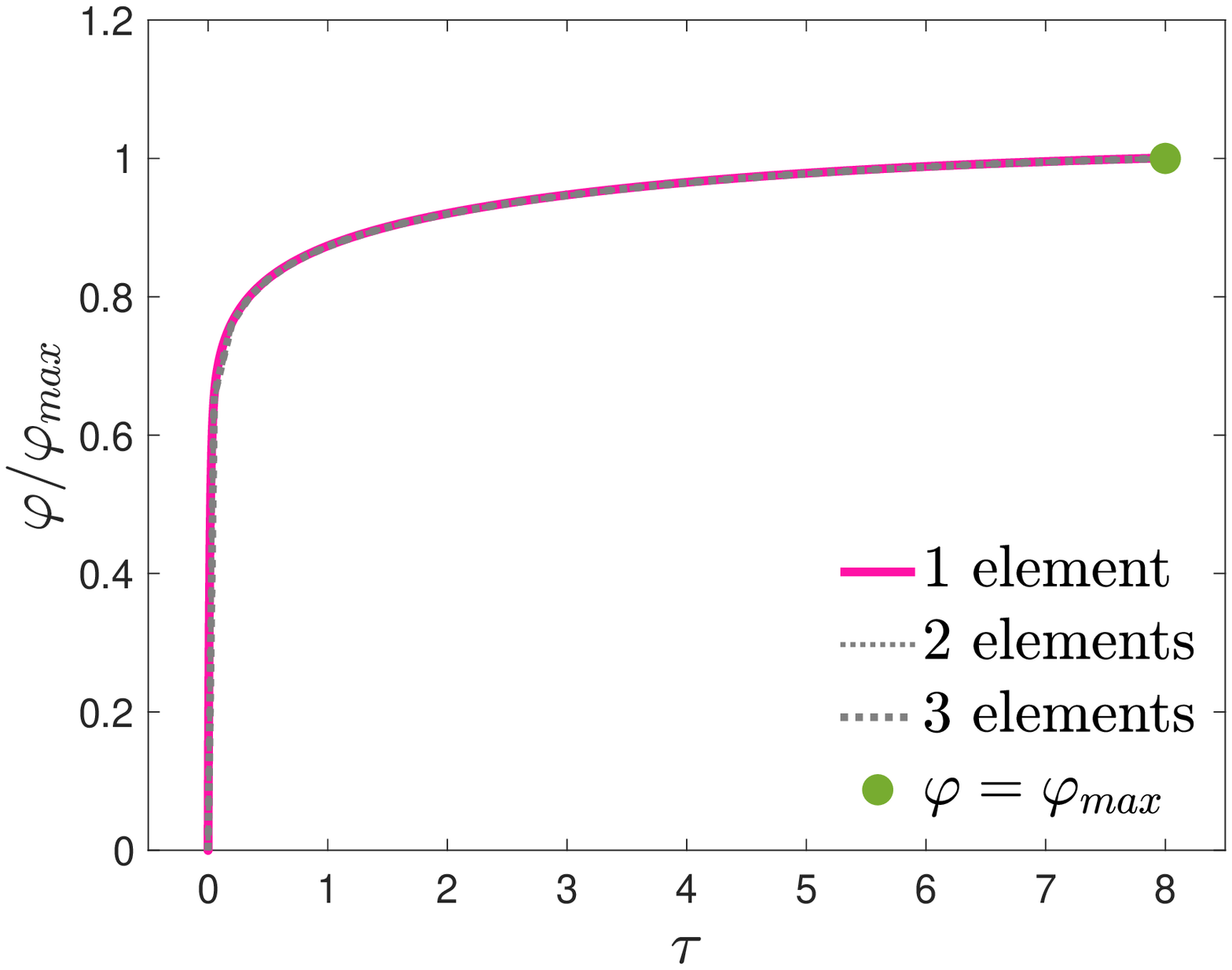}};
                

            \end{tikzpicture}
            &
            \begin{tikzpicture}
                \node[inner sep=0] (image) at (0,0){\includegraphics[width = 0.45\linewidth]{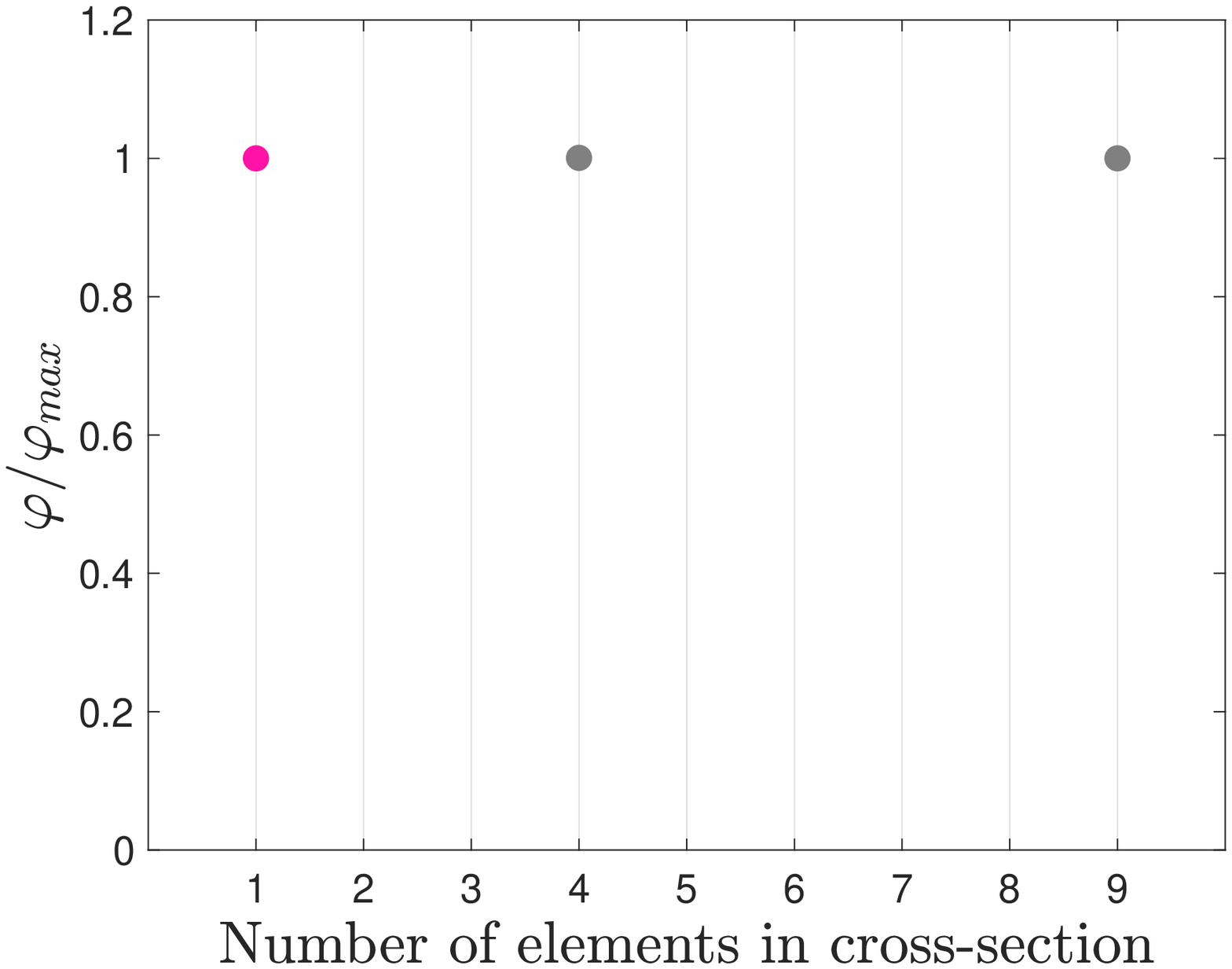}};
    
                \draw (1.95,-1.5) node {\footnotesize at $\tau = 8$};
            \end{tikzpicture}\\
            a)&b)
        \end{tabular}
        \caption{Electric potential across the junction normalized with $\varphi_{max} \left(1\text{ element},\tau = 8\right)$ when finite element mesh contains 1, 4, and 9 elements in cross-section. The green circle represents the maximum electric potential used for normalization. b) Electric potential across the junction at $\tau = 8$ for different number of finite elements in cross-section. Grid lines are added to guide the eye.}
        \label{fig:Sensitivity_numElem}
    \end{figure}
    
    
    

\end{enumerate}

\clearpage

\bibliographystyle{abbrvnat}
\bibliography{mainFile}

\begin{thebibliography}{96}
\providecommand{\natexlab}[1]{#1}
\providecommand{\url}[1]{\texttt{#1}}
\expandafter\ifx\csname urlstyle\endcsname\relax
  \providecommand{\doi}[1]{doi: #1}\else
  \providecommand{\doi}{doi: \begingroup \urlstyle{rm}\Url}\fi

\bibitem[Agmon(1995)]{agmon1995grotthuss}
N.~Agmon.
\newblock The grotthuss mechanism.
\newblock \emph{Chemical Physics Letters}, 244\penalty0 (5-6):\penalty0
  456--462, 1995.

\bibitem[Akle et~al.(2011)Akle, Habchi, Wallmersperger, Akle, and
  Leo]{akle2011high}
B.~J. Akle, W.~Habchi, T.~Wallmersperger, E.~J. Akle, and D.~J. Leo.
\newblock High surface area electrodes in ionic polymer transducers: numerical
  and experimental investigations of the electro-chemical behavior.
\newblock \emph{Journal of Applied Physics}, 109\penalty0 (7):\penalty0 074509,
  2011.

\bibitem[Aureli et~al.(2009)Aureli, Prince, Porfiri, and
  Peterson]{aureli2009energy}
M.~Aureli, C.~Prince, M.~Porfiri, and S.~D. Peterson.
\newblock Energy harvesting from base excitation of ionic polymer metal
  composites in fluid environments.
\newblock \emph{Smart materials and Structures}, 19\penalty0 (1):\penalty0
  015003, 2009.

\bibitem[Banerjee and Curtin(2004)]{banerjee2004nafion}
S.~Banerjee and D.~E. Curtin.
\newblock Nafion{\textregistered} perfluorinated membranes in fuel cells.
\newblock \emph{Journal of fluorine chemistry}, 125\penalty0 (8):\penalty0
  1211--1216, 2004.

\bibitem[Bazant et~al.(2004)Bazant, Thornton, and Ajdari]{bazant2004diffuse}
M.~Z. Bazant, K.~Thornton, and A.~Ajdari.
\newblock Diffuse-charge dynamics in electrochemical systems.
\newblock \emph{Physical review E}, 70\penalty0 (2):\penalty0 021506, 2004.

\bibitem[Bouklas et~al.(2015)Bouklas, Landis, and Huang]{bouklas2015nonlinear}
N.~Bouklas, C.~M. Landis, and R.~Huang.
\newblock A nonlinear, transient finite element method for coupled solvent
  diffusion and large deformation of hydrogels.
\newblock \emph{Journal of the Mechanics and Physics of Solids}, 79:\penalty0
  21--43, 2015.

\bibitem[Brumleve and Buck(1978)]{brumleve1978numerical}
T.~R. Brumleve and R.~P. Buck.
\newblock Numerical solution of the nernst-planck and poisson equation system
  with applications to membrane electrochemistry and solid state physics.
\newblock \emph{Journal of Electroanalytical Chemistry and Interfacial
  Electrochemistry}, 90\penalty0 (1):\penalty0 1--31, 1978.

\bibitem[Bucci et~al.(2014)Bucci, Nadimpalli, Sethuraman, Bower, and
  Guduru]{bucci2014measurement}
G.~Bucci, S.~P. Nadimpalli, V.~A. Sethuraman, A.~F. Bower, and P.~R. Guduru.
\newblock Measurement and modeling of the mechanical and electrochemical
  response of amorphous si thin film electrodes during cyclic lithiation.
\newblock \emph{Journal of the Mechanics and Physics of Solids}, 62:\penalty0
  276--294, 2014.

\bibitem[Buck(1969)]{buck1969diffuse}
R.~Buck.
\newblock Diffuse layer charge relaxation at the ideally polarized electrode.
\newblock \emph{Journal of Electroanalytical Chemistry and Interfacial
  Electrochemistry}, 23\penalty0 (2):\penalty0 219--240, 1969.

\bibitem[Cayre et~al.(2007)Cayre, Chang, and Velev]{cayre2007polyelectrolyte}
O.~J. Cayre, S.~T. Chang, and O.~D. Velev.
\newblock Polyelectrolyte diode: nonlinear current response of a junction
  between aqueous ionic gels.
\newblock \emph{Journal of the American Chemical Society}, 129\penalty0
  (35):\penalty0 10801--10806, 2007.

\bibitem[Chen et~al.(2014)Chen, Lu, Yang, Yang, Zhou, Chen, and
  Suo]{chen2014highly}
B.~Chen, J.~J. Lu, C.~H. Yang, J.~H. Yang, J.~Zhou, Y.~M. Chen, and Z.~Suo.
\newblock Highly stretchable and transparent ionogels as nonvolatile conductors
  for dielectric elastomer transducers.
\newblock \emph{ACS applied materials \& interfaces}, 6\penalty0 (10):\penalty0
  7840--7845, 2014.

\bibitem[Chen et~al.(2017)Chen, Henderson, Pan, Perdue, Cao, Hu, Wan, Han,
  Mueller, Zhang, et~al.]{chen2017improving}
J.~Chen, W.~A. Henderson, H.~Pan, B.~R. Perdue, R.~Cao, J.~Z. Hu, C.~Wan, K.~S.
  Han, K.~T. Mueller, J.-G. Zhang, et~al.
\newblock Improving lithium--sulfur battery performance under lean electrolyte
  through nanoscale confinement in soft swellable gels.
\newblock \emph{Nano letters}, 17\penalty0 (5):\penalty0 3061--3067, 2017.

\bibitem[Chester and Anand(2010)]{chester2010}
S.~A. Chester and L.~Anand.
\newblock A coupled theory of fluid permeation and large deformations for
  elastomeric materials.
\newblock \emph{Journal of the Mechanics and Physics of Solids}, 58\penalty0
  (11):\penalty0 1879--1906, 2010.

\bibitem[Chester et~al.(2015)Chester, Di~Leo, and Anand]{chester2015}
S.~A. Chester, C.~V. Di~Leo, and L.~Anand.
\newblock A finite element implementation of a coupled diffusion-deformation
  theory for elastomeric gels.
\newblock \emph{International Journal of Solids and Structures}, 52:\penalty0
  1--18, 2015.

\bibitem[Choi et~al.(2005)Choi, Jalani, and Datta]{choi2005thermodynamics}
P.~Choi, N.~H. Jalani, and R.~Datta.
\newblock Thermodynamics and proton transport in nafion: Ii. proton diffusion
  mechanisms and conductivity.
\newblock \emph{Journal of the electrochemical society}, 152\penalty0
  (3):\penalty0 E123, 2005.

\bibitem[Corry et~al.(2000)Corry, Kuyucak, and Chung]{corry2000tests}
B.~Corry, S.~Kuyucak, and S.-H. Chung.
\newblock Tests of continuum theories as models of ion channels. ii.
  poisson--nernst--planck theory versus brownian dynamics.
\newblock \emph{Biophysical Journal}, 78\penalty0 (5):\penalty0 2364--2381,
  2000.

\bibitem[Cukierman(2006)]{cukierman2006tu}
S.~Cukierman.
\newblock Et tu, grotthuss! and other unfinished stories.
\newblock \emph{Biochimica et Biophysica Acta (BBA)-Bioenergetics},
  1757\penalty0 (8):\penalty0 876--885, 2006.

\bibitem[De~Gennes et~al.(2000)De~Gennes, Okumura, Shahinpoor, and
  Kim]{de2000mechanoelectric}
P.~De~Gennes, K.~Okumura, M.~Shahinpoor, and K.~J. Kim.
\newblock Mechanoelectric effects in ionic gels.
\newblock \emph{EPL (Europhysics Letters)}, 50\penalty0 (4):\penalty0 513,
  2000.

\bibitem[Doi et~al.(1992)Doi, Matsumoto, and Hirose]{doi1992deformation}
M.~Doi, M.~Matsumoto, and Y.~Hirose.
\newblock Deformation of ionic polymer gels by electric fields.
\newblock \emph{Macromolecules}, 25\penalty0 (20):\penalty0 5504--5511, 1992.

\bibitem[Drozdov and deClaville Christiansen(2015)]{drozdov2015modeling}
A.~Drozdov and J.~deClaville Christiansen.
\newblock Modeling the effects of ph and ionic strength on swelling of
  polyelectrolyte gels.
\newblock \emph{The Journal of chemical physics}, 142\penalty0 (11):\penalty0
  114904, 2015.

\bibitem[Eisenberg et~al.(2011)Eisenberg, Hyon, and
  Liu]{eisenberg2011mathematical}
B.~Eisenberg, Y.~Hyon, and C.~Liu.
\newblock A mathematical model for the hard sphere repulsion in ionic
  solutions.
\newblock \emph{Communications in Mathematical Sciences}, 9\penalty0
  (2):\penalty0 459--475, 2011.

\bibitem[Flavell et~al.(2014)Flavell, Machen, Eisenberg, Kabre, Liu, and
  Li]{flavell2014conservative}
A.~Flavell, M.~Machen, B.~Eisenberg, J.~Kabre, C.~Liu, and X.~Li.
\newblock A conservative finite difference scheme for poisson--nernst--planck
  equations.
\newblock \emph{Journal of Computational Electronics}, 13\penalty0
  (1):\penalty0 235--249, 2014.

\bibitem[Gabrielsson et~al.(2012)Gabrielsson, Tybrandt, and
  Berggren]{gabrielsson2012ion}
E.~O. Gabrielsson, K.~Tybrandt, and M.~Berggren.
\newblock Ion diode logics for ph control.
\newblock \emph{Lab on a Chip}, 12\penalty0 (14):\penalty0 2507--2513, 2012.

\bibitem[Ganser et~al.(2019)Ganser, Hildebrand, Kamlah, and
  McMeeking]{ganser2019finite}
M.~Ganser, F.~E. Hildebrand, M.~Kamlah, and R.~M. McMeeking.
\newblock A finite strain electro-chemo-mechanical theory for ion transport
  with application to binary solid electrolytes.
\newblock \emph{Journal of the Mechanics and Physics of Solids}, 125:\penalty0
  681--713, 2019.

\bibitem[Gauss(1877)]{gauss1877theoria}
C.~F. Gauss.
\newblock Theoria attractionis corporum sphaeroidicorum ellipticorum
  homogeneorum.
\newblock In \emph{Werke}, pages 3--22. Springer, 1877.

\bibitem[Gillespie et~al.(2002)Gillespie, Nonner, and
  Eisenberg]{gillespie2002coupling}
D.~Gillespie, W.~Nonner, and R.~S. Eisenberg.
\newblock Coupling poisson--nernst--planck and density functional theory to
  calculate ion flux.
\newblock \emph{Journal of Physics: Condensed Matter}, 14\penalty0
  (46):\penalty0 12129, 2002.

\bibitem[Golmon et~al.(2009)Golmon, Maute, and Dunn]{golmon2009numerical}
S.~Golmon, K.~Maute, and M.~L. Dunn.
\newblock Numerical modeling of electrochemical--mechanical interactions in
  lithium polymer batteries.
\newblock \emph{Computers \& Structures}, 87\penalty0 (23-24):\penalty0
  1567--1579, 2009.

\bibitem[Hagiri et~al.(2021)Hagiri, Watanabe, Hiruta, and
  Kashima]{hagiri2021modification}
M.~Hagiri, R.~Watanabe, A.~Hiruta, and K.~Kashima.
\newblock Modification of copper (ii) ion-exchange properties of freestanding
  alginate membrane by embedding with zeolite.
\newblock In \emph{MATEC Web of Conferences}, volume 333, page 04009. EDP
  Sciences, 2021.

\bibitem[Han et~al.(2018)Han, Farino, Yang, Scott, Browe, Choi, Freeman, and
  Lee]{han2018}
D.~Han, C.~Farino, C.~Yang, T.~Scott, D.~Browe, W.~Choi, J.~W. Freeman, and
  H.~Lee.
\newblock Soft robotic manipulation and locomotion with a 3d printed
  electroactive hydrogel.
\newblock \emph{ACS applied materials \& interfaces}, 10\penalty0
  (21):\penalty0 17512--17518, 2018.

\bibitem[Hong et~al.(2008)Hong, Zhao, Zhou, and Suo]{hong2008}
W.~Hong, X.~Zhao, J.~Zhou, and Z.~Suo.
\newblock A theory of coupled diffusion and large deformation in polymeric
  gels.
\newblock \emph{Journal of the Mechanics and Physics of Solids}, 56\penalty0
  (5):\penalty0 1779--1793, 2008.

\bibitem[Hou et~al.(2017)Hou, Zhou, Yang, Li, Zhang, Zhu, Hickner, Zhang, and
  Wang]{hou2017flexible}
Y.~Hou, Y.~Zhou, L.~Yang, Q.~Li, Y.~Zhang, L.~Zhu, M.~A. Hickner, Q.~Zhang, and
  Q.~Wang.
\newblock Flexible ionic diodes for low-frequency mechanical energy harvesting.
\newblock \emph{Advanced Energy Materials}, 7\penalty0 (5):\penalty0 1601983,
  2017.

\bibitem[Jung et~al.(2010)Jung, Vadahanambi, and Oh]{jung2010electro}
J.-H. Jung, S.~Vadahanambi, and I.-K. Oh.
\newblock Electro-active nano-composite actuator based on fullerene-reinforced
  nafion.
\newblock \emph{Composites science and technology}, 70\penalty0 (4):\penalty0
  584--592, 2010.

\bibitem[Kaklamani et~al.(2014)Kaklamani, Cheneler, Grover, Adams, and
  Bowen]{kaklamani2014mechanical}
G.~Kaklamani, D.~Cheneler, L.~M. Grover, M.~J. Adams, and J.~Bowen.
\newblock Mechanical properties of alginate hydrogels manufactured using
  external gelation.
\newblock \emph{Journal of the mechanical behavior of biomedical materials},
  36:\penalty0 135--142, 2014.

\bibitem[Keller et~al.(2011)Keller, Wallmersperger, Kr{\"o}plin, G{\"u}nther,
  and Gerlach]{keller2011modeling}
K.~Keller, T.~Wallmersperger, B.~Kr{\"o}plin, M.~G{\"u}nther, and G.~Gerlach.
\newblock Modeling of temperature-sensitive polyelectrolyte gels by the use of
  the coupled chemo-electro-mechanical formulation.
\newblock \emph{Mechanics of Advanced Materials and Structures}, 18\penalty0
  (7):\penalty0 511--523, 2011.

\bibitem[Kim et~al.(2007)Kim, Ahn, Kim, Lee, Kim, Yu, Nuzzo, and
  Rogers]{kim2007complementary}
D.-H. Kim, J.-H. Ahn, H.-S. Kim, K.~J. Lee, T.-H. Kim, C.-J. Yu, R.~G. Nuzzo,
  and J.~A. Rogers.
\newblock Complementary logic gates and ring oscillators on plastic substrates
  by use of printed ribbons of single-crystalline silicon.
\newblock \emph{IEEE Electron Device Letters}, 29\penalty0 (1):\penalty0
  73--76, 2007.

\bibitem[Kim et~al.(2020)Kim, Chen, Suo, and Hayward]{kim2020ionoelastomer}
H.~J. Kim, B.~Chen, Z.~Suo, and R.~C. Hayward.
\newblock Ionoelastomer junctions between polymer networks of fixed anions and
  cations.
\newblock \emph{Science}, 367\penalty0 (6479):\penalty0 773--776, 2020.

\bibitem[Kim and Tadokoro(2007)]{kim2007electroactive}
K.~J. Kim and S.~Tadokoro.
\newblock Electroactive polymers for robotic applications.
\newblock \emph{Artificial Muscles and Sensors}, 23:\penalty0 291, 2007.

\bibitem[Kornyshev and Vorotyntsev(1981)]{kornyshev1981conductivity}
A.~A. Kornyshev and M.~A. Vorotyntsev.
\newblock Conductivity and space charge phenomena in solid electrolytes with
  one mobile charge carrier species, a review with original material.
\newblock \emph{Electrochimica Acta}, 26\penalty0 (3):\penalty0 303--323, 1981.

\bibitem[Lacour et~al.(2016)Lacour, Courtine, and Guck]{lacour2016materials}
S.~P. Lacour, G.~Courtine, and J.~Guck.
\newblock Materials and technologies for soft implantable neuroprostheses.
\newblock \emph{Nature Reviews Materials}, 1\penalty0 (10):\penalty0 1--14,
  2016.

\bibitem[Lee et~al.(2012)Lee, Lee, Lee, Lee, Lee, and Ko]{lee2012very}
J.~Lee, P.~Lee, H.~Lee, D.~Lee, S.~S. Lee, and S.~H. Ko.
\newblock Very long ag nanowire synthesis and its application in a highly
  transparent, conductive and flexible metal electrode touch panel.
\newblock \emph{Nanoscale}, 4\penalty0 (20):\penalty0 6408--6414, 2012.

\bibitem[Leichsenring and Wallmersperger(2017)]{leichsenring2017modeling}
P.~Leichsenring and T.~Wallmersperger.
\newblock Modeling and simulation of the chemically induced swelling behavior
  of anionic polyelectrolyte gels by applying the theory of porous media.
\newblock \emph{Smart Materials and Structures}, 26\penalty0 (3):\penalty0
  035007, 2017.

\bibitem[Li et~al.(2001)Li, Wlaschin, and Balbuena]{li2001theoretical}
T.~Li, A.~Wlaschin, and P.~B. Balbuena.
\newblock Theoretical studies of proton transfer in water and model polymer
  electrolyte systems.
\newblock \emph{Industrial \& engineering chemistry research}, 40\penalty0
  (22):\penalty0 4789--4800, 2001.

\bibitem[Li et~al.(2017)Li, Li, Liang, Cheng, Dai, Yang, Liu, Zeng, Huang, Luo,
  et~al.]{li2017fast}
T.~Li, G.~Li, Y.~Liang, T.~Cheng, J.~Dai, X.~Yang, B.~Liu, Z.~Zeng, Z.~Huang,
  Y.~Luo, et~al.
\newblock Fast-moving soft electronic fish.
\newblock \emph{Science advances}, 3\penalty0 (4):\penalty0 e1602045, 2017.

\bibitem[Liu and Wang(2014)]{liu2014free}
H.~Liu and Z.~Wang.
\newblock A free energy satisfying finite difference method for
  poisson--nernst--planck equations.
\newblock \emph{Journal of Computational Physics}, 268:\penalty0 363--376,
  2014.

\bibitem[Liu et~al.(2018)Liu, Li, Zhang, Yin, Liu, He, Dai, Shan, Guo, Liu,
  et~al.]{liu2018electrically}
H.~Liu, Q.~Li, S.~Zhang, R.~Yin, X.~Liu, Y.~He, K.~Dai, C.~Shan, J.~Guo,
  C.~Liu, et~al.
\newblock Electrically conductive polymer composites for smart flexible strain
  sensors: a critical review.
\newblock \emph{Journal of Materials Chemistry C}, 6\penalty0 (45):\penalty0
  12121--12141, 2018.

\bibitem[Liu et~al.(2020)Liu, Wang, Ren, Jin, Zhang, Chen, and
  Yan]{liu2020poly}
Z.~Liu, Y.~Wang, Y.~Ren, G.~Jin, C.~Zhang, W.~Chen, and F.~Yan.
\newblock Poly (ionic liquid) hydrogel-based anti-freezing ionic skin for a
  soft robotic gripper.
\newblock \emph{Materials Horizons}, 7\penalty0 (3):\penalty0 919--927, 2020.

\bibitem[Lobo et~al.(2001)Lobo, Valente, Polishchuk, and
  Geuskens]{lobo2001transport}
V.~Lobo, A.~J.~M. Valente, A.~Y. Polishchuk, and G.~Geuskens.
\newblock Transport of non-associated electrolytes in acrylamide hydrogels.
\newblock \emph{Journal of Molecular Liquids}, 94\penalty0 (3):\penalty0
  179--192, 2001.

\bibitem[Lopreore et~al.(2008)Lopreore, Bartol, Coggan, Keller, Sosinsky,
  Ellisman, and Sejnowski]{lopreore2008computational}
C.~L. Lopreore, T.~M. Bartol, J.~S. Coggan, D.~X. Keller, G.~E. Sosinsky, M.~H.
  Ellisman, and T.~J. Sejnowski.
\newblock Computational modeling of three-dimensional electrodiffusion in
  biological systems: application to the node of ranvier.
\newblock \emph{Biophysical journal}, 95\penalty0 (6):\penalty0 2624--2635,
  2008.

\bibitem[Lu et~al.(2010)Lu, Holst, McCammon, and Zhou]{lu2010poisson}
B.~Lu, M.~J. Holst, J.~A. McCammon, and Y.~Zhou.
\newblock Poisson--nernst--planck equations for simulating biomolecular
  diffusion--reaction processes i: Finite element solutions.
\newblock \emph{Journal of computational physics}, 229\penalty0 (19):\penalty0
  6979--6994, 2010.

\bibitem[Markvicka et~al.(2018)Markvicka, Bartlett, Huang, and
  Majidi]{markvicka2018autonomously}
E.~J. Markvicka, M.~D. Bartlett, X.~Huang, and C.~Majidi.
\newblock An autonomously electrically self-healing liquid metal--elastomer
  composite for robust soft-matter robotics and electronics.
\newblock \emph{Nature materials}, 17\penalty0 (7):\penalty0 618--624, 2018.

\bibitem[McNaught et~al.(1997)McNaught, Wilkinson,
  et~al.]{mcnaught1997compendium}
A.~D. McNaught, A.~Wilkinson, et~al.
\newblock \emph{Compendium of chemical terminology}, volume 1669.
\newblock Blackwell Science Oxford, 1997.

\bibitem[Miyake and Rolandi(2015)]{miyake2015grotthuss}
T.~Miyake and M.~Rolandi.
\newblock Grotthuss mechanisms: from proton transport in proton wires to
  bioprotonic devices.
\newblock \emph{Journal of Physics: Condensed Matter}, 28\penalty0
  (2):\penalty0 023001, 2015.

\bibitem[Mizushina(1971)]{mizushina1971electrochemical}
T.~Mizushina.
\newblock The electrochemical method in transport phenomena.
\newblock In \emph{Advances in heat transfer}, volume~7, pages 87--161.
  Elsevier, 1971.

\bibitem[Morales et~al.(2014)Morales, Palleau, Dickey, and
  Velev]{morales2014electro}
D.~Morales, E.~Palleau, M.~D. Dickey, and O.~D. Velev.
\newblock Electro-actuated hydrogel walkers with dual responsive legs.
\newblock \emph{Soft matter}, 10\penalty0 (9):\penalty0 1337--1348, 2014.

\bibitem[Narayan and Anand(2022)]{narayan2022coupled}
S.~Narayan and L.~Anand.
\newblock A coupled electro-chemo-mechanical theory for polyelectrolyte gels
  with application to modeling their chemical stimuli-driven swelling response.
\newblock \emph{Journal of the Mechanics and Physics of Solids}, 159:\penalty0
  104734, 2022.

\bibitem[Narayan et~al.(2021)Narayan, Stewart, and Anand]{narayan2021coupled}
S.~Narayan, E.~M. Stewart, and L.~Anand.
\newblock Coupled electro-chemo-elasticity: Application to modeling the
  actuation response of ionic polymer--metal composites.
\newblock \emph{Journal of the Mechanics and Physics of Solids}, 152:\penalty0
  104394, 2021.

\bibitem[Nardinocchi et~al.(2011)Nardinocchi, Pezzulla, and
  Placidi]{nardinocchi2011thermodynamically}
P.~Nardinocchi, M.~Pezzulla, and L.~Placidi.
\newblock Thermodynamically based multiphysic modeling of ionic polymer metal
  composites.
\newblock \emph{Journal of Intelligent Material Systems and Structures},
  22\penalty0 (16):\penalty0 1887--1897, 2011.

\bibitem[Nemat-Nasser(2002)]{nemat2002micromechanics}
S.~Nemat-Nasser.
\newblock Micromechanics of actuation of ionic polymer-metal composites.
\newblock \emph{Journal of applied Physics}, 92\penalty0 (5):\penalty0
  2899--2915, 2002.

\bibitem[Nernst(1888)]{nernst1888kinetik}
W.~Nernst.
\newblock Zur kinetik der in l{\"o}sung befindlichen k{\"o}rper.
\newblock \emph{Zeitschrift f{\"u}r physikalische Chemie}, 2\penalty0
  (1):\penalty0 613--637, 1888.

\bibitem[Ochi et~al.(2009)Ochi, Kamishima, Mizusaki, and
  Kawamura]{ochi2009investigation}
S.~Ochi, O.~Kamishima, J.~Mizusaki, and J.~Kawamura.
\newblock Investigation of proton diffusion in nafion{\textregistered} 117
  membrane by electrical conductivity and nmr.
\newblock \emph{Solid State Ionics}, 180\penalty0 (6-8):\penalty0 580--584,
  2009.

\bibitem[Odent et~al.(2017)Odent, Wallin, Pan, Kruemplestaedter, Shepherd, and
  Giannelis]{odent2017highly}
J.~Odent, T.~J. Wallin, W.~Pan, K.~Kruemplestaedter, R.~F. Shepherd, and E.~P.
  Giannelis.
\newblock Highly elastic, transparent, and conductive 3d-printed ionic
  composite hydrogels.
\newblock \emph{Advanced Functional Materials}, 27\penalty0 (33):\penalty0
  1701807, 2017.

\bibitem[Paddison and Paul(2002)]{paddison2002nature}
S.~J. Paddison and R.~Paul.
\newblock The nature of proton transport in fully hydrated
  nafion{\textregistered}.
\newblock \emph{Physical Chemistry Chemical Physics}, 4\penalty0 (7):\penalty0
  1158--1163, 2002.

\bibitem[Paddison et~al.(1998)Paddison, Reagor, and
  Zawodzinski~Jr]{paddison1998high}
S.~J. Paddison, D.~W. Reagor, and T.~A. Zawodzinski~Jr.
\newblock High frequency dielectric studies of hydrated
  nafion{\textregistered}.
\newblock \emph{Journal of Electroanalytical Chemistry}, 459\penalty0
  (1):\penalty0 91--97, 1998.

\bibitem[Paz-Garc{\'\i}a et~al.(2011)Paz-Garc{\'\i}a, Johannesson, Ottosen,
  Ribeiro, and Rodr{\'\i}guez-Maroto]{paz2011modeling}
J.~M. Paz-Garc{\'\i}a, B.~Johannesson, L.~M. Ottosen, A.~B. Ribeiro, and J.~M.
  Rodr{\'\i}guez-Maroto.
\newblock Modeling of electrokinetic processes by finite element integration of
  the nernst--planck--poisson system of equations.
\newblock \emph{Separation and Purification Technology}, 79\penalty0
  (2):\penalty0 183--192, 2011.

\bibitem[Planck(1890)]{planck1890ueber}
M.~Planck.
\newblock Ueber die erregung von electricit{\"a}t und w{\"a}rme in
  electrolyten.
\newblock \emph{Annalen der Physik}, 275\penalty0 (2):\penalty0 161--186, 1890.

\bibitem[Poisson(1826)]{poisson1826memoire}
S.-D. Poisson.
\newblock \emph{M{\'e}moire sur la th{\'e}orie du magn{\'e}tisme en movement}.
\newblock L'Acad{\'e}mie, 1826.

\bibitem[Rejovitzky et~al.(2015)Rejovitzky, Di~Leo, and
  Anand]{rejovitzky2015theory}
E.~Rejovitzky, C.~V. Di~Leo, and L.~Anand.
\newblock A theory and a simulation capability for the growth of a solid
  electrolyte interphase layer at an anode particle in a li-ion battery.
\newblock \emph{Journal of the Mechanics and Physics of Solids}, 78:\penalty0
  210--230, 2015.

\bibitem[Rivers et~al.(2002)Rivers, Hudson, and Schmidt]{rivers2002synthesis}
T.~J. Rivers, T.~W. Hudson, and C.~E. Schmidt.
\newblock Synthesis of a novel, biodegradable electrically conducting polymer
  for biomedical applications.
\newblock \emph{Advanced Functional Materials}, 12\penalty0 (1):\penalty0
  33--37, 2002.

\bibitem[Robinson et~al.(2015)Robinson, O’Brien, Zhao, Peele, Larson,
  Mac~Murray, Van~Meerbeek, Dunham, and Shepherd]{robinson2015integrated}
S.~S. Robinson, K.~W. O’Brien, H.~Zhao, B.~N. Peele, C.~M. Larson, B.~C.
  Mac~Murray, I.~M. Van~Meerbeek, S.~N. Dunham, and R.~F. Shepherd.
\newblock Integrated soft sensors and elastomeric actuators for tactile
  machines with kinesthetic sense.
\newblock \emph{Extreme Mechanics Letters}, 5:\penalty0 47--53, 2015.

\bibitem[Rossi et~al.(2018)Rossi, Wallmersperger, Ramirez, and
  Nardinocchi]{rossi2018thermodynamically}
M.~Rossi, T.~Wallmersperger, J.~A. Ramirez, and P.~Nardinocchi.
\newblock Thermodynamically consistent electro-chemo-mechanical model for
  polymer membranes.
\newblock In \emph{Electroactive Polymer Actuators and Devices (EAPAD) XX},
  volume 10594, page 105940K. International Society for Optics and Photonics,
  2018.

\bibitem[Rus and Tolley(2015)]{rus2015design}
D.~Rus and M.~T. Tolley.
\newblock Design, fabrication and control of soft robots.
\newblock \emph{Nature}, 521\penalty0 (7553):\penalty0 467--475, 2015.

\bibitem[Sauerteig et~al.(2018)Sauerteig, Hanselmann, Arzberger, Reinshagen,
  Ivanov, and Bund]{sauerteig2018electrochemical}
D.~Sauerteig, N.~Hanselmann, A.~Arzberger, H.~Reinshagen, S.~Ivanov, and
  A.~Bund.
\newblock Electrochemical-mechanical coupled modeling and parameterization of
  swelling and ionic transport in lithium-ion batteries.
\newblock \emph{Journal of Power Sources}, 378:\penalty0 235--247, 2018.

\bibitem[Schauser et~al.(2019)Schauser, Seshadri, and
  Segalman]{schauser2019multivalent}
N.~S. Schauser, R.~Seshadri, and R.~A. Segalman.
\newblock Multivalent ion conduction in solid polymer systems.
\newblock \emph{Molecular Systems Design \& Engineering}, 4\penalty0
  (2):\penalty0 263--279, 2019.

\bibitem[Schuszter et~al.(2017)Schuszter, Geh{\'e}r-Herczegh, Sz{\H{u}}cs,
  T{\'o}th, and Horv{\'a}th]{schuszter2017determination}
G.~Schuszter, T.~Geh{\'e}r-Herczegh, {\'A}.~Sz{\H{u}}cs, {\'A}.~T{\'o}th, and
  D.~Horv{\'a}th.
\newblock Determination of the diffusion coefficient of hydrogen ion in
  hydrogels.
\newblock \emph{Physical Chemistry Chemical Physics}, 19\penalty0
  (19):\penalty0 12136--12143, 2017.

\bibitem[Shi et~al.(2018)Shi, Zhu, Gao, Zhang, Wei, Liu, and
  Ding]{shi2018highly}
L.~Shi, T.~Zhu, G.~Gao, X.~Zhang, W.~Wei, W.~Liu, and S.~Ding.
\newblock Highly stretchable and transparent ionic conducting elastomers.
\newblock \emph{Nature communications}, 9\penalty0 (1):\penalty0 1--7, 2018.

\bibitem[Shin et~al.(2014)Shin, Song, Lim, Lim, Park, and
  Jeong]{shin2014highly}
M.~Shin, J.~H. Song, G.-H. Lim, B.~Lim, J.-J. Park, and U.~Jeong.
\newblock Highly stretchable polymer transistors consisting entirely of
  stretchable device components.
\newblock \emph{Advanced Materials}, 26\penalty0 (22):\penalty0 3706--3711,
  2014.

\bibitem[Song et~al.(2014)Song, Li, Xu, and Zeng]{song2014superstable}
J.~Song, J.~Li, J.~Xu, and H.~Zeng.
\newblock Superstable transparent conductive cu@ cu4ni nanowire elastomer
  composites against oxidation, bending, stretching, and twisting for flexible
  and stretchable optoelectronics.
\newblock \emph{Nano letters}, 14\penalty0 (11):\penalty0 6298--6305, 2014.

\bibitem[Sun et~al.(2014)Sun, Keplinger, Whitesides, and Suo]{sun2014ionic}
J.-Y. Sun, C.~Keplinger, G.~M. Whitesides, and Z.~Suo.
\newblock Ionic skin.
\newblock \emph{Advanced Materials}, 26\penalty0 (45):\penalty0 7608--7614,
  2014.

\bibitem[Suo et~al.(2008)Suo, Zhao, and Greene]{suo2008nonlinear}
Z.~Suo, X.~Zhao, and W.~H. Greene.
\newblock A nonlinear field theory of deformable dielectrics.
\newblock \emph{Journal of the Mechanics and Physics of Solids}, 56\penalty0
  (2):\penalty0 467--486, 2008.

\bibitem[Tiwari and Kim(2010)]{tiwari2010disc}
R.~Tiwari and K.~Kim.
\newblock Disc-shaped ionic polymer metal composites for use in
  mechano-electrical applications.
\newblock \emph{Smart Materials and Structures}, 19\penalty0 (6):\penalty0
  065016, 2010.

\bibitem[Toi and Kang(2005)]{toi2005finite}
Y.~Toi and S.-S. Kang.
\newblock Finite element analysis of two-dimensional
  electrochemical--mechanical response of ionic conducting polymer--metal
  composite beams.
\newblock \emph{Computers \& structures}, 83\penalty0 (31-32):\penalty0
  2573--2583, 2005.

\bibitem[Tybrandt et~al.(2010)Tybrandt, Larsson, Richter-Dahlfors, and
  Berggren]{tybrandt2010ion}
K.~Tybrandt, K.~C. Larsson, A.~Richter-Dahlfors, and M.~Berggren.
\newblock Ion bipolar junction transistors.
\newblock \emph{Proceedings of the National Academy of Sciences}, 107\penalty0
  (22):\penalty0 9929--9932, 2010.

\bibitem[Wallmersperger et~al.(2004)Wallmersperger, Kr{\"o}plin, and
  G{\"u}lch]{wallmersperger2004coupled}
T.~Wallmersperger, B.~Kr{\"o}plin, and R.~W. G{\"u}lch.
\newblock Coupled chemo-electro-mechanical formulation for ionic polymer
  gels----numerical and experimental investigations.
\newblock \emph{Mechanics of Materials}, 36\penalty0 (5-6):\penalty0 411--420,
  2004.

\bibitem[Wallmersperger et~al.(2008)Wallmersperger, Akle, Leo, and
  Kr{\"o}plin]{wallmersperger2008electrochemical}
T.~Wallmersperger, B.~J. Akle, D.~J. Leo, and B.~Kr{\"o}plin.
\newblock Electrochemical response in ionic polymer transducers: An
  experimental and theoretical study.
\newblock \emph{Composites Science and Technology}, 68\penalty0 (5):\penalty0
  1173--1180, 2008.

\bibitem[Wan et~al.(2019)Wan, Xie, Kong, Liu, Liu, Shi, Pei, Chen, Chen, Chen,
  et~al.]{wan2019ultrathin}
J.~Wan, J.~Xie, X.~Kong, Z.~Liu, K.~Liu, F.~Shi, A.~Pei, H.~Chen, W.~Chen,
  J.~Chen, et~al.
\newblock Ultrathin, flexible, solid polymer composite electrolyte enabled with
  aligned nanoporous host for lithium batteries.
\newblock \emph{Nature nanotechnology}, 14\penalty0 (7):\penalty0 705--711,
  2019.

\bibitem[Wang et~al.(2016)Wang, Decker, Henann, and Chester]{wang2016modeling}
S.~Wang, M.~Decker, D.~L. Henann, and S.~A. Chester.
\newblock Modeling of dielectric viscoelastomers with application to
  electromechanical instabilities.
\newblock \emph{Journal of the Mechanics and Physics of Solids}, 95:\penalty0
  213--229, 2016.

\bibitem[Wang et~al.(2021)Wang, Sun, Zhao, and Zuo]{wang2021highly}
S.~Wang, Z.~Sun, Y.~Zhao, and L.~Zuo.
\newblock A highly stretchable hydrogel sensor for soft robot multi-modal
  perception.
\newblock \emph{Sensors and Actuators A: Physical}, 331:\penalty0 113006, 2021.

\bibitem[Wang et~al.(2019)Wang, Wang, Su, and Cai]{wang2019stretchable}
Y.~Wang, Z.~Wang, Z.~Su, and S.~Cai.
\newblock Stretchable and transparent ionic diode and logic gates.
\newblock \emph{Extreme Mechanics Letters}, 28:\penalty0 81--86, 2019.

\bibitem[Weingaertner(2014)]{weingaertner2014static}
H.~Weingaertner.
\newblock The static dielectric permittivity of ionic liquids.
\newblock \emph{Journal of Molecular Liquids}, 192:\penalty0 185--190, 2014.

\bibitem[Wu et~al.(2009)Wu, Joseph, and Aluru]{wu2009effect}
Y.~Wu, S.~Joseph, and N.~Aluru.
\newblock Effect of cross-linking on the diffusion of water, ions, and small
  molecules in hydrogels.
\newblock \emph{The Journal of Physical Chemistry B}, 113\penalty0
  (11):\penalty0 3512--3520, 2009.

\bibitem[Yan et~al.(2019)Yan, Malakooti, Lu, Wang, Kazem, Pan, Bockstaller,
  Majidi, and Matyjaszewski]{yan2019solution}
J.~Yan, M.~H. Malakooti, Z.~Lu, Z.~Wang, N.~Kazem, C.~Pan, M.~R. Bockstaller,
  C.~Majidi, and K.~Matyjaszewski.
\newblock Solution processable liquid metal nanodroplets by surface-initiated
  atom transfer radical polymerization.
\newblock \emph{Nature nanotechnology}, 14\penalty0 (7):\penalty0 684--690,
  2019.

\bibitem[Yang and Suo(2018)]{yang2018hydrogel}
C.~Yang and Z.~Suo.
\newblock Hydrogel ionotronics.
\newblock \emph{Nature Reviews Materials}, 3\penalty0 (6):\penalty0 125--142,
  2018.

\bibitem[Yang et~al.(2015)Yang, Chen, Lu, Yang, Zhou, Chen, and
  Suo]{yang2015ionic}
C.~H. Yang, B.~Chen, J.~J. Lu, J.~H. Yang, J.~Zhou, Y.~M. Chen, and Z.~Suo.
\newblock Ionic cable.
\newblock \emph{Extreme Mechanics Letters}, 3:\penalty0 59--65, 2015.

\bibitem[Yeom and Oh(2009)]{yeom2009biomimetic}
S.-W. Yeom and I.-K. Oh.
\newblock A biomimetic jellyfish robot based on ionic polymer metal composite
  actuators.
\newblock \emph{Smart materials and structures}, 18\penalty0 (8):\penalty0
  085002, 2009.

\bibitem[Zhang et~al.(2020)Zhang, Dehghany, and Hu]{zhang2020kinetics}
H.~Zhang, M.~Dehghany, and Y.~Hu.
\newblock Kinetics of polyelectrolyte gels.
\newblock \emph{Journal of Applied Mechanics}, 87\penalty0 (6):\penalty0
  061010, 2020.

\bibitem[Zhou et~al.(2017)Zhou, Hou, Li, Yang, Cao, Choi, Wang, and
  Zhang]{zhou2017biocompatible}
Y.~Zhou, Y.~Hou, Q.~Li, L.~Yang, Y.~Cao, K.~H. Choi, Q.~Wang, and Q.~Zhang.
\newblock Biocompatible and flexible hydrogel diode-based mechanical energy
  harvesting.
\newblock \emph{Advanced Materials Technologies}, 2\penalty0 (9):\penalty0
  1700118, 2017.

\end{thebibliography}





\end{document}